\documentclass[letterpaper,twocolumn,aps,pre,floatfix,superscriptaddress,footinbib,preprintnumbers]{revtex4-2}
\usepackage{float}
\usepackage{graphicx} 
\usepackage{dcolumn}
\usepackage{setspace}
\usepackage{bm}
\usepackage{amssymb}
\usepackage{amsmath,mathtools}
\usepackage{comment}
\usepackage{color}
\usepackage{tabularx}
\usepackage{MnSymbol}
\usepackage{slashed}
\usepackage[colorlinks=true,citecolor=blue,linkcolor=black]{hyperref}
\usepackage[nameinlink]{cleveref} 
\usepackage[dvipsnames]{xcolor}
\usepackage{braket}
\usepackage[export]{adjustbox}
\usepackage{multirow}
\usepackage{rotating,booktabs} 
\usepackage[caption=false]{subfig}
\usepackage [latin1]{inputenc} 
\restylefloat{table}
\usepackage{accents} 
\usepackage{bbm} 
\usepackage{soul}
\usepackage{nicefrac} 
\usepackage[T1]{fontenc} 
\usepackage{parskip} 
\usepackage{enumitem}   

\newcommand{\sur}{{S \cup R}}
\newcommand{\Tr}{{\mathrm{Tr}}}
\newcommand{\mi}{{A}}
\newcommand{\f}{{B}}
\newcommand{\tot}{\mathrm{diff}}
\newcommand{\h}{\hat{H}}
\newcommand{\pio}{\hat{\pi}}
\newcommand{\rhoo}{\hat{\rho}}
\newcommand{\hso}{\hat{H}_S^*}
\newcommand{\hsoi}{\hat{H}_S^{*A}}
\newcommand{\hsof}{\hat{H}_S^{*B}}
\newcommand{\eso}{\hat{E}_S^*}
\newcommand{\esoi}{\hat{E}_S^{*A}}
\newcommand{\esof}{\hat{E}_S^{*B}}

\crefname{equation}{Eq.}{Eqs.}
\Crefname{equation}{Equation}{Equations} 
\crefname{pluralequation}{Eqs.}{Eqs.} 
\crefformat{pluralequation}{Eqs.~(#2#1#3)}
\Crefname{pluralequation}{Equations}{Equations}
\Crefformat{pluralequation}{Equations~(#2#1#3)}
\crefname{figure}{Fig.}{Figs.}
\Crefname{figure}{Figure}{Figures}
\crefname{tabular}{Tab.}{Tabs.}
\crefname{section}{Sec.}{Secs.}
\Crefname{section}{Section}{Sections}
\Crefname{appendix}{Appendix}{Appendices}

\crefrangeformat{equation}
  {Eqs.~(#3#1#4)--(#5#2#6)}
\Crefrangeformat{equation}
  {Equations~(#3#1#4)--(#5#2#6)}

\begin{document}

\title{Work and heat exchanged during sudden quenches
\\
of strongly coupled quantum systems}
\author{Zohreh~Davoudi}
\email{davoudi@umd.edu}
\affiliation{Department of Physics, University of Maryland, College Park, Maryland 20742, USA}
\affiliation{Maryland Center for Fundamental Physics, University of Maryland, College Park, Maryland 20742, USA}
\affiliation{Joint Center for Quantum Information and Computer Science, NIST and University of Maryland, College Park, Maryland 20742, USA}
\affiliation{NSF Institute for Robust Quantum Simulation, University of Maryland, College Park, Maryland 20742, USA}

\author{Christopher~Jarzynski}
\email{cjarzyns@umd.edu }
\affiliation{Department of Chemistry and Biochemistry, University of Maryland, College Park, Maryland 20742, USA}
\affiliation{Institute for Physical Science and Technology, University of Maryland, College Park, Maryland 20742, USA}
\affiliation{Department of Physics, University of Maryland, College Park, Maryland 20742, USA}
\affiliation{NSF Institute for Robust Quantum Simulation, University of Maryland, College Park, Maryland 20742, USA}

\author{Niklas~Mueller}
\email{niklasmu@unm.edu}
\affiliation{InQubator for Quantum Simulation (IQuS), Department of Physics,
University of Washington, Seattle, Washington 98195, USA}
\affiliation{Center for Quantum Information and Control, University of New Mexico, Albuquerque, New Mexico 87106, USA}
\affiliation{Department of Physics and Astronomy, University of New Mexico, Albuquerque, New Mexico 87106, USA}

\author{Greeshma~Oruganti}
\email{gshivali@umd.edu}
\thanks{corresponding author.}
\affiliation{Institute for Physical Science and Technology, University of Maryland, College Park, Maryland 20742, USA}
\affiliation{NSF Institute for Robust Quantum Simulation, University of Maryland, College Park, Maryland 20742, USA}

\author{Connor~Powers}
\email{cdpowers@umd.edu}
\affiliation{Department of Physics, University of Maryland, College Park, Maryland 20742, USA}
\affiliation{Maryland Center for Fundamental Physics, 
University of Maryland, College Park, Maryland 20742, USA}
\affiliation{Joint Center for Quantum Information and Computer Science, NIST and University of Maryland, College Park, Maryland 20742, USA}
\affiliation{NSF Institute for Robust Quantum Simulation, University of Maryland, College Park, Maryland 20742, USA}

\author{Nicole~Yunger~Halpern}
\email{nicoleyh@umd.edu}
\affiliation{Joint Center for Quantum Information and Computer Science, NIST and University of Maryland, College Park, Maryland 20742, USA}
\affiliation{Institute for Physical Science and Technology, University of Maryland, College Park, Maryland 20742, USA}
\affiliation{NSF Institute for Robust Quantum Simulation, University of Maryland, College Park, Maryland 20742, USA}

\preprint{UMD-PP-025-01, IQuS@UW-21-097.}

\begin{abstract}
How should one define thermodynamic quantities (internal energy, work, heat, etc.) for quantum systems coupled to their environments strongly? We examine three definitions, inspired by the literature, of a quantum system's internal energy under strong-coupling conditions. Each internal-energy definition implies a definition of work and a definition of heat. Our study focuses on quenches, common processes in which the Hamiltonian changes abruptly. In these processes, the first law of thermodynamics holds for each set of definitions by construction. However, we prove that only two sets obey the second law. We illustrate our findings using a simple spin model. Our results guide studies of thermodynamic quantities in strongly coupled quantum systems.
\end{abstract}

\maketitle

\section{Introduction}
Quantum thermodynamics generalizes 19th-century principles about energy processing to quantum systems~\cite{goold_role_2016,vinjanampathy_quantum_2016,millen_perspective_2016,alicki_introduction_2018,binder_thermodynamics_2018}. The field has advanced the studies of thermalization~\cite{gogolin_equilibration_2016,majidy_noncommuting_2023}, fluctuation theorems (extensions of the second law)~\cite{tasaki_jarzynski_2000,kurchan_quantum_2001,mukamel_quantum_2003,campisi_colloquium_2011}, thermal machines~\cite{mitchison_quantum_2019,mukherjee_many-body_2021,maringuzman_key_2024}, energetic and informational resources~\cite{chitambar_quantum_2019,lostaglio_introductory_2019}, and more. Controlled quantum systems have enabled experimental tests of quantum-thermodynamic predictions~\cite{an_experimental_2015,xiong_experimental_2018,schuckert_probing_2023,maringuzman_key_2024,kaufman_quantum_2016,kranzl_experimental_2023,hahn_quantum_2023}. Yet how to define basic quantum-thermodynamic quantities (internal energy, work, heat, etc.) remains an open question~\cite{alicki_quantum_1979,kosloff_quantum_1984,talkner_fluctuation_2009,alipour_correlations_2016,ahmadi_contribution_2023,binder_quantum_2015,colla_open-system_2022,gallego_thermal_2014,guarnieri_quantum_2019,rivas_quantum_2019,rivas_strong_2020,silva_quantum_2021,sone_quantum_2020,strasberg_non-markovianity_2019,talkner_aspects_2016,anto-sztrikacs_effective-hamiltonian_2023,dann_unification_2023,deffner_quantum_2016,webb_how_2024,kumar_work_2024}.

A standard thermodynamic setting features two subsystems: a system $S$ of interest and a thermal reservoir $R$. Typically, $S$ and $R$ satisfy the {\it weak-coupling assumption}: the subsystems' interaction energy is negligible compared to the system's and reservoir's energies. The total system-reservoir internal energy then approximately equals the sum of the subsystems' internal energies. This decomposition motivates the notion of heat as the energy lost by $R$ and gained by $S$~\cite{feynman_feynman_2010,callen_thermodynamics_1985}.

Macroscopic systems obey the weak-coupling assumption. The reason is that the interaction energy scales as the system-reservoir boundary's surface area, while the system's and reservoir's energies scale with the subsystems' volumes. When $S$ is microscopic, this argument does not apply, and the weak-coupling assumption can break down. The weak-coupling assumption can break also under long-range interactions between a system's and reservoir's degrees of freedom (DOFs). These examples fall into the \emph{strong-coupling regime}. In this regime, whether the interaction energy should be attributed to $S$ or to $R$, or somehow split between the two, is unclear~\footnote{One could attribute the interaction energy to neither $S$ nor $R$. This approach falls outside the standard thermodynamic understanding of heat as energy lost by $R$ to $S$. Therefore, we will not consider this approach.}. For classical systems, thermodynamic quantities (internal energy, entropy, heat, work, etc.) can nonetheless be defined in the strong-coupling regime consistently with the first and second laws of thermodynamics~\cite{seifert_first_2016,jarzynski_stochastic_2017,miller_hamiltonian_2018}.

Several approaches have been proposed for extending the strong-coupling framework into the quantum regime~\cite{rivas_quantum_2019,rivas_strong_2020,miller_hamiltonian_2018,campisi_thermodynamics_2009,strasberg_non-markovianity_2019,hsiang_quantum_2018,rivas_refined_2017,garciamarch_non-equilibrium_2016,perarnau_strong_2018}. We collate three candidate definitions for internal energy~\cite{seifert_first_2016,jarzynski_nonequilibrium_2004,miller_hamiltonian_2018}. Classically, these definitions lead to work and heat definitions that obey the second law. In the quantum case, we prove, only two definitions satisfy the second law; the third definition does not. (All internal-energy, heat, and work quantities in this paper are averages, i.e., expectation values.) Our proof applies to quench processes, in which the Hamiltonian changes abruptly. Such processes enable us to naturally partition the internal-energy change into work and heat. In summary, our results advance quantum thermodynamics by defining quantities consistently with thermodynamic laws.

Our paper is organized as follows: \Cref{sec:setup,sec:intenergy} specify the setup and three internal-energy definitions. \Cref{sec:quench} describes our quench processes, as well as the partitioning of internal-energy changes into heat and work. \Cref{sec:wandh} identifies the definitions that obey the second law. A spin model illustrates our findings in \cref{sec:spinmodel}.

\section{Preliminaries}\label{sec:setup}
Consider a finite quantum system $\sur$ composed of subsystems $S$ and $R$. The Hamiltonian of $\sur$ is 
\begin{align}\label{eq:hsurgen}
    \h_{\sur} \coloneqq \h_S \otimes \hat{\mathbbm{1}}_R + \hat{\mathbbm{1}}_S \otimes \h_R + \hat{V}_{\sur} \, .
\end{align}
$\h_S$ and $\hat{\mathbbm{1}}_S$ ($\h_R$ and $\hat{\mathbbm{1}}_R$) act on the Hilbert space of $S$ ($R$). $\hat{V}_{\sur}$ denotes the interaction between the subsystems. Throughout this paper, we call $S$ the system of interest or the \emph{system} and call $R$ the \textit{reservoir}. Also, we use the shorthand $\h_{\sur} = \h_S + \h_R + \hat{V}_{\sur}$. 

We denote by $\rhoo_{\sur}$ a state, pure or mixed, of ${\sur}$. The system's reduced state follows from tracing out the reservoir: $\rhoo_S \coloneqq \Tr_{R} \left( \rhoo_{\sur}\right)$. We suppose that the global equilibrium state is the Gibbs state
\begin{equation}\label{eq:pisur}
    \pio_{\sur} \coloneqq \frac{e^{-\beta \h_{\sur}}}{Z_{\sur}} \, .
\end{equation}
$\beta \coloneqq (k_{\rm B} T)^{-1}$ denotes the inverse temperature; $k_{\rm B}$, the Boltzmann factor; and $T$, the temperature. The partition function is $Z_{\sur} \coloneqq \Tr_{SR} ( e^{-\beta \h_{\sur}} )$. Consequently, the system's equilibrium state is
\begin{align}\label{eq:pisop}
    \pio_{S} \coloneqq \Tr_R \left( \pio_{\sur} \right) 
    \equiv \frac{e^{-\beta \hso}}{Z_S^{*}} \, . 
\end{align}
This equation introduces the \textit{Hamiltonian of mean force}~\cite{miller_hamiltonian_2018,campisi_thermodynamics_2009,miller_entropy_2017,philbin_thermal_2016,trushechkin_open_2022,campisi_fluctuation_2009},
\begin{equation}\label{eq:hstar}
    \hso \coloneq -\frac{1}{\beta} \ln{ \left( \frac{\Tr_R \left(e^{-\beta \h_{\sur}} \right)}{ \Tr_R \left( e^{-\beta \h_R}\right)} \right) }
\end{equation}
and partition function $Z_S^{*} \coloneqq \Tr_S ( e^{-\beta \hso}) $. Throughout this paper, we denote equilibrium states by $\pio$, as in \cref{eq:pisur,eq:pisop}. We consider thermodynamic processes that begin in the global equilibrium state $\pio_\sur$.

When $\hat{V}_{\sur}$ is negligible, one can treat $\pio_{\sur}$ as a product $\pio_S^0 \otimes \pio_R^0$ of system and reservoir Gibbs states,
\begin{align}\label{eq:bareSandRstate}
    \pio_S^{0} 
    \coloneqq \frac{e^{-\beta \h_S}}{Z_S} 
    \quad \text{and} \quad 
    \pio_R^{0} \coloneqq \frac{e^{-\beta \h_R}}{Z_R} \, ,
\end{align}
wherein $Z_S \coloneqq \Tr_{S}( e^{-\beta \h_S })$ and $Z_R \coloneqq \Tr_{R}( e^{-\beta \h_R })$. In this {\it weak-coupling} regime, $\hso \approx \h_S$. When $\hat{V}_{\sur}$ is non-negligible, $\hso$ acts as a modified system Hamiltonian in \cref{eq:pisop}, capturing the interaction's effects on the system's equilibrium state.

To the equilibrium states $\pio_{\sur}$, $\pio_{S}$, and $\pio_{R}^0$, we ascribe the free energies~\cite{miller_hamiltonian_2018} 
\begin{equation} \label{eq:F_SUR}
    F_{\sur} \coloneq - \frac{1}{\beta} \ln{ \left( Z_{\sur}  \right)} \, , 
\end{equation}
\begin{equation}\label{eq:F_S}
    F_{S} \coloneq -\frac{1}{\beta} \ln{\left( Z_S^*\right)} \, ,
\end{equation}
and
\begin{equation}\label{eq:F_R}
    F_{R} \coloneq -\frac{1}{\beta} \ln{\left( Z_R\right)} \, .
\end{equation}
By the identity $ Z_{\sur} = Z_{S}^{*} Z_R$~\cite{jarzynski_nonequilibrium_2004,campisi_thermodynamics_2009}, \cref{eq:F_SUR,eq:F_S,eq:F_R} imply $F_{\sur} = F_{S} + F_{R}$. In this additive relation, all effects of $V_{\sur}$ are bundled into $F_S$, while $F_R$ does not depend on the system-reservoir interaction.  This convention for defining system and reservoir free energies follows the classical stochastic-thermodynamics literature. There, the convention implies the second law~\cite{seifert_first_2016,jarzynski_nonequilibrium_2004}.

The total internal energy in $\rhoo_{\sur}$ is defined as the expectation value of $\h_{\sur}$:
\begin{equation}\label{eq:usur}
    U_{\sur} \coloneq \Tr_{SR} \left( \h_{\sur} \; \rhoo_{\sur} \right) \, .
\end{equation}
To state the first and the second laws of thermodynamics, we need also a definition of the system internal energy $U_S$. If the system-reservoir coupling is strong, the definition of $U_S$ is unclear. The reason is that $U_{\sur}$ does not partition neatly into system and reservoir contributions: part of the total energy resides in the interaction term, which $S$ and $R$ share. In the next section, we discuss three possible definitions of $U_S$.

\section{Three definitions of internal energy} \label{sec:intenergy}

We draw three candidate definitions for the system's internal energy, $U_S$, from the strong-coupling-thermodynamics literature~\cite{jarzynski_nonequilibrium_2004,miller_hamiltonian_2018,seifert_first_2016,jarzynski_stochastic_2017}. These definitions are expressed in terms of expectation values of energy operators, representing the average internal energy of $S$.

The first definition comes from classical stochastic thermodynamics~\cite{seifert_first_2016}. The system's internal energy equals the difference between the total internal energy and an isolated reservoir's energy:
\begin{equation}\label{eq:utot}
    U_{\tot} \coloneqq U_{\sur} - U_R^0 \, .
\end{equation}
$ U_R^{0} \coloneqq \Tr_{R} \left( \h_R  \pio_R^{0} \right)$ is the reservoir's equilibrium internal energy in the system's absence.~\Cref{eq:utot} portrays $U_R^{0}$ as a fixed reference energy for the reservoir. The remaining energy of $\sur$, including contributions from $\hat{V}_{\sur}$, is assigned to $S$. When $\hat{V}_{\sur}$ is non-negligible, $U_{\tot}$ differs from $U_S^0 \coloneqq \Tr_S \left(\hat{H}_S \; \pio_S^0\right)$, the system's equilibrium internal energy in the reservoir's absence.

The second $U_S$ definition portrays $\hso$ as an effective energy operator. Thus, the system's internal energy is the expectation value of this operator~\cite{jarzynski_nonequilibrium_2004,jarzynski_stochastic_2017}:
\begin{equation}\label{eq:uhstar}
    U_{H^*} \coloneqq \Tr_S \left( \hso \; \rhoo_S \right) \, .
\end{equation}

Finally, Refs.~\cite{seifert_first_2016,miller_hamiltonian_2018} introduce another effective system Hamiltonian, 
\begin{align}
    \label{eq:estar}
    \eso \coloneqq \partial_{\beta} ( \beta \hso ) \, .
\end{align}
This operator leads to the third definition of the system's internal energy,
\begin{equation}\label{eq:uestar}
    U_{E^{*}} \coloneqq \Tr_S \left(\eso \; \rhoo_S \right) \, .
\end{equation}

In summary, $U_{\tot}$, $U_{H^*}$, and $U_{E^*}$ represent plausible definitions of the internal energy of a system in any state $\rhoo_S$. These definitions can lead to different internal-energy values. Under certain conditions, however, the definitions are equivalent. We show in \Cref{app:weakcoupling} that $U_{\tot}=U_{H^*}=U_{E^*}=U_S^0$ if two conditions are met: (i) $\sur$ is in a Gibbs state, $\pio_\sur$, and (ii) $V_\sur$ is negligible. If (ii) is satisfied but (i) is not, then $U_{H^*} = U_{E^*}$, but both can differ from $U_{\tot}$ (\Cref{app:weakcoupling}). Conversely, if (i) is satisfied but (ii) is not, then $U_{\tot} = U_{E^*}$, but both can differ from $U_{H^*}$ (\Cref{app:utotequestar})~\footnote{Reference~\cite{seifert_first_2016} introduced the classical analog of $\eso$ as an internal-energy observable. The justification relied on the equivalence of $U_{\tot}$ and $U_{E^*}$ under condition (i).}.

Later, we focus on processes in which $\sur$ begins in a global Gibbs state, $\pio_{\sur}$, at a given temperature. This temperature---often the problem's only well-defined temperature~\cite{jarzynski_nonequilibrium_2004}---appears implicitly in the definitions of $U_{\tot}$, $U_{H^*}$, and $U_{E^*}$. This approach follows the one in classical stochastic thermodynamics. There, the second law of thermodynamics and variations thereon, for systems driven away from an initial equilibrium state, depend on the initial state's temperature~\cite{jarzynski_equalities_2011,seifert_stochastic_2012}.

\section{Quench processes}
\label{sec:quench}
We now consider quench processes in which $\sur$ begins in a global Gibbs state at time $t =0^-$. Each such process consists of two stages. First, at $t=0$, a quench occurs: the Hamiltonian, $\h_{\sur}$, is changed abruptly. Then, from $t=0^+$ to $t_{\rm{f}}$, $\sur$ evolves under the new, fixed, Hamiltonian. The quench process allows us to naturally separate the system's internal-energy change into work and heat, as we show shortly. Furthermore, quench experiments have verified quantum-thermodynamic predictions~\cite{joshi_exploring_2023,fusco_assessing_2014,jurcevic_direct_2017,eisert_quantum_2015,qi_measuring_2019,kranzl_experimental_2023,alba_entanglement_2017,dorner_emergent_2012,joshi_quantum_2013,canovi_quantum_2011}, including about nonequilibrium phenomena~\cite{eisert_quantum_2015,zvyagin_dynamical_2016,mitra_quantum_2018,heyl_dynamical_2018}, quantum ergodicity~\cite{arrais_quantum_2018}, and quantum phases~\cite{degrandi_quench_2010,zhang_obersvation_2017}.

\begin{figure*} 
    \centering
    \includegraphics[scale=0.25]{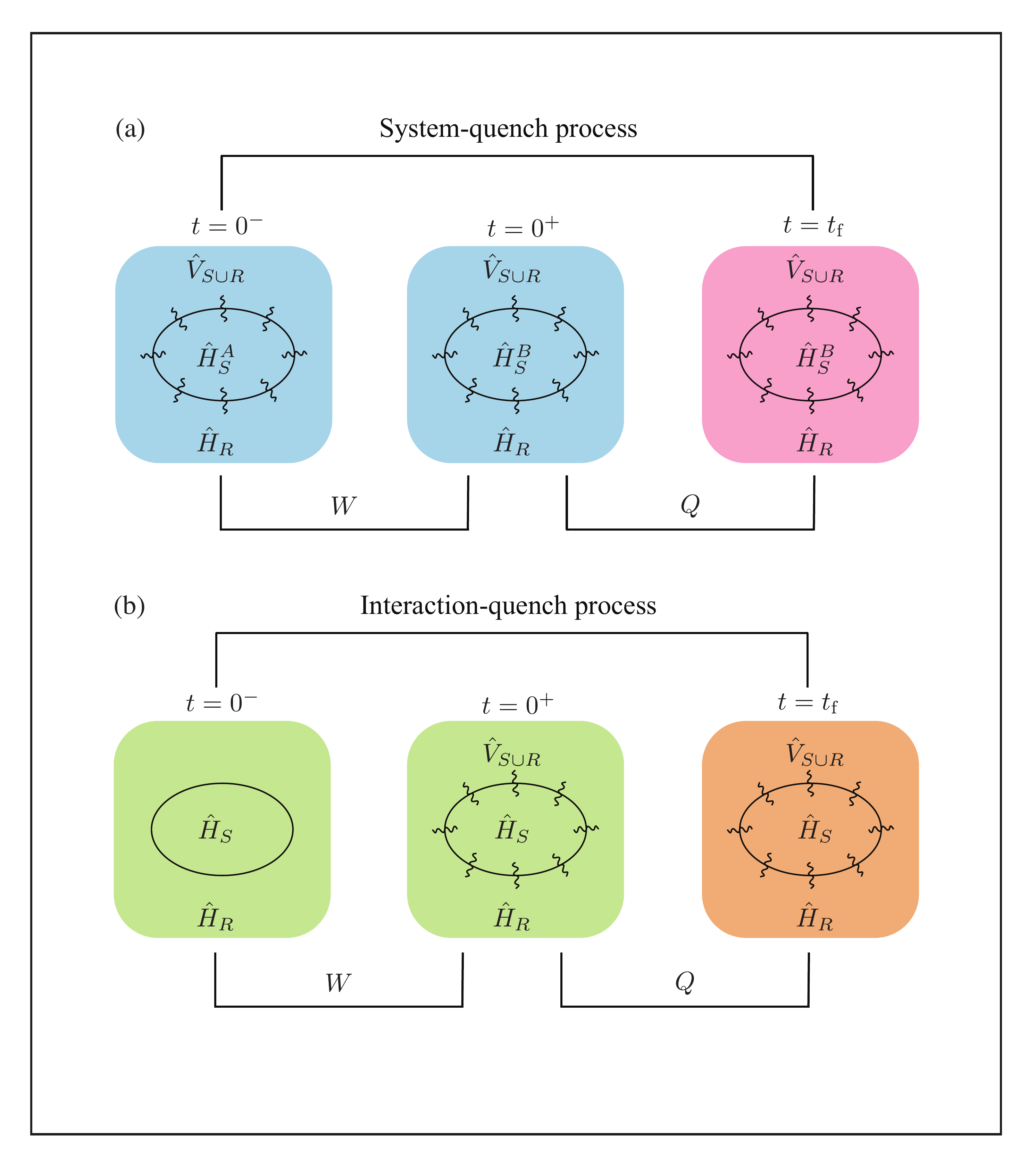}
    \caption{\textit{Quench processes.} 
    (a) A \textit{system-quench} process begins with a Gibbs state relative to the Hamiltonian $\h_\sur = \h_S^{\mi} + \h_R + \hat{V}_{\sur}$. The system's Hamiltonian is quenched to $\h_S^{\f}$ at $t=0$. $\sur$ then evolves until time $t_{\rm f}$. (b) An \textit{interaction-quench} process begins with a separable Gibbs state relative to the Hamiltonian $\h_{\sur} = \h_{S} + \h_R$. The interaction, $\hat V_\sur$, is turned on at $t=0$. In both processes, the quench injects work $W$ into the system. The system absorbs heat $Q$ during the subsequent evolution. A colored box represents $\sur$, and a black ellipse encloses $S$. Identical colors represent identical states. Squiggly lines indicate the interaction, $\hat{V}_{\sur}$.}
    \label{fig:quench}
\end{figure*}

To describe the quench, we introduce a parameter $\lambda$ whose value changes from $A$ to $B$ at $t=0$. We analyze two quench scenarios. In the first, $\h_{\sur}$ depends on $\lambda$ through the system Hamiltonian:
\begin{equation}\label{eq:qp1hsur}
    \h_{\sur}^{\lambda} \coloneq \h_{S}^{\lambda} + \h_R + \hat{V}_{\sur} \, .
\end{equation}
We call this scenario a \emph{system quench}. In the second scenario, $\h_{\sur}$ depends on $\lambda$ through the interaction term: 
\begin{equation}\label{eq:qp2hsur}
    \h_{\sur}^\lambda \coloneq \h_S + \h_R + \lambda\hat{V}_{\sur} \, .
\end{equation}
Here, $\lambda$ changes from $\mi=0$ to $\f=1$, turning on the system-reservoir interaction. We call this scenario an \emph{interaction quench}.~\Cref{tab:quench} summarizes, and~\cref{fig:quench} illustrates, the two quench processes.
\par 
\begin{table} 
    \centering 
    \begin{tabular}{ccc} 
        \toprule \toprule
         \addlinespace
        \multicolumn{3}{c}{System-quench process} \\
        \addlinespace
        \midrule
        Time & $\h_{\sur}(t)$ & $\rhoo_S(t)$ \\
        \midrule
        $t=0^{-}$ & $\h_S^{A} + \h_R + \hat{V}_{\sur}$ &  $\pio_{S}^{A}$ \\
        $t=0^{+}$ & $\h_S^{B} + \h_R + \hat{V}_{\sur}$ & $\pio _{S}^{A}$ \\
        $t = t_{\rm f} $ & $\h_S^{B} + \h_R + \hat{V}_{\sur}$ & $\rhoo_{S}(t_{\rm f})$  \\
        \bottomrule
        \addlinespace
        \multicolumn{3}{c}{Interaction-quench process} \\ 
        \addlinespace
        \midrule
        Time & $\h_{\sur}(t)$ & $\rhoo_S(t)$ \\
        \midrule
        $t=0^{-}$ & $\h_S + \h_R$ & $\pio_{S}^{A} = \pio _{S}^{0}$ \\
        $t=0^{+}$ & $\h_S + \h_R + \hat{V}_{\sur}$ & $\pio_{S}^{A} = \pio_{S}^{0}$ \\
        $t = t_{\rm f} $ & $\h_S + \h_R + \hat{V}_{\sur}$ & $\rhoo_{S}(t_{\rm f})$ \\
        \bottomrule \bottomrule
    \end{tabular}
    \caption{Overview of system- and interaction-quench processes.} 
    \label{tab:quench}
\end{table}

These processes involve the following states. The global Gibbs state, relative to the Hamiltonian $\h_{\sur}^{\lambda}$, is $\pio_{\sur}^{\lambda}$. The system's corresponding equilibrium state is
\begin{equation}
    \pio_{S}^{\lambda} \coloneqq \Tr_R(\pio_{\sur}^{\lambda}) 
    \equiv \frac{e^{-\beta \hat{H}_S^{*\lambda}}}{Z_S^{*\lambda}} \, .
\end{equation}
At time $t$, $\sur$ and $S$ are in the states $\rhoo_{\sur}(t)$ and $\rhoo_S(t)$, respectively. By assumption, $\sur$ begins in equilibrium: $\rhoo_{\sur}(0^-)= \pio_{\sur}^{\mi}$. In the interaction quench, $\pio_{\sur}^{\mi}=\pio_S^{0} \otimes \pio_R^{0}$, wherein \cref{eq:bareSandRstate} specifies $\pio_S^0$ and $\pio_R^0$. The state of $\sur$ does not change during either quench, by the sudden approximation~\footnote{The quench satisfies the sudden approximation~\cite{sakurai_modern_1993} because it occurs instantaneously.}:
\begin{equation}\label{eq:qpsuddenapprox}
    \rhoo_S (0^+) = \rhoo_S (0^-) = \pio_S^A \, .
\end{equation} 
From $t=0^+$ to $t_{\mathrm{f}}$, $\sur$ evolves under a fixed Hamiltonian: $\h_\sur(t) = \h_{\sur}^{\f}$. We discuss this evolution at the end of \cref{sec:wandh}.

\section{Work, heat, and the second law} \label{sec:wandh}
We now identify the work $W$ performed on, and the heat $Q$ absorbed by, the system during a quench process. The three internal-energy definitions (\cref{sec:intenergy}) imply corresponding definitions of work and heat. As in classical contexts~\cite{jarzynski_nonequilibrium_1997,jarzynski_nonequilibrium_2004,jarzynski_stochastic_2017,seifert_first_2016}, we determine which definitions satisfy the first and second laws of thermodynamics.

$U_S(t)$ denotes the system's internal energy at time $t$, wherein $U_S$ denotes one of the candidate definitions: $U_{\tot}$, $U_{H*}$, or $U_{E^*}$. During a quench process, the system's internal energy changes by an amount
\begin{equation}\label{eq:delu}
    \Delta U_S \coloneqq U_S\left(t_{\rm{f}} \right) - U_S\left(0^- \right) \, .
\end{equation}
The first law of thermodynamics attributes the energy change to work and heat:
\begin{equation}\label{eq:firstlaw}
    \Delta U_S = W + Q \, .
\end{equation}
During the quench, no energy flows between the system and the reservoir, as neither subsystem's state changes. Consequently, $S$ absorbs no heat; any change in the system's internal energy (due to the sudden change in the Hamiltonian) is interpreted as \emph{work}: 
\begin{equation}\label{eq:wdef}
    W \coloneqq  U_S \left(0^+ \right) - U_S \left(0^- \right) \, .
\end{equation}
After the quench (from $t=0^+$ to $t_{\rm{f}}$), the state $\rhoo_\sur$ evolves under a fixed Hamiltonian, $\h_{\sur}^{\f}$. The energy transferred between $S$ and $R$ is interpreted as \emph{heat}~\cite{seifert_stochastic_2012,peliti_stochastic_2021}: 
\begin{equation}\label{eq:qdef}
    Q \coloneqq U_S \left(t_{\rm{f}} \right) - U_S \left(0^+ \right)  \, .
\end{equation}
\Cref{eq:wdef,eq:qdef} satisfy the first law [\cref{eq:firstlaw}] by construction. They are motivated by the standard definitions used in weak-coupling quantum thermodynamics~\cite{schrodinger_statistical_1989,alicki_quantum_1979,peliti_stochastic_2021}.

Different choices for $U_S$ in \cref{eq:wdef}---$U_{\tot}$, $U_{H^*}$, and $U_{E^*}$---lead to different expressions for the work performed on $S$:
\begin{equation}\label{eq:wtotoperator}
    W_{\tot} \coloneqq \Tr_{SR} \left( \left[ \h_{\sur}^{B} - \h_{\sur}^{A} \right] \; \pio_{\sur}^{A} \right) \, ,
\end{equation}
\begin{equation}\label{eq:whstaroperator}
    W_{H^*} \coloneqq \Tr_{S} \left( \left[ \hsof  - \hsoi \right] \; \pi_{S}^{A} \right) \, ,
\end{equation}
and 
\begin{equation}\label{eq:westaroperator}
    W_{E^*} \coloneqq \Tr_S \left( \left[ \esof - \esoi \right] \; \pio_S^{A} \right) \, .
\end{equation}
In each of these definitions, work equals the internal-energy change arising from the sudden change in an energy operator. \Cref{eq:wtotoperator} equates the work performed on $S$ with the work performed on $\sur$, consistently with the treatment of $\sur$ as one system of interest.
If $R$ is much larger than $S$, however, $W_\tot$ is often a small difference between two large energies. Additionally, measuring $W_\tot$ requires access to the $S$ and $R$ degrees of freedom.
In these respects, \cref{eq:whstaroperator,eq:westaroperator} are closer in spirit to the notion of work in classical thermodynamics:
to measure work, one does not subtract system-and-reservoir energies.

Similarly, different $U_S$ definitions yield different definitions for the heat absorbed by $S$:
\begin{equation}\label{eq:qtotoperator}
    Q_{\tot} \coloneqq \Tr_{SR} \left( \h_{\sur}^{B} \left[ \rhoo_{\sur}(t_{\rm f}) - \pio_{\sur}^{A} \right] \right) \, ,
\end{equation}
\begin{equation}\label{eq:qhstaroperator}
    Q_{H^*} \coloneqq \Tr_S \left( \hsof \left[ \rhoo_{S}(t_{\rm f}) - \pio_S^{A} \right] \right) \, ,
\end{equation}
and 
\begin{equation}\label{eq:qestaroperator}
    Q_{E^*} \coloneqq \Tr_S \left( \esof \left[ \rhoo_{S}(t_{\rm f}) - \pio_S^{A} \right] \right) \, . 
\end{equation}
%
In classical thermodynamics, heat is the energy exchanged by $S$ and $R$ during the change in $S$'s state. 
\Cref{eq:qhstaroperator,eq:qestaroperator} are consistent with this notion, whereas \cref{eq:qtotoperator} is not. If $\sur$ is isolated, \cref{eq:qtotoperator} implies $Q_\tot=0$, even though $S$ and $R$ can exchange energy throughout the time interval $t \in (0, t_{\rm f}]$. Thus, representing $S$'s energy with $U_\tot$ may preclude a heat definition that simultaneously satisfies the first law and represents heat as the energy exchanged by $S$ and $R$~\footnote{If $\sur$ is not isolated, then $Q_\tot$ is the change in $\sur$'s energy between $t=0^+$ and $t=t_{\rm f}$. This change represents an exchange of heat not by $S$ and $R$, but by $\sur$ and an external heat bath.}. 

In summary, all three sets of work [\cref{eq:wtotoperator,eq:whstaroperator,eq:westaroperator}] and heat definitions [\cref{eq:qtotoperator,eq:qhstaroperator,eq:qestaroperator}] satisfy the first law by construction: within each set, work plus heat equals the corresponding internal-energy change. We now identify which definitions obey the second law.

To state the second law, we introduce further notation. Denote by $F_S^{\mi}\coloneqq -\beta^{-1}\ln Z_S^{A*}$ the free energy of $S$ in the equilibrium state relative to the initial Hamiltonian. Define $F_S^{\f}$ analogously relative to the final Hamiltonian. Because $S$ begins in the equilibrium state $\pio_S^{\mi}$, the second law of thermodynamics assumes the form~\footnote{\Cref{eq:genwsecondlaw} commonly expresses the second law when a system (in contact with one thermal reservoir) begins and ends in equilibrium. This statement remains true if the system ends in a nonequilibrium state, upon beginning in equilibrium. See comment 2 at the end of Sec.~2.1 in Ref.~\cite{jarzynski_equalities_2011}.}
\begin{equation}\label{eq:genwsecondlaw}
    W \geq \Delta F_{S} \, , 
\end{equation}
wherein $\Delta F_S \coloneqq F_S^{\f} - F_S^{\mi}$. In classical thermodynamics, \cref{eq:genwsecondlaw} is the statement of the second law for a system in contact with one thermal reservoir~\cite{callen_thermodynamics_1985}.~\Cref{eq:genwsecondlaw} remains valid even under strong system-reservoir coupling~\cite{seifert_first_2016,jarzynski_stochastic_2017}, and even if $S$ does not end in equilibrium~\cite{jarzynski_equalities_2011}. Rearranging~\cref{eq:genwsecondlaw} yields $W-\Delta F_S \geq 0$. The greater the {\it dissipated work} $W-\Delta F_S$, the more irreversible the process~\cite{seifert_stochastic_2012}.

We now show that two work definitions [\cref{eq:wtotoperator,eq:whstaroperator}] satisfy the second law. In each case, we calculate $W - \Delta F_S$ and prove the inequality in \cref{eq:genwsecondlaw}. First consider $W_{\tot}$:
\begin{widetext}
\begin{subequations}\label{eq:wtotgeqf}
    \begin{align}
        \beta \left( W_{\tot} - \Delta F_{S} \right ) 
        & = \beta \left\{ \Tr_{SR} 
        \left(  \left[ \h_{\sur}^{\f} - \h_{\sur}^{\mi} \right] \pio_{\sur}^{\mi} \right)
        - \left ( F_{S}^{\f} - F_{S}^{\mi} \right ) \right\} \label{eq:wtotgeqf1}  \\ 
        & = \Tr_{SR} \left( \left\{  \beta \left[\h_{\sur}^{\f} - F_{\sur}^{\f} \right] - \beta \left[ \h_{\sur}^{\mi} - F_{\sur}^{\mi} \right] \; \right\} \pio_{\sur}^{A} \right) \label{eq:wtotgeqf2}  \\
        & = \Tr_{SR} \left( \left[ \ln \pio_{\sur}^{\mi} - \ln \pio_{\sur}^{\f} \right] \; \pio_{\sur}^{\mi} \right) \geq 0 \, \label{eq:wtotgeqf3} .
        \end{align}
    \end{subequations}
\Cref{eq:wtotgeqf2} follows from $F_{\sur} = F_S + F_R$ [from below \cref{eq:F_SUR,eq:F_S,eq:F_R}] and from $\Tr_{SR}(\pio^{A/B}_{S \cup R})=1$; the equality in \cref{eq:wtotgeqf3}, from $ \pio_{\sur} = e^{-\beta \left( \h_{\sur} - F_{\sur} \right)}$; and the inequality, from the quantum relative entropy's non-negativity~\footnote{The quantum relative entropy between density matrices $\rho$ and $\rho^{\prime}$ is $D(\rho || \rho^{\prime}) \coloneq \Tr(\rho [ \ln{\rho} - \ln{ \rho^{\prime}} ] ) \geq 0$~\cite{uhlmann_relative_1977,lindblad_expectations_1974,donald_relative_1986}.}. Similarly, for $W_{H^*}$,
%
\begin{subequations}\label{eq:whstargeqf}
    \begin{align}
        \beta \left( W_{H^*} - \Delta F_{S} \right ) 
        & = \beta \left\{ \Tr_{S} \left( \left[ \hsof - \hsoi \right] \; \pio_S^{\mi}  \right) - \left ( F_{S}^{\f} - F_{S}^{\mi} \right ) \right\} \label{eq:whstargeqf1} \\ 
        & = \Tr_{S} \left( \left\{  \beta \left[ \hsof - F_{S}^{\f} \right] 
        - \beta \left[ \hsoi - F_{S}^{\mi} \right] \right\} \; \pio_S^{\mi}  \right)  \label{eq:whstargeqf2} \\ 
        & = \Tr_{S} \left( \left[ \ln \pio_S^{\mi} - \ln \pio_S^{\f} \right] \; \pio_S^{\mi} \right)  \geq 0 \, . \label{eq:whstargeqf3}
        \end{align}
    \end{subequations}
\end{widetext}
The equality in \cref{eq:whstargeqf3} follows from $\pio_S=e^{-\beta(\hat{H}_S^*-F_S)}$. We lack an analogous derivation for $W_{E^*}$, and \cref{sec:spinmodel} illustrates numerically that $W_{E*}$ can violate \cref{eq:genwsecondlaw}~\footnote{Because $\pio_S$ is a Gibbs state relative to $\hso$, one can rewrite $\beta (\hso - \Delta F_S)$ as $\ln{\pio_S}$. Combining this relation with the relative entropy's non-negativity yields the second law, $\beta (W_{H^*} -\Delta F_S)\geq0$. Since $\pio_S$ is not a Gibbs state relative to $\eso$, one cannot similarly derive a second law for $W_{E^*}$.}.
Thus, of the three quantum work definitions, only $W_{\tot}$ and $W_{H^*}$ satisfy the second law for our quench processes.

The values of $W_{\tot}$, $W_{H^*}$, and $W_{E^*}$ can differ. For system quenches, however, we show in \Cref{app:commutation} that, if
\begin{equation}\label{eq:commutationrels}
    \left[ \h_{S}^{A}, \hat{V}_{\sur} \right]  =  \left[ \h_{S}^{B}, \hat{V}_{\sur} \right]
    = 0 \, ,
\end{equation}
then
\begin{equation}
\label{eq:WWW}
    W_{\tot}=W_{H^*}=W_{E^*} \, .
\end{equation}
For classical systems undergoing system quenches, $W_{\tot}$, $W_{H^*}$, and $W_{E^*}$ are identical~\cite{seifert_first_2016,jarzynski_stochastic_2017}.

If the system begins and ends in equilibrium, one can express the second law in terms of heat and entropy. Define the (thermal) entropy of an equilibrium state $\pio_S$ [\cref{eq:pisop}] using the thermodynamic relation
\begin{equation}\label{eq:genentropy}
    \mathcal{S} \coloneq \beta (U_S -F_S) \, .
\end{equation}
$U_S$ evaluates to $U_{\tot}$, $U_{H^*}$, or $U_{E^*}$; and \cref{eq:F_S} specifies $F_S$. Suppose $S$ ends in an equilibrium state at inverse temperature $\beta$,
\begin{equation}
\label{eq:finaleqstate}
    \rhoo_{S}(t_{\mathrm{f}}) = \pio_S^{\f} \, .
\end{equation}
\Cref{eq:genentropy}, with the first law [\cref{eq:firstlaw}], implies that $W - \Delta F_{S} = \beta^{-1}\Delta \mathcal{S} - Q$. Hence the second law [\cref{eq:genwsecondlaw}] is equivalent to
\begin{equation}\label{eq:genqsecondlaw}
    Q \leq \beta^{-1} \Delta \mathcal{S} \, .
\end{equation}
Since \cref{eq:genwsecondlaw,eq:genqsecondlaw} are equivalent when $S$ ends in equilibrium, \cref{eq:wtotgeqf,eq:whstargeqf} imply
\begin{equation} \label{eq:delstotgeqqtot}
    Q_{\tot} \leq \beta^{-1} \Delta \mathcal{S}_{\tot} \, 
\end{equation}
and
\begin{equation}\label{eq:delshsgeqqhs}
    Q_{H^*} \leq \beta^{-1} \Delta \mathcal{S}_{H^*} \, , 
\end{equation}
respectively. We show numerically in \cref{sec:spinmodel} that $Q_{E^*}$ can violate \cref{eq:genqsecondlaw}.

We derived \cref{eq:delstotgeqqtot,eq:delshsgeqqhs} by assuming that $S$ ends in the equilibrium state $\pio_S^{\f}$ [\cref{eq:finaleqstate}]. At least two realistic scenarios motivate this assumption:
\begin{enumerate}[label=(\roman*)]
    \item Suppose that $\sur$ is a closed quantum system evolving unitarily under $\h_{\sur}^{\f}$. Let $\sur$ satisfy the eigenstate thermalization hypothesis~\cite{srednicki_chaos_1994}, and let $R$ have a thermal reservoir's generic properties: macroscopically many DOFs and a heat capacity much greater than the system's. For sufficiently large $t_{\mathrm{f}}$, one expects $\sur$ to equilibrate to a temperature essentially identical to the initial temperature, $\beta^{-1}$. In this case, $S$ ends in the state $\pio_S^{\f}$.
    \item \label{item:equil2} Suppose that $\sur$ couples weakly to a much larger, thermal \emph{super-reservoir} at a temperature $\beta^{-1}$. If $t_{\mathrm{f}}$ is sufficiently large, then $\sur$ relaxes to the global Gibbs state $\pio_\sur^B$, which implies \cref{eq:finaleqstate}.
\end{enumerate}

\section{Two-spin model} \label{sec:spinmodel}
\begin{figure}[btp]
   \centering
   \includegraphics[scale=0.75]{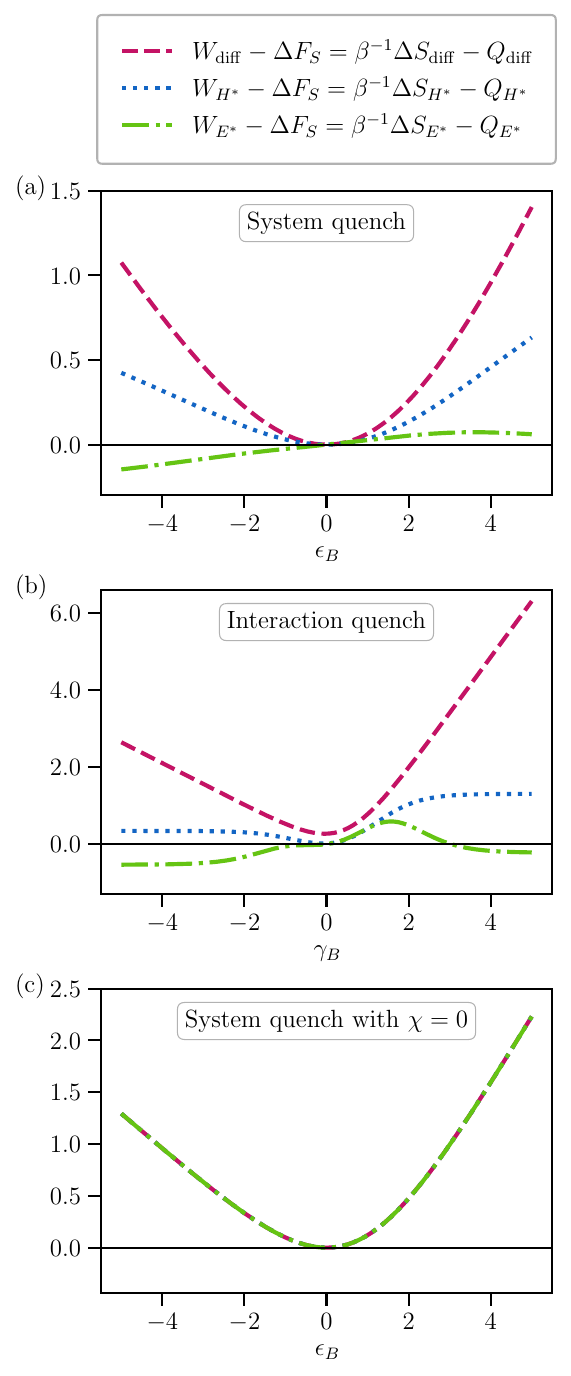}
    \caption{\textit{Thermodynamic quantities of the two-spin model.} (a) During a system quench, the $\epsilon$ in \cref{eq:toyham} changes from $\epsilon_{\mi} =0$ to $\epsilon_{\f}$. The dissipated work is plotted as a function of $\epsilon_{\f}$, with $\alpha=0.8$, $\gamma=1.2$, and $\chi = 1.8$. $W_{\tot}$ and $W_{H^*}$ satisfy the second law [\cref{eq:genwsecondlaw}]; consequently, so do $Q_{\tot}$ and $Q_{H^*}$ [\cref{eq:genqsecondlaw}]. $W_{E^*}$ and $Q_{E^*}$ do not. (b) During an interaction quench, the coupling constants $\gamma$ and $\chi$ change from $\gamma_A=\chi_A=0$ to $\gamma_B$ and $\chi_B$. Dissipated work is plotted as a function of $\gamma_B$, with $\epsilon=1.0$, $\alpha=5.0$, and $\chi_B = 1.2$. Qualitatively similar results follow from plotting the dissipated work as a function of $\chi_B$ at fixed $\epsilon$, $\alpha$, and $\gamma_B$. (c) We set $\chi = 0$ in~\cref{eq:toyham}, such that $\h_S$ commutes with $\hat{V}_{\sur}$ before and after the quench [Eqs.~\eqref{eq:commutationrels}]. The work dissipated during a system quench is plotted as a function of $\epsilon_{\f}$, with fixed $\epsilon_{\mi} =0$, $\alpha=0.8$, and $\gamma=1.2$. The three curves coincide, as expected [\cref{eq:WWW}].}
    \label{fig:toy_quench}
\end{figure}
A simple, illustrative model features spins $S$ and $R$. Denote by $\sigma_a^i$ the Pauli operator with $a\in \{S,R\}$ and $i\in \{x,z\}$. The spins evolve under the Hamiltonian
\begin{equation}\label{eq:toyham}
    \h_{\sur} = \frac{\epsilon}{2} \sigma_{S}^{z} + \frac{\alpha}{2}  \sigma_{R}^{z} + \gamma \sigma_{S}^{z} \sigma_{R}^{z}  + \chi \sigma_{S}^{x} \sigma_{R}^{x} \,  ,
\end{equation}
wherein $\gamma$ and $\chi$ denote coupling strengths. Treating one spin as a reservoir may seem odd. In previous sections, however, we did not assume that $R$ contains many degrees of freedom. The Hamiltonian~\eqref{eq:toyham} has the form in \cref{eq:hsurgen} and enables us to calculate work and heat analytically.

We assume $\sur$ couples weakly to a thermal super-reservoir at a temperature $\beta^{-1}$. Moreover, we assume $t_{\rm f}$ is sufficiently large for $\sur$ to equilibrate with the super-reservoir, postquench. (We do not model the super-reservoir explicitly.) These assumptions realize scenario \ref{item:equil2}, sketched at the end of \cref{sec:wandh}. We compute the work, heat, free-energy change, and entropy change relevant to each quench described in \Cref{tab:quench}.~\Cref{app:wandhforspin} lists analytical expressions for these quantities. Here, we illustrate the conclusions of \cref{sec:wandh} by evaluating these expressions at specific $\epsilon$, $\alpha$, $\gamma$ and $\chi$ values. Throughout this section, $\beta=1$.

During a system quench, $\h_S \coloneqq \frac{\epsilon}{2} \, \sigma_{S}^{z}$ changes abruptly: $\epsilon$ switches from  $\epsilon_{\mi}=0$ to $\epsilon_{\f}$. $\alpha$, $\gamma$, and $\chi$ remain fixed.~\Cref{fig:toy_quench}(a) displays the dissipated work, $W - \Delta F_S$, as a function of $\epsilon_{\f}$. $W_{\tot}$ and $W_{H^*}$ satisfy the second law, in agreement with \cref{eq:wtotgeqf,eq:whstargeqf}. In contrast, $W_{E^*}$ violates the second law at certain $\epsilon_\f$ values. These conclusions hold also if $\epsilon_{\mi}$ is nonzero.

During an interaction quench, $\gamma$ and $\chi$ change suddenly from $\gamma_{\mi}=\chi_{\mi}=0$ to $\gamma_{\f}$ and $\chi_{\f}$.~\Cref{fig:toy_quench}(b) displays the dissipated work as a function of $\gamma_\f$. The other parameters remain fixed. Again, $W_{\tot}$ and $W_{H^*}$ satisfy the second law. However, $W_{E^*}$ violates it at some $\gamma_\f$ values.

Turning to \cref{fig:toy_quench}(c), we return to the system quench. However, we now set $\chi=0$ in~\cref{eq:toyham}, to satisfy the commutation relations in Eqs.~\eqref{eq:commutationrels}. In agreement with \cref{eq:WWW}, the three dissipated-work quantities equal each other. Additionally, they obey the second law. 

Strong-coupling quantum thermodynamics naturally applies to lattice gauge theories, as proposed in Ref.~\cite{davoudi_quantum_2024}. Gauge theories and their lattice formulations are crucial to high-energy and nuclear physics~\cite{aitchison_gauge_2012,quigg_gauge_2021}, condensed and synthetic quantum matter~\cite{fradkin_field_2013,kleinert_gauge_1989,wen_topological_1990,levin_string_2005}, and quantum information science~\cite{chen_exact_2018,chen_exact_2020,chen_equivalence_2023,kitaev_fault_2003,kitaev_anyons_2006,sarma_topological_2006,nayak_non_2008,lahtinen_short_2017}. Local constraints among DOFs define lattice gauge theories. Physical systems can only be in states consistent with those constraints. A slight modification converts our two-spin model into a simple lattice gauge theory. We introduce an LGT-type constraint by setting $k=-\gamma >0$ and adding a term $k \mathbbm{1}_{\sur}$ to $\h_{\sur}$ in \cref{eq:toyham}: 
\begin{align}\label{eq:toylgtham}
\h_{\sur} = \frac{\epsilon}{2} \sigma_{S}^{z} + \frac{\alpha}{2}  \sigma_{R}^{z} + \chi \sigma_{S}^{x} \sigma_{R}^{x} +
k ( \mathbbm{1}_{\sur} - \sigma_{S}^{z} \sigma_{R}^{z} ) \, .
\end{align}
In the limit as $k \to \infty$ (at a fixed, finite temperature), the last term acts as an energy penalty: it constrains the system to the eigenvalue-1 eigenspace of $\sigma^z_S \sigma^z_R$. This operator serves as a \emph{Gauss-law operator}. It specifies which states satisfy the constraint. Since the interactions in \cref{eq:toylgtham} are non-negligible, strong-coupling quantum thermodynamics applies to this model, as to other lattice gauge theories~\cite{davoudi_quantum_2024}. To test our results, we subject the two-spin model~\cref{eq:toylgtham} (with $k \rightarrow \infty$) to the system-quench process [Fig.~\ref{fig:toy_quench}(a)]~\footnote{We do not analyze an interaction-quench process [defined in \cref{sec:quench,fig:quench}(b)] because it would change $k$. In the LGT-type model, $k$ maintains a large, constant value.}. We find that $W_\tot$ and $W_{H^*}$ satisfy the second law, whereas $W_{E^*}$ does not [see \cref{app:wandhforspin}]. 

\section{Discussion and outlook}
We have scrutinized definitions of thermodynamic properties of strongly coupled quantum systems. We compared three definitions of the system's internal energy---$U_{\tot}$, $U_{H^*}$, and $U_{E^*}$ [\cref{eq:utot,eq:uhstar,eq:uestar}]---during quench processes. All three lead to work and heat definitions that satisfy the first law of thermodynamics, by construction. However, we found, only the definitions based on $U_{\tot}$ and $U_{H^*}$ satisfy the second law. These conclusions hold independently of the final states when the second law is expressed in terms of work and free energy. If the system equilibrates after the quench, the same conclusions hold when the second law is expressed in terms of heat and entropy. We illustrated these general results with a simple model of two coupled spins. These conclusions distinguish quantum from classical thermodynamics. Our work can therefore guide the thermodynamics of strongly coupled quantum systems of relevance to condensed matter, high-energy and nuclear physics, quantum chemistry, and quantum error correction (see, e.g., Refs.~\cite{ferraz_strongly_2020,davoudi_quantum_2024,sun_quantum_2024,bilokur_thermodynamic_2024}).

This work opens the door to further opportunities:
\begin{itemize}
\item[$\circ$] We focused on quench processes because they allow for naturally partitioning the system's internal-energy change into work and heat, even when the system and reservoir couple strongly. Our results may extend to thermodynamic processes beyond quenches, where the partitioning is less clear. Examples include quantum-adiabatic processes, used to prepare quantum states in quantum simulations, and particle collisions, relevant to hydrodynamics~\cite{tsubota_quantum_2013} and nuclear and high-energy physics~\cite{davoudi_quantum_2024,surace_stringbreaking_2024,jacob_universal_2024}. 

\item[$\circ$] In this work, we have defined average work and heat quantities. One can define classical work and heat as fluctuating quantities, whose values differ from realization to realization of a process. Fluctuation theorems (extensions of the second law), derived in this setting~\cite{jarzynski_nonequilibrium_1997,jarzynski_hamiltonian_2000,jarzynski_nonequilibrium_2004,jarzynski_equalities_2011, jarzynski_stochastic_2017,crooks_entropy_1999,esposito_entropy_2010,korbel_nonequilibrium_2024,miller_entropy_2017,seifert_first_2016,esposito_three_2010,horowitz_comparison_2007,bertini_thermodynamic_2012,bertini_macroscopic_2002}, have been extended to strong system-reservoir couplings~\cite{jarzynski_nonequilibrium_2004,seifert_first_2016,jarzynski_stochastic_2017,miller_entropy_2017}. Can one define fluctuating work and heat exchanged within strongly coupled quantum systems? For example, the two-point-measurement definition of work supports a fluctuation theorem for closed quantum systems~\cite{tasaki_jarzynski_2000,kurchan_quantum_2001,mukamel_quantum_2003,campisi_fluctuation_2009,talkner_fluctuation_2007,talkner_fluctuation_2009,funo_quantum_2018}. One can apply this scheme in strong-coupling contexts by treating $\sur$ as a closed quantum system. Yet can one infer the fluctuating work performed on $S$ from measurements of $S$ alone? 

\item[$\circ$] Of the two internal-energy definitions that obey the second law, one [\cref{eq:uhstar}] stems from the Hamiltonian of mean force, $\hso$ [\cref{eq:hstar}]. This operator is related to the system's entanglement Hamiltonian if $\sur$ is in equilibrium~\cite{davoudi_quantum_2024}. Entanglement-Hamiltonian tomography enables efficient experimental measurements, or numerical determinations, of a subsystem's quantum state~\cite{dalmonte_entanglement_2022,elben_statistical_2019,huang_predicting_2020,huang_quantum_2022,elben_randomized_2023}. Such tomography has been applied to ground, excited, and nonequilibrium states of isolated systems~\cite{pichler_measurement_2016,dalmonte_quantum_2018,kokail_entanglement_2021,kokail_quantum_2021,joshi_exploring_2023,mueller_quantum_2023,bringewatt_randomized_2024,mueller_quantum_2024}. Future work will extend entanglement-Hamiltonian tomography to thermal states. This extension will allow one to access thermodynamic quantities in the strong-coupling regime of quantum simulations.
\end{itemize}
%


\section{Acknowledgments}
We thank Sherry Wang and {\'A}ngel Rivas for valuable feedback on the manuscript. 
Z.D., G.O., and C.P. were supported by the National Science Foundation (NSF) Quantum Leap Challenge Institutes (QLCI) (Award No.~OMA-2120757).
Z.D. further acknowledges funding by the Department of Energy (DOE), Office of Science, Early Career Award (Award No.~DESC0020271), as well as by the Department of Physics; Maryland Center for Fundamental Physics; and College of Computer, Mathematical, and Natural Sciences at the University of Maryland, College Park. She is grateful for the hospitality of Perimeter Institute, and of Kavli Institute for Theoretical Physics (KITP), where part of this work was carried out. Research at Perimeter Institute is supported in part by the Government of Canada through the Department of Innovation, Science, and Economic Development and by the Province of Ontario through the Ministry of Colleges and Universities. Z.D. was also supported in part by the Simons Foundation through the Simons Foundation Emmy Noether Fellows Program at Perimeter Institute. Research at the KITP was supported in part by the NSF award PHY-2309135. 
C.J. and N.Y.H. further acknowledge support from John Templeton Foundation (Award No.~62422).
N.Y.H. thanks Harry Miller for conversations about strong-coupling thermodynamics.
N.M. acknowledges funding by the DOE, Office of Science, Office of Nuclear Physics, IQuS, via the program on Quantum Horizons: QIS Research and Innovation for Nuclear Science under Award No. DE-SC0020970.
C.P. is grateful for discussions about thermal gauge theories with Robert Pisarski and for the hospitality of Brookhaven National Laboratory, which hosted C.P. as part of an Office of Science Graduate Student Research Fellowship.
G.O. further acknowledges support from the American Association of University Women through an International Fellowship. 
Finally, N.M., G.O., C.P., and N.Y.H. thank the participants of the InQubator for Quantum Simulation (IQuS) workshop ``\href{https://iqus.uw.edu/events/iqus-workshop-thermalization/}{Thermalization, from Cold Atoms to Hot Quantum Chromodynamics}'' which took place at the University of Washington in September 2023, for valuable discussions.
\bibliography{bibliography}

\begin{thebibliography}{134}%
\makeatletter
\providecommand \@ifxundefined [1]{%
 \@ifx{#1\undefined}
}%
\providecommand \@ifnum [1]{%
 \ifnum #1\expandafter \@firstoftwo
 \else \expandafter \@secondoftwo
 \fi
}%
\providecommand \@ifx [1]{%
 \ifx #1\expandafter \@firstoftwo
 \else \expandafter \@secondoftwo
 \fi
}%
\providecommand \natexlab [1]{#1}%
\providecommand \enquote  [1]{``#1''}%
\providecommand \bibnamefont  [1]{#1}%
\providecommand \bibfnamefont [1]{#1}%
\providecommand \citenamefont [1]{#1}%
\providecommand \href@noop [0]{\@secondoftwo}%
\providecommand \href [0]{\begingroup \@sanitize@url \@href}%
\providecommand \@href[1]{\@@startlink{#1}\@@href}%
\providecommand \@@href[1]{\endgroup#1\@@endlink}%
\providecommand \@sanitize@url [0]{\catcode `\\12\catcode `\$12\catcode
  `\&12\catcode `\#12\catcode `\^12\catcode `\_12\catcode `\%12\relax}%
\providecommand \@@startlink[1]{}%
\providecommand \@@endlink[0]{}%
\providecommand \url  [0]{\begingroup\@sanitize@url \@url }%
\providecommand \@url [1]{\endgroup\@href {#1}{\urlprefix }}%
\providecommand \urlprefix  [0]{URL }%
\providecommand \Eprint [0]{\href }%
\providecommand \doibase [0]{https://doi.org/}%
\providecommand \selectlanguage [0]{\@gobble}%
\providecommand \bibinfo  [0]{\@secondoftwo}%
\providecommand \bibfield  [0]{\@secondoftwo}%
\providecommand \translation [1]{[#1]}%
\providecommand \BibitemOpen [0]{}%
\providecommand \bibitemStop [0]{}%
\providecommand \bibitemNoStop [0]{.\EOS\space}%
\providecommand \EOS [0]{\spacefactor3000\relax}%
\providecommand \BibitemShut  [1]{\csname bibitem#1\endcsname}%
\let\auto@bib@innerbib\@empty
\bibitem [{\citenamefont {Goold}\ \emph {et~al.}(2016)\citenamefont {Goold},
  \citenamefont {Huber}, \citenamefont {Riera}, \citenamefont {del Rio},\ and\
  \citenamefont {Skrzypczyk}}]{goold_role_2016}%
  \BibitemOpen
  \bibfield  {author} {\bibinfo {author} {\bibfnamefont {J.}~\bibnamefont
  {Goold}}, \bibinfo {author} {\bibfnamefont {M.}~\bibnamefont {Huber}},
  \bibinfo {author} {\bibfnamefont {A.}~\bibnamefont {Riera}}, \bibinfo
  {author} {\bibfnamefont {L.}~\bibnamefont {del Rio}},\ and\ \bibinfo {author}
  {\bibfnamefont {P.}~\bibnamefont {Skrzypczyk}},\ }\href
  {https://doi.org/10.1088/1751-8113/49/14/143001} {\bibfield  {journal}
  {\bibinfo  {journal} {Journal of Physics A: Mathematical and Theoretical}\
  }\textbf {\bibinfo {volume} {49}},\ \bibinfo {pages} {143001} (\bibinfo
  {year} {2016})}\BibitemShut {NoStop}%
\bibitem [{\citenamefont {Vinjanampathy}\ and\ \citenamefont
  {Anders}(2016)}]{vinjanampathy_quantum_2016}%
  \BibitemOpen
  \bibfield  {author} {\bibinfo {author} {\bibfnamefont {S.}~\bibnamefont
  {Vinjanampathy}}\ and\ \bibinfo {author} {\bibfnamefont {J.}~\bibnamefont
  {Anders}},\ }\href {https://doi.org/10.1080/00107514.2016.1201896} {\bibfield
   {journal} {\bibinfo  {journal} {Contemporary Physics}\ }\textbf {\bibinfo
  {volume} {57}},\ \bibinfo {pages} {545} (\bibinfo {year} {2016})},\ \Eprint
  {https://arxiv.org/abs/https://doi.org/10.1080/00107514.2016.1201896}
  {https://doi.org/10.1080/00107514.2016.1201896} \BibitemShut {NoStop}%
\bibitem [{\citenamefont {Millen}\ and\ \citenamefont
  {Xuereb}(2016)}]{millen_perspective_2016}%
  \BibitemOpen
  \bibfield  {author} {\bibinfo {author} {\bibfnamefont {J.}~\bibnamefont
  {Millen}}\ and\ \bibinfo {author} {\bibfnamefont {A.}~\bibnamefont
  {Xuereb}},\ }\href {https://doi.org/10.1088/1367-2630/18/1/011002} {\bibfield
   {journal} {\bibinfo  {journal} {New Journal of Physics}\ }\textbf {\bibinfo
  {volume} {18}},\ \bibinfo {pages} {011002} (\bibinfo {year}
  {2016})}\BibitemShut {NoStop}%
\bibitem [{\citenamefont {Alicki}\ and\ \citenamefont
  {Kosloff}(2018)}]{alicki_introduction_2018}%
  \BibitemOpen
  \bibfield  {author} {\bibinfo {author} {\bibfnamefont {R.}~\bibnamefont
  {Alicki}}\ and\ \bibinfo {author} {\bibfnamefont {R.}~\bibnamefont
  {Kosloff}},\ }in\ \href {https://doi.org/10.1007/978-3-319-99046-0_1} {\emph
  {\bibinfo {booktitle} {Thermodynamics in the {Quantum} {Regime}:
  {Fundamental} {Aspects} and {New} {Directions}}}},\ \bibinfo {series and
  number} {Fundamental {Theories} of {Physics}},\ \bibinfo {editor} {edited by\
  \bibinfo {editor} {\bibfnamefont {F.}~\bibnamefont {Binder}}, \bibinfo
  {editor} {\bibfnamefont {L.~A.}\ \bibnamefont {Correa}}, \bibinfo {editor}
  {\bibfnamefont {C.}~\bibnamefont {Gogolin}}, \bibinfo {editor} {\bibfnamefont
  {J.}~\bibnamefont {Anders}},\ and\ \bibinfo {editor} {\bibfnamefont
  {G.}~\bibnamefont {Adesso}}}\ (\bibinfo  {publisher} {Springer International
  Publishing},\ \bibinfo {address} {Cham},\ \bibinfo {year} {2018})\ pp.\
  \bibinfo {pages} {1--33}\BibitemShut {NoStop}%
\bibitem [{\citenamefont {Binder}\ \emph {et~al.}(2018)\citenamefont {Binder},
  \citenamefont {Correa}, \citenamefont {Gogolin}, \citenamefont {Anders},\
  and\ \citenamefont {Adesso}}]{binder_thermodynamics_2018}%
  \BibitemOpen
  \bibinfo {editor} {\bibfnamefont {F.}~\bibnamefont {Binder}}, \bibinfo
  {editor} {\bibfnamefont {L.~A.}\ \bibnamefont {Correa}}, \bibinfo {editor}
  {\bibfnamefont {C.}~\bibnamefont {Gogolin}}, \bibinfo {editor} {\bibfnamefont
  {J.}~\bibnamefont {Anders}},\ and\ \bibinfo {editor} {\bibfnamefont
  {G.}~\bibnamefont {Adesso}},\ eds.,\ \href
  {https://doi.org/10.1007/978-3-319-99046-0} {\emph {\bibinfo {title}
  {Thermodynamics in the {Quantum} {Regime}: {Fundamental} {Aspects} and {New}
  {Directions}}}},\ \bibinfo {series} {Fundamental {Theories} of {Physics}},
  Vol.\ \bibinfo {volume} {195}\ (\bibinfo  {publisher} {Springer International
  Publishing},\ \bibinfo {address} {Cham},\ \bibinfo {year} {2018})\BibitemShut
  {NoStop}%
\bibitem [{\citenamefont {Gogolin}\ and\ \citenamefont
  {Eisert}(2016)}]{gogolin_equilibration_2016}%
  \BibitemOpen
  \bibfield  {author} {\bibinfo {author} {\bibfnamefont {C.}~\bibnamefont
  {Gogolin}}\ and\ \bibinfo {author} {\bibfnamefont {J.}~\bibnamefont
  {Eisert}},\ }\href {https://doi.org/10.1088/0034-4885/79/5/056001} {\bibfield
   {journal} {\bibinfo  {journal} {Reports on Progress in Physics}\ }\textbf
  {\bibinfo {volume} {79}},\ \bibinfo {pages} {056001} (\bibinfo {year}
  {2016})}\BibitemShut {NoStop}%
\bibitem [{\citenamefont {Majidy}\ \emph {et~al.}(2023)\citenamefont {Majidy},
  \citenamefont {Braasch~Jr}, \citenamefont {Lasek}, \citenamefont {Upadhyaya},
  \citenamefont {Kalev},\ and\ \citenamefont
  {Yunger~Halpern}}]{majidy_noncommuting_2023}%
  \BibitemOpen
  \bibfield  {author} {\bibinfo {author} {\bibfnamefont {S.}~\bibnamefont
  {Majidy}}, \bibinfo {author} {\bibfnamefont {W.~F.}\ \bibnamefont
  {Braasch~Jr}}, \bibinfo {author} {\bibfnamefont {A.}~\bibnamefont {Lasek}},
  \bibinfo {author} {\bibfnamefont {T.}~\bibnamefont {Upadhyaya}}, \bibinfo
  {author} {\bibfnamefont {A.}~\bibnamefont {Kalev}},\ and\ \bibinfo {author}
  {\bibfnamefont {N.}~\bibnamefont {Yunger~Halpern}},\ }\href@noop {}
  {\bibfield  {journal} {\bibinfo  {journal} {Nature Reviews Physics}\ }\textbf
  {\bibinfo {volume} {5}},\ \bibinfo {pages} {689} (\bibinfo {year}
  {2023})}\BibitemShut {NoStop}%
\bibitem [{\citenamefont {Tasaki}(2000)}]{tasaki_jarzynski_2000}%
  \BibitemOpen
  \bibfield  {author} {\bibinfo {author} {\bibfnamefont {H.}~\bibnamefont
  {Tasaki}},\ }\href@noop {} {\bibinfo {title} {Jarzynski relations for quantum
  systems and some applications}} (\bibinfo {year} {2000}),\ \Eprint
  {https://arxiv.org/abs/cond-mat/0009244} {arXiv:cond-mat/0009244
  [cond-mat.stat-mech]} \BibitemShut {NoStop}%
\bibitem [{\citenamefont {Kurchan}(2001)}]{kurchan_quantum_2001}%
  \BibitemOpen
  \bibfield  {author} {\bibinfo {author} {\bibfnamefont {J.}~\bibnamefont
  {Kurchan}},\ }\href@noop {} {\bibinfo {title} {A quantum fluctuation
  theorem}} (\bibinfo {year} {2001}),\ \Eprint
  {https://arxiv.org/abs/cond-mat/0007360} {arXiv:cond-mat/0007360
  [cond-mat.stat-mech]} \BibitemShut {NoStop}%
\bibitem [{\citenamefont {Mukamel}(2003)}]{mukamel_quantum_2003}%
  \BibitemOpen
  \bibfield  {author} {\bibinfo {author} {\bibfnamefont {S.}~\bibnamefont
  {Mukamel}},\ }\href {https://doi.org/10.1103/PhysRevLett.90.170604}
  {\bibfield  {journal} {\bibinfo  {journal} {Phys. Rev. Lett.}\ }\textbf
  {\bibinfo {volume} {90}},\ \bibinfo {pages} {170604} (\bibinfo {year}
  {2003})}\BibitemShut {NoStop}%
\bibitem [{\citenamefont {Campisi}\ \emph {et~al.}(2011)\citenamefont
  {Campisi}, \citenamefont {H\"anggi},\ and\ \citenamefont
  {Talkner}}]{campisi_colloquium_2011}%
  \BibitemOpen
  \bibfield  {author} {\bibinfo {author} {\bibfnamefont {M.}~\bibnamefont
  {Campisi}}, \bibinfo {author} {\bibfnamefont {P.}~\bibnamefont {H\"anggi}},\
  and\ \bibinfo {author} {\bibfnamefont {P.}~\bibnamefont {Talkner}},\ }\href
  {https://doi.org/10.1103/RevModPhys.83.771} {\bibfield  {journal} {\bibinfo
  {journal} {Rev. Mod. Phys.}\ }\textbf {\bibinfo {volume} {83}},\ \bibinfo
  {pages} {771} (\bibinfo {year} {2011})}\BibitemShut {NoStop}%
\bibitem [{\citenamefont {Mitchison}(2019)}]{mitchison_quantum_2019}%
  \BibitemOpen
  \bibfield  {author} {\bibinfo {author} {\bibfnamefont {M.~T.}\ \bibnamefont
  {Mitchison}},\ }\href {https://doi.org/10.1080/00107514.2019.1631555}
  {\bibfield  {journal} {\bibinfo  {journal} {Contemporary Physics}\ }\textbf
  {\bibinfo {volume} {60}},\ \bibinfo {pages} {164} (\bibinfo {year} {2019})},\
  \Eprint {https://arxiv.org/abs/https://doi.org/10.1080/00107514.2019.1631555}
  {https://doi.org/10.1080/00107514.2019.1631555} \BibitemShut {NoStop}%
\bibitem [{\citenamefont {Mukherjee}\ and\ \citenamefont
  {Divakaran}(2021)}]{mukherjee_many-body_2021}%
  \BibitemOpen
  \bibfield  {author} {\bibinfo {author} {\bibfnamefont {V.}~\bibnamefont
  {Mukherjee}}\ and\ \bibinfo {author} {\bibfnamefont {U.}~\bibnamefont
  {Divakaran}},\ }\href {https://doi.org/10.1088/1361-648X/ac1b60} {\bibfield
  {journal} {\bibinfo  {journal} {Journal of Physics: Condensed Matter}\
  }\textbf {\bibinfo {volume} {33}},\ \bibinfo {pages} {454001} (\bibinfo
  {year} {2021})}\BibitemShut {NoStop}%
\bibitem [{\citenamefont {{Mar{\'\i}n Guzm{\'a}n}}\ \emph
  {et~al.}(2024)\citenamefont {{Mar{\'\i}n Guzm{\'a}n}}, \citenamefont
  {{Erker}}, \citenamefont {{Gasparinetti}}, \citenamefont {{Huber}},\ and\
  \citenamefont {{Yunger Halpern}}}]{maringuzman_key_2024}%
  \BibitemOpen
  \bibfield  {author} {\bibinfo {author} {\bibfnamefont {J.~A.}\ \bibnamefont
  {{Mar{\'\i}n Guzm{\'a}n}}}, \bibinfo {author} {\bibfnamefont
  {P.}~\bibnamefont {{Erker}}}, \bibinfo {author} {\bibfnamefont
  {S.}~\bibnamefont {{Gasparinetti}}}, \bibinfo {author} {\bibfnamefont
  {M.}~\bibnamefont {{Huber}}},\ and\ \bibinfo {author} {\bibfnamefont
  {N.}~\bibnamefont {{Yunger Halpern}}},\ }\href
  {https://doi.org/10.1088/1361-6633/ad8803} {\bibfield  {journal} {\bibinfo
  {journal} {Reports on Progress in Physics}\ }\textbf {\bibinfo {volume}
  {87}},\ \bibinfo {pages} {122001} (\bibinfo {year} {2024})}\BibitemShut
  {NoStop}%
\bibitem [{\citenamefont {Chitambar}\ and\ \citenamefont
  {Gour}(2019)}]{chitambar_quantum_2019}%
  \BibitemOpen
  \bibfield  {author} {\bibinfo {author} {\bibfnamefont {E.}~\bibnamefont
  {Chitambar}}\ and\ \bibinfo {author} {\bibfnamefont {G.}~\bibnamefont
  {Gour}},\ }\href {https://doi.org/10.1103/RevModPhys.91.025001} {\bibfield
  {journal} {\bibinfo  {journal} {Rev. Mod. Phys.}\ }\textbf {\bibinfo {volume}
  {91}},\ \bibinfo {pages} {025001} (\bibinfo {year} {2019})}\BibitemShut
  {NoStop}%
\bibitem [{\citenamefont {Lostaglio}(2019)}]{lostaglio_introductory_2019}%
  \BibitemOpen
  \bibfield  {author} {\bibinfo {author} {\bibfnamefont {M.}~\bibnamefont
  {Lostaglio}},\ }\href {https://doi.org/10.1088/1361-6633/ab46e5} {\bibfield
  {journal} {\bibinfo  {journal} {Reports on Progress in Physics}\ }\textbf
  {\bibinfo {volume} {82}},\ \bibinfo {pages} {114001} (\bibinfo {year}
  {2019})}\BibitemShut {NoStop}%
\bibitem [{\citenamefont {An}\ \emph {et~al.}(2015)\citenamefont {An},
  \citenamefont {Zhang}, \citenamefont {Um}, \citenamefont {Lv}, \citenamefont
  {Lu}, \citenamefont {Zhang}, \citenamefont {Yin}, \citenamefont {Quan},\ and\
  \citenamefont {Kim}}]{an_experimental_2015}%
  \BibitemOpen
  \bibfield  {author} {\bibinfo {author} {\bibfnamefont {S.}~\bibnamefont
  {An}}, \bibinfo {author} {\bibfnamefont {J.-N.}\ \bibnamefont {Zhang}},
  \bibinfo {author} {\bibfnamefont {M.}~\bibnamefont {Um}}, \bibinfo {author}
  {\bibfnamefont {D.}~\bibnamefont {Lv}}, \bibinfo {author} {\bibfnamefont
  {Y.}~\bibnamefont {Lu}}, \bibinfo {author} {\bibfnamefont {J.}~\bibnamefont
  {Zhang}}, \bibinfo {author} {\bibfnamefont {Z.-Q.}\ \bibnamefont {Yin}},
  \bibinfo {author} {\bibfnamefont {H.~T.}\ \bibnamefont {Quan}},\ and\
  \bibinfo {author} {\bibfnamefont {K.}~\bibnamefont {Kim}},\ }\href
  {https://doi.org/10.1038/nphys3197} {\bibfield  {journal} {\bibinfo
  {journal} {Nature Physics}\ }\textbf {\bibinfo {volume} {11}},\ \bibinfo
  {pages} {193} (\bibinfo {year} {2015})}\BibitemShut {NoStop}%
\bibitem [{\citenamefont {Xiong}\ \emph {et~al.}(2018)\citenamefont {Xiong},
  \citenamefont {Yan}, \citenamefont {Zhou}, \citenamefont {Rehan},
  \citenamefont {Liang}, \citenamefont {Chen}, \citenamefont {Yang},
  \citenamefont {Ma}, \citenamefont {Feng},\ and\ \citenamefont
  {Vedral}}]{xiong_experimental_2018}%
  \BibitemOpen
  \bibfield  {author} {\bibinfo {author} {\bibfnamefont {T.~P.}\ \bibnamefont
  {Xiong}}, \bibinfo {author} {\bibfnamefont {L.~L.}\ \bibnamefont {Yan}},
  \bibinfo {author} {\bibfnamefont {F.}~\bibnamefont {Zhou}}, \bibinfo {author}
  {\bibfnamefont {K.}~\bibnamefont {Rehan}}, \bibinfo {author} {\bibfnamefont
  {D.~F.}\ \bibnamefont {Liang}}, \bibinfo {author} {\bibfnamefont
  {L.}~\bibnamefont {Chen}}, \bibinfo {author} {\bibfnamefont {W.~L.}\
  \bibnamefont {Yang}}, \bibinfo {author} {\bibfnamefont {Z.~H.}\ \bibnamefont
  {Ma}}, \bibinfo {author} {\bibfnamefont {M.}~\bibnamefont {Feng}},\ and\
  \bibinfo {author} {\bibfnamefont {V.}~\bibnamefont {Vedral}},\ }\href
  {https://doi.org/10.1103/PhysRevLett.120.010601} {\bibfield  {journal}
  {\bibinfo  {journal} {Phys. Rev. Lett.}\ }\textbf {\bibinfo {volume} {120}},\
  \bibinfo {pages} {010601} (\bibinfo {year} {2018})}\BibitemShut {NoStop}%
\bibitem [{\citenamefont {Schuckert}\ \emph {et~al.}(2023)\citenamefont
  {Schuckert}, \citenamefont {Bohrdt}, \citenamefont {Crane},\ and\
  \citenamefont {Knap}}]{schuckert_probing_2023}%
  \BibitemOpen
  \bibfield  {author} {\bibinfo {author} {\bibfnamefont {A.}~\bibnamefont
  {Schuckert}}, \bibinfo {author} {\bibfnamefont {A.}~\bibnamefont {Bohrdt}},
  \bibinfo {author} {\bibfnamefont {E.}~\bibnamefont {Crane}},\ and\ \bibinfo
  {author} {\bibfnamefont {M.}~\bibnamefont {Knap}},\ }\href
  {https://doi.org/10.1103/PhysRevB.107.L140410} {\bibfield  {journal}
  {\bibinfo  {journal} {Phys. Rev. B}\ }\textbf {\bibinfo {volume} {107}},\
  \bibinfo {pages} {L140410} (\bibinfo {year} {2023})}\BibitemShut {NoStop}%
\bibitem [{\citenamefont {Kaufman}\ \emph {et~al.}(2016)\citenamefont
  {Kaufman}, \citenamefont {Tai}, \citenamefont {Lukin}, \citenamefont
  {Rispoli}, \citenamefont {Schittko}, \citenamefont {Preiss},\ and\
  \citenamefont {Greiner}}]{kaufman_quantum_2016}%
  \BibitemOpen
  \bibfield  {author} {\bibinfo {author} {\bibfnamefont {A.~M.}\ \bibnamefont
  {Kaufman}}, \bibinfo {author} {\bibfnamefont {M.~E.}\ \bibnamefont {Tai}},
  \bibinfo {author} {\bibfnamefont {A.}~\bibnamefont {Lukin}}, \bibinfo
  {author} {\bibfnamefont {M.}~\bibnamefont {Rispoli}}, \bibinfo {author}
  {\bibfnamefont {R.}~\bibnamefont {Schittko}}, \bibinfo {author}
  {\bibfnamefont {P.~M.}\ \bibnamefont {Preiss}},\ and\ \bibinfo {author}
  {\bibfnamefont {M.}~\bibnamefont {Greiner}},\ }\href
  {https://doi.org/10.1126/science.aaf6725} {\bibfield  {journal} {\bibinfo
  {journal} {Science}\ }\textbf {\bibinfo {volume} {353}},\ \bibinfo {pages}
  {794} (\bibinfo {year} {2016})},\ \Eprint
  {https://arxiv.org/abs/https://www.science.org/doi/pdf/10.1126/science.aaf6725}
  {https://www.science.org/doi/pdf/10.1126/science.aaf6725} \BibitemShut
  {NoStop}%
\bibitem [{\citenamefont {Kranzl}\ \emph {et~al.}(2023)\citenamefont {Kranzl},
  \citenamefont {Lasek}, \citenamefont {Joshi}, \citenamefont {Kalev},
  \citenamefont {Blatt}, \citenamefont {Roos},\ and\ \citenamefont {{Yunger
  Halpern}}}]{kranzl_experimental_2023}%
  \BibitemOpen
  \bibfield  {author} {\bibinfo {author} {\bibfnamefont {F.}~\bibnamefont
  {Kranzl}}, \bibinfo {author} {\bibfnamefont {A.}~\bibnamefont {Lasek}},
  \bibinfo {author} {\bibfnamefont {M.~K.}\ \bibnamefont {Joshi}}, \bibinfo
  {author} {\bibfnamefont {A.}~\bibnamefont {Kalev}}, \bibinfo {author}
  {\bibfnamefont {R.}~\bibnamefont {Blatt}}, \bibinfo {author} {\bibfnamefont
  {C.~F.}\ \bibnamefont {Roos}},\ and\ \bibinfo {author} {\bibfnamefont
  {N.}~\bibnamefont {{Yunger Halpern}}},\ }\href
  {https://doi.org/10.1103/PRXQuantum.4.020318} {\bibfield  {journal} {\bibinfo
   {journal} {PRX Quantum}\ }\textbf {\bibinfo {volume} {4}},\ \bibinfo {pages}
  {020318} (\bibinfo {year} {2023})}\BibitemShut {NoStop}%
\bibitem [{\citenamefont {Hahn}\ \emph {et~al.}(2023)\citenamefont {Hahn},
  \citenamefont {Dupont}, \citenamefont {Schmitt}, \citenamefont {Luitz},\ and\
  \citenamefont {Bukov}}]{hahn_quantum_2023}%
  \BibitemOpen
  \bibfield  {author} {\bibinfo {author} {\bibfnamefont {D.}~\bibnamefont
  {Hahn}}, \bibinfo {author} {\bibfnamefont {M.}~\bibnamefont {Dupont}},
  \bibinfo {author} {\bibfnamefont {M.}~\bibnamefont {Schmitt}}, \bibinfo
  {author} {\bibfnamefont {D.~J.}\ \bibnamefont {Luitz}},\ and\ \bibinfo
  {author} {\bibfnamefont {M.}~\bibnamefont {Bukov}},\ }\href
  {https://doi.org/10.1103/PhysRevX.13.041023} {\bibfield  {journal} {\bibinfo
  {journal} {Phys. Rev. X}\ }\textbf {\bibinfo {volume} {13}},\ \bibinfo
  {pages} {041023} (\bibinfo {year} {2023})}\BibitemShut {NoStop}%
\bibitem [{\citenamefont {Alicki}(1979)}]{alicki_quantum_1979}%
  \BibitemOpen
  \bibfield  {author} {\bibinfo {author} {\bibfnamefont {R.}~\bibnamefont
  {Alicki}},\ }\href {https://doi.org/10.1088/0305-4470/12/5/007} {\bibfield
  {journal} {\bibinfo  {journal} {Journal of Physics A: Mathematical and
  General}\ }\textbf {\bibinfo {volume} {12}},\ \bibinfo {pages} {L103}
  (\bibinfo {year} {1979})}\BibitemShut {NoStop}%
\bibitem [{\citenamefont {Kosloff}(1984)}]{kosloff_quantum_1984}%
  \BibitemOpen
  \bibfield  {author} {\bibinfo {author} {\bibfnamefont {R.}~\bibnamefont
  {Kosloff}},\ }\href {https://doi.org/10.1063/1.446862} {\bibfield  {journal}
  {\bibinfo  {journal} {The Journal of Chemical Physics}\ }\textbf {\bibinfo
  {volume} {80}},\ \bibinfo {pages} {1625} (\bibinfo {year}
  {1984})}\BibitemShut {NoStop}%
\bibitem [{\citenamefont {Talkner}\ \emph {et~al.}(2009)\citenamefont
  {Talkner}, \citenamefont {Campisi},\ and\ \citenamefont
  {H\"anggi}}]{talkner_fluctuation_2009}%
  \BibitemOpen
  \bibfield  {author} {\bibinfo {author} {\bibfnamefont {P.}~\bibnamefont
  {Talkner}}, \bibinfo {author} {\bibfnamefont {M.}~\bibnamefont {Campisi}},\
  and\ \bibinfo {author} {\bibfnamefont {P.}~\bibnamefont {H\"anggi}},\ }\href
  {https://doi.org/10.1088/1742-5468/2009/02/P02025} {\bibfield  {journal}
  {\bibinfo  {journal} {Journal of Statistical Mechanics: Theory and
  Experiment}\ }\textbf {\bibinfo {volume} {2009}},\ \bibinfo {pages} {P02025}
  (\bibinfo {year} {2009})}\BibitemShut {NoStop}%
\bibitem [{\citenamefont {Alipour}\ \emph {et~al.}(2016)\citenamefont
  {Alipour}, \citenamefont {Benatti}, \citenamefont {Bakhshinezhad},
  \citenamefont {Afsary}, \citenamefont {Marcantoni},\ and\ \citenamefont
  {Rezakhani}}]{alipour_correlations_2016}%
  \BibitemOpen
  \bibfield  {author} {\bibinfo {author} {\bibfnamefont {S.}~\bibnamefont
  {Alipour}}, \bibinfo {author} {\bibfnamefont {F.}~\bibnamefont {Benatti}},
  \bibinfo {author} {\bibfnamefont {F.}~\bibnamefont {Bakhshinezhad}}, \bibinfo
  {author} {\bibfnamefont {M.}~\bibnamefont {Afsary}}, \bibinfo {author}
  {\bibfnamefont {S.}~\bibnamefont {Marcantoni}},\ and\ \bibinfo {author}
  {\bibfnamefont {A.~T.}\ \bibnamefont {Rezakhani}},\ }\href
  {https://doi.org/10.1038/srep35568} {\bibfield  {journal} {\bibinfo
  {journal} {Scientific Reports}\ }\textbf {\bibinfo {volume} {6}},\ \bibinfo
  {pages} {35568} (\bibinfo {year} {2016})}\BibitemShut {NoStop}%
\bibitem [{\citenamefont {Ahmadi}\ \emph {et~al.}(2023)\citenamefont {Ahmadi},
  \citenamefont {Salimi},\ and\ \citenamefont
  {Khorashad}}]{ahmadi_contribution_2023}%
  \BibitemOpen
  \bibfield  {author} {\bibinfo {author} {\bibfnamefont {B.}~\bibnamefont
  {Ahmadi}}, \bibinfo {author} {\bibfnamefont {S.}~\bibnamefont {Salimi}},\
  and\ \bibinfo {author} {\bibfnamefont {A.~S.}\ \bibnamefont {Khorashad}},\
  }\href {https://doi.org/10.1038/s41598-022-27156-0} {\bibfield  {journal}
  {\bibinfo  {journal} {Scientific Reports}\ }\textbf {\bibinfo {volume}
  {13}},\ \bibinfo {pages} {160} (\bibinfo {year} {2023})},\ \bibinfo {note}
  {publisher: Nature Publishing Group}\BibitemShut {NoStop}%
\bibitem [{\citenamefont {Binder}\ \emph {et~al.}(2015)\citenamefont {Binder},
  \citenamefont {Vinjanampathy}, \citenamefont {Modi},\ and\ \citenamefont
  {Goold}}]{binder_quantum_2015}%
  \BibitemOpen
  \bibfield  {author} {\bibinfo {author} {\bibfnamefont {F.}~\bibnamefont
  {Binder}}, \bibinfo {author} {\bibfnamefont {S.}~\bibnamefont
  {Vinjanampathy}}, \bibinfo {author} {\bibfnamefont {K.}~\bibnamefont
  {Modi}},\ and\ \bibinfo {author} {\bibfnamefont {J.}~\bibnamefont {Goold}},\
  }\href {https://doi.org/10.1103/PhysRevE.91.032119} {\bibfield  {journal}
  {\bibinfo  {journal} {Phys. Rev. E}\ }\textbf {\bibinfo {volume} {91}},\
  \bibinfo {pages} {032119} (\bibinfo {year} {2015})}\BibitemShut {NoStop}%
\bibitem [{\citenamefont {Colla}\ and\ \citenamefont
  {Breuer}(2022)}]{colla_open-system_2022}%
  \BibitemOpen
  \bibfield  {author} {\bibinfo {author} {\bibfnamefont {A.}~\bibnamefont
  {Colla}}\ and\ \bibinfo {author} {\bibfnamefont {H.-P.}\ \bibnamefont
  {Breuer}},\ }\href {https://doi.org/10.1103/PhysRevA.105.052216} {\bibfield
  {journal} {\bibinfo  {journal} {Physical Review A}\ }\textbf {\bibinfo
  {volume} {105}},\ \bibinfo {pages} {052216} (\bibinfo {year}
  {2022})}\BibitemShut {NoStop}%
\bibitem [{\citenamefont {Gallego}\ \emph {et~al.}(2014)\citenamefont
  {Gallego}, \citenamefont {Riera},\ and\ \citenamefont
  {Eisert}}]{gallego_thermal_2014}%
  \BibitemOpen
  \bibfield  {author} {\bibinfo {author} {\bibfnamefont {R.}~\bibnamefont
  {Gallego}}, \bibinfo {author} {\bibfnamefont {A.}~\bibnamefont {Riera}},\
  and\ \bibinfo {author} {\bibfnamefont {J.}~\bibnamefont {Eisert}},\ }\href
  {https://doi.org/10.1088/1367-2630/16/12/125009} {\bibfield  {journal}
  {\bibinfo  {journal} {New Journal of Physics}\ }\textbf {\bibinfo {volume}
  {16}},\ \bibinfo {pages} {125009} (\bibinfo {year} {2014})},\ \bibinfo {note}
  {publisher: IOP Publishing}\BibitemShut {NoStop}%
\bibitem [{\citenamefont {Guarnieri}\ \emph {et~al.}(2019)\citenamefont
  {Guarnieri}, \citenamefont {Ng}, \citenamefont {Modi}, \citenamefont
  {Eisert}, \citenamefont {Paternostro},\ and\ \citenamefont
  {Goold}}]{guarnieri_quantum_2019}%
  \BibitemOpen
  \bibfield  {author} {\bibinfo {author} {\bibfnamefont {G.}~\bibnamefont
  {Guarnieri}}, \bibinfo {author} {\bibfnamefont {N.~H.~Y.}\ \bibnamefont
  {Ng}}, \bibinfo {author} {\bibfnamefont {K.}~\bibnamefont {Modi}}, \bibinfo
  {author} {\bibfnamefont {J.}~\bibnamefont {Eisert}}, \bibinfo {author}
  {\bibfnamefont {M.}~\bibnamefont {Paternostro}},\ and\ \bibinfo {author}
  {\bibfnamefont {J.}~\bibnamefont {Goold}},\ }\href
  {https://doi.org/10.1103/PhysRevE.99.050101} {\bibfield  {journal} {\bibinfo
  {journal} {Physical Review E}\ }\textbf {\bibinfo {volume} {99}},\ \bibinfo
  {pages} {050101} (\bibinfo {year} {2019})}\BibitemShut {NoStop}%
\bibitem [{\citenamefont {Rivas}(2019)}]{rivas_quantum_2019}%
  \BibitemOpen
  \bibfield  {author} {\bibinfo {author} {\bibfnamefont {{\'A}.}~\bibnamefont
  {Rivas}},\ }\bibfield  {journal} {\bibinfo  {journal} {Entropy}\ }\textbf
  {\bibinfo {volume} {21}},\ \href {https://doi.org/10.3390/e21080725}
  {10.3390/e21080725} (\bibinfo {year} {2019})\BibitemShut {NoStop}%
\bibitem [{\citenamefont {Rivas}(2020)}]{rivas_strong_2020}%
  \BibitemOpen
  \bibfield  {author} {\bibinfo {author} {\bibfnamefont {{\'A}.}~\bibnamefont
  {Rivas}},\ }\href {https://doi.org/10.1103/PhysRevLett.124.160601} {\bibfield
   {journal} {\bibinfo  {journal} {Phys. Rev. Lett.}\ }\textbf {\bibinfo
  {volume} {124}},\ \bibinfo {pages} {160601} (\bibinfo {year}
  {2020})}\BibitemShut {NoStop}%
\bibitem [{\citenamefont {Silva}\ and\ \citenamefont
  {Angelo}(2021)}]{silva_quantum_2021}%
  \BibitemOpen
  \bibfield  {author} {\bibinfo {author} {\bibfnamefont {T.~A. B.~P.}\
  \bibnamefont {Silva}}\ and\ \bibinfo {author} {\bibfnamefont {R.~M.}\
  \bibnamefont {Angelo}},\ }\href {https://doi.org/10.1103/PhysRevA.104.042215}
  {\bibfield  {journal} {\bibinfo  {journal} {Phys. Rev. A}\ }\textbf {\bibinfo
  {volume} {104}},\ \bibinfo {pages} {042215} (\bibinfo {year}
  {2021})}\BibitemShut {NoStop}%
\bibitem [{\citenamefont {Sone}\ \emph {et~al.}(2020)\citenamefont {Sone},
  \citenamefont {Liu},\ and\ \citenamefont {Cappellaro}}]{sone_quantum_2020}%
  \BibitemOpen
  \bibfield  {author} {\bibinfo {author} {\bibfnamefont {A.}~\bibnamefont
  {Sone}}, \bibinfo {author} {\bibfnamefont {Y.-X.}\ \bibnamefont {Liu}},\ and\
  \bibinfo {author} {\bibfnamefont {P.}~\bibnamefont {Cappellaro}},\ }\href
  {https://doi.org/10.1103/PhysRevLett.125.060602} {\bibfield  {journal}
  {\bibinfo  {journal} {Phys. Rev. Lett.}\ }\textbf {\bibinfo {volume} {125}},\
  \bibinfo {pages} {060602} (\bibinfo {year} {2020})}\BibitemShut {NoStop}%
\bibitem [{\citenamefont {Strasberg}\ and\ \citenamefont
  {Esposito}(2019)}]{strasberg_non-markovianity_2019}%
  \BibitemOpen
  \bibfield  {author} {\bibinfo {author} {\bibfnamefont {P.}~\bibnamefont
  {Strasberg}}\ and\ \bibinfo {author} {\bibfnamefont {M.}~\bibnamefont
  {Esposito}},\ }\href {https://doi.org/10.1103/PhysRevE.99.012120} {\bibfield
  {journal} {\bibinfo  {journal} {Phys. Rev. E}\ }\textbf {\bibinfo {volume}
  {99}},\ \bibinfo {pages} {012120} (\bibinfo {year} {2019})}\BibitemShut
  {NoStop}%
\bibitem [{\citenamefont {Talkner}\ and\ \citenamefont
  {H\"anggi}(2016)}]{talkner_aspects_2016}%
  \BibitemOpen
  \bibfield  {author} {\bibinfo {author} {\bibfnamefont {P.}~\bibnamefont
  {Talkner}}\ and\ \bibinfo {author} {\bibfnamefont {P.}~\bibnamefont
  {H\"anggi}},\ }\href {https://doi.org/10.1103/PhysRevE.93.022131} {\bibfield
  {journal} {\bibinfo  {journal} {Phys. Rev. E}\ }\textbf {\bibinfo {volume}
  {93}},\ \bibinfo {pages} {022131} (\bibinfo {year} {2016})}\BibitemShut
  {NoStop}%
\bibitem [{\citenamefont {Anto-Sztrikacs}\ \emph {et~al.}(2023)\citenamefont
  {Anto-Sztrikacs}, \citenamefont {Nazir},\ and\ \citenamefont
  {Segal}}]{anto-sztrikacs_effective-hamiltonian_2023}%
  \BibitemOpen
  \bibfield  {author} {\bibinfo {author} {\bibfnamefont {N.}~\bibnamefont
  {Anto-Sztrikacs}}, \bibinfo {author} {\bibfnamefont {A.}~\bibnamefont
  {Nazir}},\ and\ \bibinfo {author} {\bibfnamefont {D.}~\bibnamefont {Segal}},\
  }\href {https://doi.org/10.1103/PRXQuantum.4.020307} {\bibfield  {journal}
  {\bibinfo  {journal} {PRX Quantum}\ }\textbf {\bibinfo {volume} {4}},\
  \bibinfo {pages} {020307} (\bibinfo {year} {2023})},\ \bibinfo {note}
  {publisher: American Physical Society}\BibitemShut {NoStop}%
\bibitem [{\citenamefont {Dann}\ and\ \citenamefont
  {Kosloff}(2023)}]{dann_unification_2023}%
  \BibitemOpen
  \bibfield  {author} {\bibinfo {author} {\bibfnamefont {R.}~\bibnamefont
  {Dann}}\ and\ \bibinfo {author} {\bibfnamefont {R.}~\bibnamefont {Kosloff}},\
  }\href {https://doi.org/10.1088/1367-2630/acc967} {\bibfield  {journal}
  {\bibinfo  {journal} {New Journal of Physics}\ }\textbf {\bibinfo {volume}
  {25}},\ \bibinfo {pages} {043019} (\bibinfo {year} {2023})}\BibitemShut
  {NoStop}%
\bibitem [{\citenamefont {Deffner}\ \emph {et~al.}(2016)\citenamefont
  {Deffner}, \citenamefont {Paz},\ and\ \citenamefont
  {Zurek}}]{deffner_quantum_2016}%
  \BibitemOpen
  \bibfield  {author} {\bibinfo {author} {\bibfnamefont {S.}~\bibnamefont
  {Deffner}}, \bibinfo {author} {\bibfnamefont {J.~P.}\ \bibnamefont {Paz}},\
  and\ \bibinfo {author} {\bibfnamefont {W.~H.}\ \bibnamefont {Zurek}},\ }\href
  {https://doi.org/10.1103/PhysRevE.94.010103} {\bibfield  {journal} {\bibinfo
  {journal} {Phys. Rev. E}\ }\textbf {\bibinfo {volume} {94}},\ \bibinfo
  {pages} {010103} (\bibinfo {year} {2016})}\BibitemShut {NoStop}%
\bibitem [{\citenamefont {Webb}\ and\ \citenamefont
  {Stafford}(2024)}]{webb_how_2024}%
  \BibitemOpen
  \bibfield  {author} {\bibinfo {author} {\bibfnamefont {C.~M.}\ \bibnamefont
  {Webb}}\ and\ \bibinfo {author} {\bibfnamefont {C.~A.}\ \bibnamefont
  {Stafford}},\ }\bibfield  {journal} {\bibinfo  {journal} {Entropy}\ }\textbf
  {\bibinfo {volume} {26}},\ \href {https://doi.org/10.3390/e26070611}
  {10.3390/e26070611} (\bibinfo {year} {2024})\BibitemShut {NoStop}%
\bibitem [{\citenamefont {Kumar}\ \emph {et~al.}(2024)\citenamefont {Kumar},
  \citenamefont {Webb},\ and\ \citenamefont {Stafford}}]{kumar_work_2024}%
  \BibitemOpen
  \bibfield  {author} {\bibinfo {author} {\bibfnamefont {P.}~\bibnamefont
  {Kumar}}, \bibinfo {author} {\bibfnamefont {C.~M.}\ \bibnamefont {Webb}},\
  and\ \bibinfo {author} {\bibfnamefont {C.~A.}\ \bibnamefont {Stafford}},\
  }\href {https://doi.org/10.1103/PhysRevLett.133.070404} {\bibfield  {journal}
  {\bibinfo  {journal} {Phys. Rev. Lett.}\ }\textbf {\bibinfo {volume} {133}},\
  \bibinfo {pages} {070404} (\bibinfo {year} {2024})}\BibitemShut {NoStop}%
\bibitem [{\citenamefont {Feynman}\ \emph {et~al.}(1963)\citenamefont
  {Feynman}, \citenamefont {Leighton},\ and\ \citenamefont
  {Sands}}]{feynman_feynman_2010}%
  \BibitemOpen
  \bibfield  {author} {\bibinfo {author} {\bibfnamefont {R.~P.}\ \bibnamefont
  {Feynman}}, \bibinfo {author} {\bibfnamefont {R.~B.}\ \bibnamefont
  {Leighton}},\ and\ \bibinfo {author} {\bibfnamefont {M.}~\bibnamefont
  {Sands}},\ }\href@noop {} {\emph {\bibinfo {title} {The Feynman Lectures on
  Physics: Volume 1}}},\ \bibinfo {edition} {2nd}\ ed.,\ \bibinfo {series} {The
  Feynman Lectures on Physics}, Vol.~\bibinfo {volume} {1}\ (\bibinfo
  {publisher} {Addison-Wesley},\ \bibinfo {address} {Boston},\ \bibinfo {year}
  {1963})\BibitemShut {NoStop}%
\bibitem [{\citenamefont {Callen}(1985)}]{callen_thermodynamics_1985}%
  \BibitemOpen
  \bibfield  {author} {\bibinfo {author} {\bibfnamefont {H.~B.}\ \bibnamefont
  {Callen}},\ }\href@noop {} {\emph {\bibinfo {title} {Thermodynamics and an
  introduction to thermostatistics; 2nd ed.}}}\ (\bibinfo  {publisher}
  {Wiley},\ \bibinfo {address} {New York, NY},\ \bibinfo {year}
  {1985})\BibitemShut {NoStop}%
\bibitem [{Note1()}]{Note1}%
  \BibitemOpen
  \bibinfo {note} {One could attribute the interaction energy to neither $S$
  nor $R$. This approach falls outside the standard thermodynamic understanding
  of heat as energy lost by $R$ to $S$. Therefore, we will not consider this
  approach.}\BibitemShut {Stop}%
\bibitem [{\citenamefont {Seifert}(2016)}]{seifert_first_2016}%
  \BibitemOpen
  \bibfield  {author} {\bibinfo {author} {\bibfnamefont {U.}~\bibnamefont
  {Seifert}},\ }\href {https://doi.org/10.1103/PhysRevLett.116.020601}
  {\bibfield  {journal} {\bibinfo  {journal} {Phys. Rev. Lett.}\ }\textbf
  {\bibinfo {volume} {116}},\ \bibinfo {pages} {020601} (\bibinfo {year}
  {2016})}\BibitemShut {NoStop}%
\bibitem [{\citenamefont {Jarzynski}(2017)}]{jarzynski_stochastic_2017}%
  \BibitemOpen
  \bibfield  {author} {\bibinfo {author} {\bibfnamefont {C.}~\bibnamefont
  {Jarzynski}},\ }\href {https://doi.org/10.1103/PhysRevX.7.011008} {\bibfield
  {journal} {\bibinfo  {journal} {Phys. Rev. X}\ }\textbf {\bibinfo {volume}
  {7}},\ \bibinfo {pages} {011008} (\bibinfo {year} {2017})}\BibitemShut
  {NoStop}%
\bibitem [{\citenamefont {Miller}(2018)}]{miller_hamiltonian_2018}%
  \BibitemOpen
  \bibfield  {author} {\bibinfo {author} {\bibfnamefont {H.~J.~D.}\
  \bibnamefont {Miller}},\ }in\ \href
  {https://doi.org/10.1007/978-3-319-99046-0_22} {\emph {\bibinfo {booktitle}
  {Thermodynamics in the {Quantum} {Regime}: {Fundamental} {Aspects} and {New}
  {Directions}}}},\ \bibinfo {series and number} {Fundamental {Theories} of
  {Physics}},\ \bibinfo {editor} {edited by\ \bibinfo {editor} {\bibfnamefont
  {F.}~\bibnamefont {Binder}}, \bibinfo {editor} {\bibfnamefont {L.~A.}\
  \bibnamefont {Correa}}, \bibinfo {editor} {\bibfnamefont {C.}~\bibnamefont
  {Gogolin}}, \bibinfo {editor} {\bibfnamefont {J.}~\bibnamefont {Anders}},\
  and\ \bibinfo {editor} {\bibfnamefont {G.}~\bibnamefont {Adesso}}}\ (\bibinfo
   {publisher} {Springer International Publishing},\ \bibinfo {address}
  {Cham},\ \bibinfo {year} {2018})\ pp.\ \bibinfo {pages}
  {531--549}\BibitemShut {NoStop}%
\bibitem [{\citenamefont {Campisi}\ \emph
  {et~al.}(2009{\natexlab{a}})\citenamefont {Campisi}, \citenamefont
  {Talkner},\ and\ \citenamefont {H\"anggi}}]{campisi_thermodynamics_2009}%
  \BibitemOpen
  \bibfield  {author} {\bibinfo {author} {\bibfnamefont {M.}~\bibnamefont
  {Campisi}}, \bibinfo {author} {\bibfnamefont {P.}~\bibnamefont {Talkner}},\
  and\ \bibinfo {author} {\bibfnamefont {P.}~\bibnamefont {H\"anggi}},\ }\href
  {https://doi.org/10.1088/1751-8113/42/39/392002} {\bibfield  {journal}
  {\bibinfo  {journal} {Journal of Physics A: Mathematical and Theoretical}\
  }\textbf {\bibinfo {volume} {42}},\ \bibinfo {pages} {392002} (\bibinfo
  {year} {2009}{\natexlab{a}})}\BibitemShut {NoStop}%
\bibitem [{\citenamefont {Hsiang}\ and\ \citenamefont
  {Hu}(2018)}]{hsiang_quantum_2018}%
  \BibitemOpen
  \bibfield  {author} {\bibinfo {author} {\bibfnamefont {J.-T.}\ \bibnamefont
  {Hsiang}}\ and\ \bibinfo {author} {\bibfnamefont {B.-L.}\ \bibnamefont
  {Hu}},\ }\bibfield  {journal} {\bibinfo  {journal} {Entropy}\ }\textbf
  {\bibinfo {volume} {20}},\ \href {https://doi.org/10.3390/e20060423}
  {10.3390/e20060423} (\bibinfo {year} {2018})\BibitemShut {NoStop}%
\bibitem [{\citenamefont {Rivas}(2017)}]{rivas_refined_2017}%
  \BibitemOpen
  \bibfield  {author} {\bibinfo {author} {\bibfnamefont {{\'A}.}~\bibnamefont
  {Rivas}},\ }\href {https://doi.org/10.1103/PhysRevA.95.042104} {\bibfield
  {journal} {\bibinfo  {journal} {Phys. Rev. A}\ }\textbf {\bibinfo {volume}
  {95}},\ \bibinfo {pages} {042104} (\bibinfo {year} {2017})}\BibitemShut
  {NoStop}%
\bibitem [{\citenamefont {Garc{\'i}a-March}\ \emph {et~al.}(2016)\citenamefont
  {Garc{\'i}a-March}, \citenamefont {Fogarty}, \citenamefont {Campbell},
  \citenamefont {Busch},\ and\ \citenamefont
  {Paternostro}}]{garciamarch_non-equilibrium_2016}%
  \BibitemOpen
  \bibfield  {author} {\bibinfo {author} {\bibfnamefont {M.~{\'A}.}\
  \bibnamefont {Garc{\'i}a-March}}, \bibinfo {author} {\bibfnamefont
  {T.}~\bibnamefont {Fogarty}}, \bibinfo {author} {\bibfnamefont
  {S.}~\bibnamefont {Campbell}}, \bibinfo {author} {\bibfnamefont
  {T.}~\bibnamefont {Busch}},\ and\ \bibinfo {author} {\bibfnamefont
  {M.}~\bibnamefont {Paternostro}},\ }\href
  {https://doi.org/10.1088/1367-2630/18/10/103035} {\bibfield  {journal}
  {\bibinfo  {journal} {New Journal of Physics}\ }\textbf {\bibinfo {volume}
  {18}},\ \bibinfo {pages} {103035} (\bibinfo {year} {2016})}\BibitemShut
  {NoStop}%
\bibitem [{\citenamefont {Perarnau-Llobet}\ \emph {et~al.}(2018)\citenamefont
  {Perarnau-Llobet}, \citenamefont {Wilming}, \citenamefont {Riera},
  \citenamefont {Gallego},\ and\ \citenamefont
  {Eisert}}]{perarnau_strong_2018}%
  \BibitemOpen
  \bibfield  {author} {\bibinfo {author} {\bibfnamefont {M.}~\bibnamefont
  {Perarnau-Llobet}}, \bibinfo {author} {\bibfnamefont {H.}~\bibnamefont
  {Wilming}}, \bibinfo {author} {\bibfnamefont {A.}~\bibnamefont {Riera}},
  \bibinfo {author} {\bibfnamefont {R.}~\bibnamefont {Gallego}},\ and\ \bibinfo
  {author} {\bibfnamefont {J.}~\bibnamefont {Eisert}},\ }\href
  {https://doi.org/10.1103/PhysRevLett.120.120602} {\bibfield  {journal}
  {\bibinfo  {journal} {Phys. Rev. Lett.}\ }\textbf {\bibinfo {volume} {120}},\
  \bibinfo {pages} {120602} (\bibinfo {year} {2018})}\BibitemShut {NoStop}%
\bibitem [{\citenamefont {Jarzynski}(2004)}]{jarzynski_nonequilibrium_2004}%
  \BibitemOpen
  \bibfield  {author} {\bibinfo {author} {\bibfnamefont {C.}~\bibnamefont
  {Jarzynski}},\ }\href {https://doi.org/10.1088/1742-5468/2004/09/P09005}
  {\bibfield  {journal} {\bibinfo  {journal} {Journal of Statistical Mechanics:
  Theory and Experiment}\ }\textbf {\bibinfo {volume} {2004}},\ \bibinfo
  {pages} {P09005} (\bibinfo {year} {2004})}\BibitemShut {NoStop}%
\bibitem [{\citenamefont {Miller}\ and\ \citenamefont
  {Anders}(2017)}]{miller_entropy_2017}%
  \BibitemOpen
  \bibfield  {author} {\bibinfo {author} {\bibfnamefont {H.~J.~D.}\
  \bibnamefont {Miller}}\ and\ \bibinfo {author} {\bibfnamefont
  {J.}~\bibnamefont {Anders}},\ }\href
  {https://doi.org/10.1103/PhysRevE.95.062123} {\bibfield  {journal} {\bibinfo
  {journal} {Physical Review E}\ }\textbf {\bibinfo {volume} {95}},\ \bibinfo
  {pages} {062123} (\bibinfo {year} {2017})}\BibitemShut {NoStop}%
\bibitem [{\citenamefont {Philbin}\ and\ \citenamefont
  {Anders}(2016)}]{philbin_thermal_2016}%
  \BibitemOpen
  \bibfield  {author} {\bibinfo {author} {\bibfnamefont {T.~G.}\ \bibnamefont
  {Philbin}}\ and\ \bibinfo {author} {\bibfnamefont {J.}~\bibnamefont
  {Anders}},\ }\href {https://doi.org/10.1088/1751-8113/49/21/215303}
  {\bibfield  {journal} {\bibinfo  {journal} {Journal of Physics A:
  Mathematical and Theoretical}\ }\textbf {\bibinfo {volume} {49}},\ \bibinfo
  {pages} {215303} (\bibinfo {year} {2016})}\BibitemShut {NoStop}%
\bibitem [{\citenamefont {Trushechkin}\ \emph {et~al.}(2022)\citenamefont
  {Trushechkin}, \citenamefont {Merkli}, \citenamefont {Cresser},\ and\
  \citenamefont {Anders}}]{trushechkin_open_2022}%
  \BibitemOpen
  \bibfield  {author} {\bibinfo {author} {\bibfnamefont {A.~S.}\ \bibnamefont
  {Trushechkin}}, \bibinfo {author} {\bibfnamefont {M.}~\bibnamefont {Merkli}},
  \bibinfo {author} {\bibfnamefont {J.~D.}\ \bibnamefont {Cresser}},\ and\
  \bibinfo {author} {\bibfnamefont {J.}~\bibnamefont {Anders}},\ }\href
  {https://doi.org/10.1116/5.0073853} {\bibfield  {journal} {\bibinfo
  {journal} {AVS Quantum Science}\ }\textbf {\bibinfo {volume} {4}},\ \bibinfo
  {pages} {012301} (\bibinfo {year} {2022})}\BibitemShut {NoStop}%
\bibitem [{\citenamefont {Campisi}\ \emph
  {et~al.}(2009{\natexlab{b}})\citenamefont {Campisi}, \citenamefont
  {Talkner},\ and\ \citenamefont {H\"anggi}}]{campisi_fluctuation_2009}%
  \BibitemOpen
  \bibfield  {author} {\bibinfo {author} {\bibfnamefont {M.}~\bibnamefont
  {Campisi}}, \bibinfo {author} {\bibfnamefont {P.}~\bibnamefont {Talkner}},\
  and\ \bibinfo {author} {\bibfnamefont {P.}~\bibnamefont {H\"anggi}},\ }\href
  {https://doi.org/10.1103/PhysRevLett.102.210401} {\bibfield  {journal}
  {\bibinfo  {journal} {Phys. Rev. Lett.}\ }\textbf {\bibinfo {volume} {102}},\
  \bibinfo {pages} {210401} (\bibinfo {year} {2009}{\natexlab{b}})}\BibitemShut
  {NoStop}%
\bibitem [{Note2()}]{Note2}%
  \BibitemOpen
  \bibinfo {note} {Reference~\cite {seifert_first_2016} introduced the
  classical analog of $\protect \cc@accent {"705E}{E}_S^*$ as an
  internal-energy observable. The justification relied on the equivalence of
  $U_{\protect \mathrm {diff}}$ and $U_{E^*}$ under condition (i).}\BibitemShut
  {Stop}%
\bibitem [{\citenamefont {Jarzynski}(2011)}]{jarzynski_equalities_2011}%
  \BibitemOpen
  \bibfield  {author} {\bibinfo {author} {\bibfnamefont {C.}~\bibnamefont
  {Jarzynski}},\ }\href
  {https://doi.org/https://doi.org/10.1146/annurev-conmatphys-062910-140506}
  {\bibfield  {journal} {\bibinfo  {journal} {Annual Review of Condensed Matter
  Physics}\ }\textbf {\bibinfo {volume} {2}},\ \bibinfo {pages} {329} (\bibinfo
  {year} {2011})}\BibitemShut {NoStop}%
\bibitem [{\citenamefont {Seifert}(2012)}]{seifert_stochastic_2012}%
  \BibitemOpen
  \bibfield  {author} {\bibinfo {author} {\bibfnamefont {U.}~\bibnamefont
  {Seifert}},\ }\href {https://doi.org/10.1088/0034-4885/75/12/126001}
  {\bibfield  {journal} {\bibinfo  {journal} {Rep. Prog. Phys.}\ }\textbf
  {\bibinfo {volume} {75}},\ \bibinfo {pages} {126001} (\bibinfo {year}
  {2012})}\BibitemShut {NoStop}%
\bibitem [{\citenamefont {Joshi}\ \emph {et~al.}(2023)\citenamefont {Joshi},
  \citenamefont {Kokail}, \citenamefont {van Bijnen}, \citenamefont {Kranzl},
  \citenamefont {Zache}, \citenamefont {Blatt}, \citenamefont {Roos},\ and\
  \citenamefont {Zoller}}]{joshi_exploring_2023}%
  \BibitemOpen
  \bibfield  {author} {\bibinfo {author} {\bibfnamefont {M.~K.}\ \bibnamefont
  {Joshi}}, \bibinfo {author} {\bibfnamefont {C.}~\bibnamefont {Kokail}},
  \bibinfo {author} {\bibfnamefont {R.}~\bibnamefont {van Bijnen}}, \bibinfo
  {author} {\bibfnamefont {F.}~\bibnamefont {Kranzl}}, \bibinfo {author}
  {\bibfnamefont {T.~V.}\ \bibnamefont {Zache}}, \bibinfo {author}
  {\bibfnamefont {R.}~\bibnamefont {Blatt}}, \bibinfo {author} {\bibfnamefont
  {C.~F.}\ \bibnamefont {Roos}},\ and\ \bibinfo {author} {\bibfnamefont
  {P.}~\bibnamefont {Zoller}},\ }\href
  {https://doi.org/10.1038/s41586-023-06768-0} {\bibfield  {journal} {\bibinfo
  {journal} {Nature}\ }\textbf {\bibinfo {volume} {624}},\ \bibinfo {pages}
  {539} (\bibinfo {year} {2023})}\BibitemShut {NoStop}%
\bibitem [{\citenamefont {Fusco}\ \emph {et~al.}(2014)\citenamefont {Fusco},
  \citenamefont {Pigeon}, \citenamefont {Apollaro}, \citenamefont {Xuereb},
  \citenamefont {Mazzola}, \citenamefont {Campisi}, \citenamefont {Ferraro},
  \citenamefont {Paternostro},\ and\ \citenamefont
  {De~Chiara}}]{fusco_assessing_2014}%
  \BibitemOpen
  \bibfield  {author} {\bibinfo {author} {\bibfnamefont {L.}~\bibnamefont
  {Fusco}}, \bibinfo {author} {\bibfnamefont {S.}~\bibnamefont {Pigeon}},
  \bibinfo {author} {\bibfnamefont {T.~J.~G.}\ \bibnamefont {Apollaro}},
  \bibinfo {author} {\bibfnamefont {A.}~\bibnamefont {Xuereb}}, \bibinfo
  {author} {\bibfnamefont {L.}~\bibnamefont {Mazzola}}, \bibinfo {author}
  {\bibfnamefont {M.}~\bibnamefont {Campisi}}, \bibinfo {author} {\bibfnamefont
  {A.}~\bibnamefont {Ferraro}}, \bibinfo {author} {\bibfnamefont
  {M.}~\bibnamefont {Paternostro}},\ and\ \bibinfo {author} {\bibfnamefont
  {G.}~\bibnamefont {De~Chiara}},\ }\href
  {https://doi.org/10.1103/PhysRevX.4.031029} {\bibfield  {journal} {\bibinfo
  {journal} {Phys. Rev. X}\ }\textbf {\bibinfo {volume} {4}},\ \bibinfo {pages}
  {031029} (\bibinfo {year} {2014})}\BibitemShut {NoStop}%
\bibitem [{\citenamefont {Jurcevic}\ \emph {et~al.}(2017)\citenamefont
  {Jurcevic}, \citenamefont {Shen}, \citenamefont {Hauke}, \citenamefont
  {Maier}, \citenamefont {Brydges}, \citenamefont {Hempel}, \citenamefont
  {Lanyon}, \citenamefont {Heyl}, \citenamefont {Blatt},\ and\ \citenamefont
  {Roos}}]{jurcevic_direct_2017}%
  \BibitemOpen
  \bibfield  {author} {\bibinfo {author} {\bibfnamefont {P.}~\bibnamefont
  {Jurcevic}}, \bibinfo {author} {\bibfnamefont {H.}~\bibnamefont {Shen}},
  \bibinfo {author} {\bibfnamefont {P.}~\bibnamefont {Hauke}}, \bibinfo
  {author} {\bibfnamefont {C.}~\bibnamefont {Maier}}, \bibinfo {author}
  {\bibfnamefont {T.}~\bibnamefont {Brydges}}, \bibinfo {author} {\bibfnamefont
  {C.}~\bibnamefont {Hempel}}, \bibinfo {author} {\bibfnamefont {B.~P.}\
  \bibnamefont {Lanyon}}, \bibinfo {author} {\bibfnamefont {M.}~\bibnamefont
  {Heyl}}, \bibinfo {author} {\bibfnamefont {R.}~\bibnamefont {Blatt}},\ and\
  \bibinfo {author} {\bibfnamefont {C.~F.}\ \bibnamefont {Roos}},\ }\href
  {https://doi.org/10.1103/PhysRevLett.119.080501} {\bibfield  {journal}
  {\bibinfo  {journal} {Phys. Rev. Lett.}\ }\textbf {\bibinfo {volume} {119}},\
  \bibinfo {pages} {080501} (\bibinfo {year} {2017})}\BibitemShut {NoStop}%
\bibitem [{\citenamefont {Eisert}\ \emph {et~al.}(2015)\citenamefont {Eisert},
  \citenamefont {Friesdorf},\ and\ \citenamefont
  {Gogolin}}]{eisert_quantum_2015}%
  \BibitemOpen
  \bibfield  {author} {\bibinfo {author} {\bibfnamefont {J.}~\bibnamefont
  {Eisert}}, \bibinfo {author} {\bibfnamefont {M.}~\bibnamefont {Friesdorf}},\
  and\ \bibinfo {author} {\bibfnamefont {C.}~\bibnamefont {Gogolin}},\ }\href
  {https://doi.org/10.1038/nphys3215} {\bibfield  {journal} {\bibinfo
  {journal} {Nature Physics}\ }\textbf {\bibinfo {volume} {11}},\ \bibinfo
  {pages} {124} (\bibinfo {year} {2015})}\BibitemShut {NoStop}%
\bibitem [{\citenamefont {Qi}\ \emph {et~al.}(2019)\citenamefont {Qi},
  \citenamefont {Davis}, \citenamefont {Periwal},\ and\ \citenamefont
  {Schleier-Smith}}]{qi_measuring_2019}%
  \BibitemOpen
  \bibfield  {author} {\bibinfo {author} {\bibfnamefont {X.-L.}\ \bibnamefont
  {Qi}}, \bibinfo {author} {\bibfnamefont {E.~J.}\ \bibnamefont {Davis}},
  \bibinfo {author} {\bibfnamefont {A.}~\bibnamefont {Periwal}},\ and\ \bibinfo
  {author} {\bibfnamefont {M.}~\bibnamefont {Schleier-Smith}},\ }\href@noop {}
  {\bibinfo {title} {Measuring operator size growth in quantum quench
  experiments}} (\bibinfo {year} {2019}),\ \Eprint
  {https://arxiv.org/abs/1906.00524} {arXiv:1906.00524 [quant-ph]} \BibitemShut
  {NoStop}%
\bibitem [{\citenamefont {Alba}\ and\ \citenamefont
  {Calabrese}(2017)}]{alba_entanglement_2017}%
  \BibitemOpen
  \bibfield  {author} {\bibinfo {author} {\bibfnamefont {V.}~\bibnamefont
  {Alba}}\ and\ \bibinfo {author} {\bibfnamefont {P.}~\bibnamefont
  {Calabrese}},\ }\href {https://doi.org/10.1073/pnas.1703516114} {\bibfield
  {journal} {\bibinfo  {journal} {Proceedings of the National Academy of
  Sciences}\ }\textbf {\bibinfo {volume} {114}},\ \bibinfo {pages} {7947}
  (\bibinfo {year} {2017})}\BibitemShut {NoStop}%
\bibitem [{\citenamefont {Dorner}\ \emph {et~al.}(2012)\citenamefont {Dorner},
  \citenamefont {Goold}, \citenamefont {Cormick}, \citenamefont {Paternostro},\
  and\ \citenamefont {Vedral}}]{dorner_emergent_2012}%
  \BibitemOpen
  \bibfield  {author} {\bibinfo {author} {\bibfnamefont {R.}~\bibnamefont
  {Dorner}}, \bibinfo {author} {\bibfnamefont {J.}~\bibnamefont {Goold}},
  \bibinfo {author} {\bibfnamefont {C.}~\bibnamefont {Cormick}}, \bibinfo
  {author} {\bibfnamefont {M.}~\bibnamefont {Paternostro}},\ and\ \bibinfo
  {author} {\bibfnamefont {V.}~\bibnamefont {Vedral}},\ }\href
  {https://doi.org/10.1103/PhysRevLett.109.160601} {\bibfield  {journal}
  {\bibinfo  {journal} {Phys. Rev. Lett.}\ }\textbf {\bibinfo {volume} {109}},\
  \bibinfo {pages} {160601} (\bibinfo {year} {2012})}\BibitemShut {NoStop}%
\bibitem [{\citenamefont {Joshi}\ and\ \citenamefont
  {Campisi}(2013)}]{joshi_quantum_2013}%
  \BibitemOpen
  \bibfield  {author} {\bibinfo {author} {\bibfnamefont {D.~G.}\ \bibnamefont
  {Joshi}}\ and\ \bibinfo {author} {\bibfnamefont {M.}~\bibnamefont
  {Campisi}},\ }\href {https://doi.org/10.1140/epjb/e2013-40003-x} {\bibfield
  {journal} {\bibinfo  {journal} {The European Physical Journal B}\ }\textbf
  {\bibinfo {volume} {86}},\ \bibinfo {pages} {157} (\bibinfo {year}
  {2013})}\BibitemShut {NoStop}%
\bibitem [{\citenamefont {Canovi}\ \emph {et~al.}(2011)\citenamefont {Canovi},
  \citenamefont {Rossini}, \citenamefont {Fazio}, \citenamefont {Santoro},\
  and\ \citenamefont {Silva}}]{canovi_quantum_2011}%
  \BibitemOpen
  \bibfield  {author} {\bibinfo {author} {\bibfnamefont {E.}~\bibnamefont
  {Canovi}}, \bibinfo {author} {\bibfnamefont {D.}~\bibnamefont {Rossini}},
  \bibinfo {author} {\bibfnamefont {R.}~\bibnamefont {Fazio}}, \bibinfo
  {author} {\bibfnamefont {G.~E.}\ \bibnamefont {Santoro}},\ and\ \bibinfo
  {author} {\bibfnamefont {A.}~\bibnamefont {Silva}},\ }\href
  {https://doi.org/10.1103/PhysRevB.83.094431} {\bibfield  {journal} {\bibinfo
  {journal} {Phys. Rev. B}\ }\textbf {\bibinfo {volume} {83}},\ \bibinfo
  {pages} {094431} (\bibinfo {year} {2011})}\BibitemShut {NoStop}%
\bibitem [{\citenamefont {Zvyagin}(2016)}]{zvyagin_dynamical_2016}%
  \BibitemOpen
  \bibfield  {author} {\bibinfo {author} {\bibfnamefont {A.~A.}\ \bibnamefont
  {Zvyagin}},\ }\href {https://doi.org/10.1063/1.4969869} {\bibfield  {journal}
  {\bibinfo  {journal} {Low Temperature Physics}\ }\textbf {\bibinfo {volume}
  {42}},\ \bibinfo {pages} {971} (\bibinfo {year} {2016})}\BibitemShut
  {NoStop}%
\bibitem [{\citenamefont {Mitra}(2018)}]{mitra_quantum_2018}%
  \BibitemOpen
  \bibfield  {author} {\bibinfo {author} {\bibfnamefont {A.}~\bibnamefont
  {Mitra}},\ }\href
  {https://doi.org/https://doi.org/10.1146/annurev-conmatphys-031016-025451}
  {\bibfield  {journal} {\bibinfo  {journal} {Annual Review of Condensed Matter
  Physics}\ }\textbf {\bibinfo {volume} {9}},\ \bibinfo {pages} {245} (\bibinfo
  {year} {2018})}\BibitemShut {NoStop}%
\bibitem [{\citenamefont {Heyl}(2018)}]{heyl_dynamical_2018}%
  \BibitemOpen
  \bibfield  {author} {\bibinfo {author} {\bibfnamefont {M.}~\bibnamefont
  {Heyl}},\ }\href {https://doi.org/10.1088/1361-6633/aaaf9a} {\bibfield
  {journal} {\bibinfo  {journal} {Reports on Progress in Physics}\ }\textbf
  {\bibinfo {volume} {81}},\ \bibinfo {pages} {054001} (\bibinfo {year}
  {2018})}\BibitemShut {NoStop}%
\bibitem [{\citenamefont {Arrais}\ \emph {et~al.}(2018)\citenamefont {Arrais},
  \citenamefont {Wisniacki}, \citenamefont {C\'eleri}, \citenamefont
  {de~Almeida}, \citenamefont {Roncaglia},\ and\ \citenamefont
  {Toscano}}]{arrais_quantum_2018}%
  \BibitemOpen
  \bibfield  {author} {\bibinfo {author} {\bibfnamefont {E.~G.}\ \bibnamefont
  {Arrais}}, \bibinfo {author} {\bibfnamefont {D.~A.}\ \bibnamefont
  {Wisniacki}}, \bibinfo {author} {\bibfnamefont {L.~C.}\ \bibnamefont
  {C\'eleri}}, \bibinfo {author} {\bibfnamefont {N.~G.}\ \bibnamefont
  {de~Almeida}}, \bibinfo {author} {\bibfnamefont {A.~J.}\ \bibnamefont
  {Roncaglia}},\ and\ \bibinfo {author} {\bibfnamefont {F.}~\bibnamefont
  {Toscano}},\ }\href {https://doi.org/10.1103/PhysRevE.98.012106} {\bibfield
  {journal} {\bibinfo  {journal} {Phys. Rev. E}\ }\textbf {\bibinfo {volume}
  {98}},\ \bibinfo {pages} {012106} (\bibinfo {year} {2018})}\BibitemShut
  {NoStop}%
\bibitem [{\citenamefont {De~Grandi}\ \emph {et~al.}(2010)\citenamefont
  {De~Grandi}, \citenamefont {Gritsev},\ and\ \citenamefont
  {Polkovnikov}}]{degrandi_quench_2010}%
  \BibitemOpen
  \bibfield  {author} {\bibinfo {author} {\bibfnamefont {C.}~\bibnamefont
  {De~Grandi}}, \bibinfo {author} {\bibfnamefont {V.}~\bibnamefont {Gritsev}},\
  and\ \bibinfo {author} {\bibfnamefont {A.}~\bibnamefont {Polkovnikov}},\
  }\href {https://doi.org/10.1103/PhysRevB.81.224301} {\bibfield  {journal}
  {\bibinfo  {journal} {Phys. Rev. B}\ }\textbf {\bibinfo {volume} {81}},\
  \bibinfo {pages} {224301} (\bibinfo {year} {2010})}\BibitemShut {NoStop}%
\bibitem [{\citenamefont {Zhang}\ \emph {et~al.}(2017)\citenamefont {Zhang},
  \citenamefont {Pagano}, \citenamefont {Hess}, \citenamefont {Kyprianidis},
  \citenamefont {Becker}, \citenamefont {Kaplan}, \citenamefont {Gorshkov},
  \citenamefont {Gong},\ and\ \citenamefont {Monroe}}]{zhang_obersvation_2017}%
  \BibitemOpen
  \bibfield  {author} {\bibinfo {author} {\bibfnamefont {J.}~\bibnamefont
  {Zhang}}, \bibinfo {author} {\bibfnamefont {G.}~\bibnamefont {Pagano}},
  \bibinfo {author} {\bibfnamefont {P.~W.}\ \bibnamefont {Hess}}, \bibinfo
  {author} {\bibfnamefont {A.}~\bibnamefont {Kyprianidis}}, \bibinfo {author}
  {\bibfnamefont {P.}~\bibnamefont {Becker}}, \bibinfo {author} {\bibfnamefont
  {H.}~\bibnamefont {Kaplan}}, \bibinfo {author} {\bibfnamefont {A.~V.}\
  \bibnamefont {Gorshkov}}, \bibinfo {author} {\bibfnamefont {Z.-X.}\
  \bibnamefont {Gong}},\ and\ \bibinfo {author} {\bibfnamefont
  {C.}~\bibnamefont {Monroe}},\ }\href@noop {} {\bibfield  {journal} {\bibinfo
  {journal} {Nature}\ }\textbf {\bibinfo {volume} {551}},\ \bibinfo {pages}
  {601} (\bibinfo {year} {2017})}\BibitemShut {NoStop}%
\bibitem [{Note3()}]{Note3}%
  \BibitemOpen
  \bibinfo {note} {The quench satisfies the sudden approximation~\cite
  {sakurai_modern_1993} because it occurs instantaneously.}\BibitemShut {Stop}%
\bibitem [{\citenamefont {Jarzynski}(1997)}]{jarzynski_nonequilibrium_1997}%
  \BibitemOpen
  \bibfield  {author} {\bibinfo {author} {\bibfnamefont {C.}~\bibnamefont
  {Jarzynski}},\ }\href {https://doi.org/10.1103/PhysRevLett.78.2690}
  {\bibfield  {journal} {\bibinfo  {journal} {Phys. Rev. Lett.}\ }\textbf
  {\bibinfo {volume} {78}},\ \bibinfo {pages} {2690} (\bibinfo {year}
  {1997})}\BibitemShut {NoStop}%
\bibitem [{\citenamefont {Peliti}\ and\ \citenamefont
  {Pigolotti}(2021)}]{peliti_stochastic_2021}%
  \BibitemOpen
  \bibfield  {author} {\bibinfo {author} {\bibfnamefont {L.}~\bibnamefont
  {Peliti}}\ and\ \bibinfo {author} {\bibfnamefont {S.}~\bibnamefont
  {Pigolotti}},\ }\href@noop {} {\emph {\bibinfo {title} {Stochastic
  Thermodynamics: An Introduction}}}\ (\bibinfo  {publisher} {Princeton
  University Press},\ \bibinfo {year} {2021})\BibitemShut {NoStop}%
\bibitem [{\citenamefont
  {Schr\"{o}dinger}(1989)}]{schrodinger_statistical_1989}%
  \BibitemOpen
  \bibfield  {author} {\bibinfo {author} {\bibfnamefont {E.}~\bibnamefont
  {Schr\"{o}dinger}},\ }\href@noop {} {\emph {\bibinfo {title} {Statistical
  Thermodynamics}}}\ (\bibinfo  {publisher} {Dover Publications},\ \bibinfo
  {year} {1989})\BibitemShut {NoStop}%
\bibitem [{Note4()}]{Note4}%
  \BibitemOpen
  \bibinfo {note} {If ${S \cup R}$ is not isolated, then $Q_\protect \mathrm
  {diff}$ is the change in ${S \cup R}$'s energy between $t=0^+$ and
  $t=t_{\protect \rm f}$. This change represents an exchange of heat not by $S$
  and $R$, but by ${S \cup R}$ and an external heat bath.}\BibitemShut {Stop}%
\bibitem [{Note5()}]{Note5}%
  \BibitemOpen
  \bibinfo {note} {\protect \Cref {eq:genwsecondlaw} commonly expresses the
  second law when a system (in contact with one thermal reservoir) begins and
  ends in equilibrium. This statement remains true if the system ends in a
  nonequilibrium state, upon beginning in equilibrium. See comment 2 at the end
  of Sec.~2.1 in Ref.~\cite {jarzynski_equalities_2011}.}\BibitemShut {Stop}%
\bibitem [{Note6()}]{Note6}%
  \BibitemOpen
  \bibinfo {note} {The quantum relative entropy between density matrices $\rho
  $ and $\rho ^{\prime }$ is $D(\rho || \rho ^{\prime }) \protect \coloneq
  {\protect \mathrm {Tr}}(\rho [ \ln {\rho } - \ln { \rho ^{\prime }} ] ) \geq
  0$~\cite
  {uhlmann_relative_1977,lindblad_expectations_1974,donald_relative_1986}.}\BibitemShut
  {Stop}%
\bibitem [{Note7()}]{Note7}%
  \BibitemOpen
  \bibinfo {note} {Because $\protect \cc@accent {"705E}{\pi }_S$ is a Gibbs
  state relative to $\protect \cc@accent {"705E}{H}_S^*$, one can rewrite
  $\beta (\protect \cc@accent {"705E}{H}_S^*- \Delta F_S)$ as $\ln {\protect
  \cc@accent {"705E}{\pi }_S}$. Combining this relation with the relative
  entropy's non-negativity yields the second law, $\beta (W_{H^*} -\Delta
  F_S)\geq 0$. Since $\protect \cc@accent {"705E}{\pi }_S$ is not a Gibbs state
  relative to $\protect \cc@accent {"705E}{E}_S^*$, one cannot similarly derive
  a second law for $W_{E^*}$.}\BibitemShut {Stop}%
\bibitem [{\citenamefont {Srednicki}(1994)}]{srednicki_chaos_1994}%
  \BibitemOpen
  \bibfield  {author} {\bibinfo {author} {\bibfnamefont {M.}~\bibnamefont
  {Srednicki}},\ }\href {https://doi.org/10.1103/PhysRevE.50.888} {\bibfield
  {journal} {\bibinfo  {journal} {Phys. Rev. E}\ }\textbf {\bibinfo {volume}
  {50}},\ \bibinfo {pages} {888} (\bibinfo {year} {1994})}\BibitemShut
  {NoStop}%
\bibitem [{\citenamefont {Davoudi}\ \emph {et~al.}(2024)\citenamefont
  {Davoudi}, \citenamefont {Jarzynski}, \citenamefont {Mueller}, \citenamefont
  {Oruganti}, \citenamefont {Powers},\ and\ \citenamefont {{Yunger
  Halpern}}}]{davoudi_quantum_2024}%
  \BibitemOpen
  \bibfield  {author} {\bibinfo {author} {\bibfnamefont {Z.}~\bibnamefont
  {Davoudi}}, \bibinfo {author} {\bibfnamefont {C.}~\bibnamefont {Jarzynski}},
  \bibinfo {author} {\bibfnamefont {N.}~\bibnamefont {Mueller}}, \bibinfo
  {author} {\bibfnamefont {G.}~\bibnamefont {Oruganti}}, \bibinfo {author}
  {\bibfnamefont {C.}~\bibnamefont {Powers}},\ and\ \bibinfo {author}
  {\bibfnamefont {N.}~\bibnamefont {{Yunger Halpern}}},\ }\href
  {https://doi.org/10.1103/PhysRevLett.133.250402} {\bibfield  {journal}
  {\bibinfo  {journal} {Phys. Rev. Lett.}\ }\textbf {\bibinfo {volume} {133}},\
  \bibinfo {pages} {250402} (\bibinfo {year} {2024})}\BibitemShut {NoStop}%
\bibitem [{\citenamefont {Aitchison}\ and\ \citenamefont
  {Hey}(2012)}]{aitchison_gauge_2012}%
  \BibitemOpen
  \bibfield  {author} {\bibinfo {author} {\bibfnamefont {I.~J.}\ \bibnamefont
  {Aitchison}}\ and\ \bibinfo {author} {\bibfnamefont {A.~J.}\ \bibnamefont
  {Hey}},\ }\href@noop {} {\emph {\bibinfo {title} {Gauge Theories in Particle
  Physics: A Practical Introduction, -2 Volume set}}}\ (\bibinfo  {publisher}
  {Taylor \& Francis},\ \bibinfo {year} {2012})\BibitemShut {NoStop}%
\bibitem [{\citenamefont {Quigg}(2021)}]{quigg_gauge_2021}%
  \BibitemOpen
  \bibfield  {author} {\bibinfo {author} {\bibfnamefont {C.}~\bibnamefont
  {Quigg}},\ }\href@noop {} {\emph {\bibinfo {title} {Gauge theories of strong,
  weak, and electromagnetic interactions}}}\ (\bibinfo  {publisher} {CRC
  Press},\ \bibinfo {year} {2021})\BibitemShut {NoStop}%
\bibitem [{\citenamefont {Fradkin}(2013)}]{fradkin_field_2013}%
  \BibitemOpen
  \bibfield  {author} {\bibinfo {author} {\bibfnamefont {E.}~\bibnamefont
  {Fradkin}},\ }\href@noop {} {\emph {\bibinfo {title} {Field theories of
  condensed matter physics}}}\ (\bibinfo  {publisher} {Cambridge University
  Press},\ \bibinfo {year} {2013})\BibitemShut {NoStop}%
\bibitem [{\citenamefont {Kleinert}(1989)}]{kleinert_gauge_1989}%
  \BibitemOpen
  \bibfield  {author} {\bibinfo {author} {\bibfnamefont {H.}~\bibnamefont
  {Kleinert}},\ }\href@noop {} {\emph {\bibinfo {title} {Gauge Fields in
  Condensed Matter: Vol. 1: Superflow and Vortex Lines (Disorder Fields, Phase
  Transitions) Vol. 2: Stresses and Defects (Differential Geometry, Crystal
  Melting)}}}\ (\bibinfo  {publisher} {World Scientific},\ \bibinfo {year}
  {1989})\BibitemShut {NoStop}%
\bibitem [{\citenamefont {Wen}(1990)}]{wen_topological_1990}%
  \BibitemOpen
  \bibfield  {author} {\bibinfo {author} {\bibfnamefont {X.-G.}\ \bibnamefont
  {Wen}},\ }\href@noop {} {\bibfield  {journal} {\bibinfo  {journal}
  {International Journal of Modern Physics B}\ }\textbf {\bibinfo {volume}
  {4}},\ \bibinfo {pages} {239} (\bibinfo {year} {1990})}\BibitemShut {NoStop}%
\bibitem [{\citenamefont {Levin}\ and\ \citenamefont
  {Wen}(2005)}]{levin_string_2005}%
  \BibitemOpen
  \bibfield  {author} {\bibinfo {author} {\bibfnamefont {M.~A.}\ \bibnamefont
  {Levin}}\ and\ \bibinfo {author} {\bibfnamefont {X.-G.}\ \bibnamefont
  {Wen}},\ }\href@noop {} {\bibfield  {journal} {\bibinfo  {journal} {Physical
  Review B}\ }\textbf {\bibinfo {volume} {71}},\ \bibinfo {pages} {045110}
  (\bibinfo {year} {2005})}\BibitemShut {NoStop}%
\bibitem [{\citenamefont {Chen}\ \emph {et~al.}(2018)\citenamefont {Chen},
  \citenamefont {Kapustin},\ and\ \citenamefont
  {Radi{\v{c}}evi{\'c}}}]{chen_exact_2018}%
  \BibitemOpen
  \bibfield  {author} {\bibinfo {author} {\bibfnamefont {Y.-A.}\ \bibnamefont
  {Chen}}, \bibinfo {author} {\bibfnamefont {A.}~\bibnamefont {Kapustin}},\
  and\ \bibinfo {author} {\bibfnamefont {D.}~\bibnamefont
  {Radi{\v{c}}evi{\'c}}},\ }\href@noop {} {\bibfield  {journal} {\bibinfo
  {journal} {Annals of Physics}\ }\textbf {\bibinfo {volume} {393}},\ \bibinfo
  {pages} {234} (\bibinfo {year} {2018})}\BibitemShut {NoStop}%
\bibitem [{\citenamefont {Chen}(2020)}]{chen_exact_2020}%
  \BibitemOpen
  \bibfield  {author} {\bibinfo {author} {\bibfnamefont {Y.-A.}\ \bibnamefont
  {Chen}},\ }\href@noop {} {\bibfield  {journal} {\bibinfo  {journal} {Physical
  Review Research}\ }\textbf {\bibinfo {volume} {2}},\ \bibinfo {pages}
  {033527} (\bibinfo {year} {2020})}\BibitemShut {NoStop}%
\bibitem [{\citenamefont {Chen}\ and\ \citenamefont
  {Xu}(2023)}]{chen_equivalence_2023}%
  \BibitemOpen
  \bibfield  {author} {\bibinfo {author} {\bibfnamefont {Y.-A.}\ \bibnamefont
  {Chen}}\ and\ \bibinfo {author} {\bibfnamefont {Y.}~\bibnamefont {Xu}},\
  }\href@noop {} {\bibfield  {journal} {\bibinfo  {journal} {PRX Quantum}\
  }\textbf {\bibinfo {volume} {4}},\ \bibinfo {pages} {010326} (\bibinfo {year}
  {2023})}\BibitemShut {NoStop}%
\bibitem [{\citenamefont {Kitaev}(2003)}]{kitaev_fault_2003}%
  \BibitemOpen
  \bibfield  {author} {\bibinfo {author} {\bibfnamefont {A.~Y.}\ \bibnamefont
  {Kitaev}},\ }\href@noop {} {\bibfield  {journal} {\bibinfo  {journal} {Annals
  of physics}\ }\textbf {\bibinfo {volume} {303}},\ \bibinfo {pages} {2}
  (\bibinfo {year} {2003})}\BibitemShut {NoStop}%
\bibitem [{\citenamefont {Kitaev}(2006)}]{kitaev_anyons_2006}%
  \BibitemOpen
  \bibfield  {author} {\bibinfo {author} {\bibfnamefont {A.}~\bibnamefont
  {Kitaev}},\ }\href {https://doi.org/10.1016/j.aop.2005.10.005} {\bibfield
  {journal} {\bibinfo  {journal} {Annals of Physics}\ }\textbf {\bibinfo
  {volume} {321}},\ \bibinfo {pages} {2} (\bibinfo {year} {2006})}\BibitemShut
  {NoStop}%
\bibitem [{\citenamefont {Das~Sarma}\ \emph {et~al.}(2006)\citenamefont
  {Das~Sarma}, \citenamefont {Freedman},\ and\ \citenamefont
  {Nayak}}]{sarma_topological_2006}%
  \BibitemOpen
  \bibfield  {author} {\bibinfo {author} {\bibfnamefont {S.}~\bibnamefont
  {Das~Sarma}}, \bibinfo {author} {\bibfnamefont {M.}~\bibnamefont
  {Freedman}},\ and\ \bibinfo {author} {\bibfnamefont {C.}~\bibnamefont
  {Nayak}},\ }\href {https://doi.org/10.1063/1.2337825} {\bibfield  {journal}
  {\bibinfo  {journal} {Physics Today}\ }\textbf {\bibinfo {volume} {59}},\
  \bibinfo {pages} {32} (\bibinfo {year} {2006})}\BibitemShut {NoStop}%
\bibitem [{\citenamefont {Nayak}\ \emph {et~al.}(2008)\citenamefont {Nayak},
  \citenamefont {Simon}, \citenamefont {Stern}, \citenamefont {Freedman},\ and\
  \citenamefont {Das~Sarma}}]{nayak_non_2008}%
  \BibitemOpen
  \bibfield  {author} {\bibinfo {author} {\bibfnamefont {C.}~\bibnamefont
  {Nayak}}, \bibinfo {author} {\bibfnamefont {S.~H.}\ \bibnamefont {Simon}},
  \bibinfo {author} {\bibfnamefont {A.}~\bibnamefont {Stern}}, \bibinfo
  {author} {\bibfnamefont {M.}~\bibnamefont {Freedman}},\ and\ \bibinfo
  {author} {\bibfnamefont {S.}~\bibnamefont {Das~Sarma}},\ }\href
  {https://doi.org/10.1103/RevModPhys.80.1083} {\bibfield  {journal} {\bibinfo
  {journal} {Rev. Mod. Phys.}\ }\textbf {\bibinfo {volume} {80}},\ \bibinfo
  {pages} {1083} (\bibinfo {year} {2008})}\BibitemShut {NoStop}%
\bibitem [{\citenamefont {Lahtinen}\ and\ \citenamefont
  {Pachos}(2017)}]{lahtinen_short_2017}%
  \BibitemOpen
  \bibfield  {author} {\bibinfo {author} {\bibfnamefont {V.}~\bibnamefont
  {Lahtinen}}\ and\ \bibinfo {author} {\bibfnamefont {J.~K.}\ \bibnamefont
  {Pachos}},\ }\href {https://doi.org/10.21468/SciPostPhys.3.3.021} {\bibfield
  {journal} {\bibinfo  {journal} {SciPost Phys.}\ }\textbf {\bibinfo {volume}
  {3}},\ \bibinfo {pages} {021} (\bibinfo {year} {2017})}\BibitemShut {NoStop}%
\bibitem [{Note8()}]{Note8}%
  \BibitemOpen
  \bibinfo {note} {We do not analyze an interaction-quench process [defined in
  \protect \cref {sec:quench,fig:quench}(b)] because it would change $k$. In
  the LGT-type model, $k$ maintains a large, constant value.}\BibitemShut
  {Stop}%
\bibitem [{\citenamefont {Ferraz}\ \emph {et~al.}(2020)\citenamefont {Ferraz},
  \citenamefont {Gupta}, \citenamefont {Semenoff},\ and\ \citenamefont
  {Sodano}}]{ferraz_strongly_2020}%
  \BibitemOpen
  \bibinfo {editor} {\bibfnamefont {A.}~\bibnamefont {Ferraz}}, \bibinfo
  {editor} {\bibfnamefont {K.~S.}\ \bibnamefont {Gupta}}, \bibinfo {editor}
  {\bibfnamefont {G.~W.}\ \bibnamefont {Semenoff}},\ and\ \bibinfo {editor}
  {\bibfnamefont {P.}~\bibnamefont {Sodano}},\ eds.,\ \href
  {https://doi.org/https://doi.org/10.1007/978-3-030-35473-2} {\emph {\bibinfo
  {title} {Strongly Coupled Field Theories for Condensed Matter and Quantum
  Information Theory}}}\ (\bibinfo  {publisher} {Springer Cham},\ \bibinfo
  {year} {2020})\BibitemShut {NoStop}%
\bibitem [{\citenamefont {Sun}\ \emph {et~al.}(2024)\citenamefont {Sun},
  \citenamefont {Kang}, \citenamefont {Nuomin}, \citenamefont {Schwartz},
  \citenamefont {Beratan}, \citenamefont {Brown},\ and\ \citenamefont
  {Kim}}]{sun_quantum_2024}%
  \BibitemOpen
  \bibfield  {author} {\bibinfo {author} {\bibfnamefont {K.}~\bibnamefont
  {Sun}}, \bibinfo {author} {\bibfnamefont {M.}~\bibnamefont {Kang}}, \bibinfo
  {author} {\bibfnamefont {H.}~\bibnamefont {Nuomin}}, \bibinfo {author}
  {\bibfnamefont {G.}~\bibnamefont {Schwartz}}, \bibinfo {author}
  {\bibfnamefont {D.~N.}\ \bibnamefont {Beratan}}, \bibinfo {author}
  {\bibfnamefont {K.~R.}\ \bibnamefont {Brown}},\ and\ \bibinfo {author}
  {\bibfnamefont {J.}~\bibnamefont {Kim}},\ }\href
  {https://arxiv.org/abs/2405.14624} {\bibinfo {title} {Quantum simulation of
  spin-boson models with structured bath}} (\bibinfo {year} {2024}),\ \Eprint
  {https://arxiv.org/abs/2405.14624} {arXiv:2405.14624 [quant-ph]} \BibitemShut
  {NoStop}%
\bibitem [{\citenamefont {Bilokur}\ \emph {et~al.}(2024)\citenamefont
  {Bilokur}, \citenamefont {Gopalakrishnan},\ and\ \citenamefont
  {Majidy}}]{bilokur_thermodynamic_2024}%
  \BibitemOpen
  \bibfield  {author} {\bibinfo {author} {\bibfnamefont {M.}~\bibnamefont
  {Bilokur}}, \bibinfo {author} {\bibfnamefont {S.}~\bibnamefont
  {Gopalakrishnan}},\ and\ \bibinfo {author} {\bibfnamefont {S.}~\bibnamefont
  {Majidy}},\ }\href {https://arxiv.org/abs/2411.12805} {\bibinfo {title}
  {Thermodynamic limitations on fault-tolerant quantum computing}} (\bibinfo
  {year} {2024}),\ \Eprint {https://arxiv.org/abs/2411.12805} {arXiv:2411.12805
  [quant-ph]} \BibitemShut {NoStop}%
\bibitem [{\citenamefont {Tsubota}\ \emph {et~al.}(2013)\citenamefont
  {Tsubota}, \citenamefont {Kobayashi},\ and\ \citenamefont
  {Takeuchi}}]{tsubota_quantum_2013}%
  \BibitemOpen
  \bibfield  {author} {\bibinfo {author} {\bibfnamefont {M.}~\bibnamefont
  {Tsubota}}, \bibinfo {author} {\bibfnamefont {M.}~\bibnamefont {Kobayashi}},\
  and\ \bibinfo {author} {\bibfnamefont {H.}~\bibnamefont {Takeuchi}},\ }\href
  {https://doi.org/https://doi.org/10.1016/j.physrep.2012.09.007} {\bibfield
  {journal} {\bibinfo  {journal} {Physics Reports}\ }\textbf {\bibinfo {volume}
  {522}},\ \bibinfo {pages} {191} (\bibinfo {year} {2013})}\BibitemShut
  {NoStop}%
\bibitem [{\citenamefont {Surace}\ \emph {et~al.}(2024)\citenamefont {Surace},
  \citenamefont {Lerose}, \citenamefont {Katz}, \citenamefont {Bennewitz},
  \citenamefont {Schuckert}, \citenamefont {Luo}, \citenamefont {De},
  \citenamefont {Ware}, \citenamefont {Morong}, \citenamefont {Collins},
  \citenamefont {Monroe}, \citenamefont {Davoudi},\ and\ \citenamefont
  {Gorshkov}}]{surace_stringbreaking_2024}%
  \BibitemOpen
  \bibfield  {author} {\bibinfo {author} {\bibfnamefont {F.~M.}\ \bibnamefont
  {Surace}}, \bibinfo {author} {\bibfnamefont {A.}~\bibnamefont {Lerose}},
  \bibinfo {author} {\bibfnamefont {O.}~\bibnamefont {Katz}}, \bibinfo {author}
  {\bibfnamefont {E.~R.}\ \bibnamefont {Bennewitz}}, \bibinfo {author}
  {\bibfnamefont {A.}~\bibnamefont {Schuckert}}, \bibinfo {author}
  {\bibfnamefont {D.}~\bibnamefont {Luo}}, \bibinfo {author} {\bibfnamefont
  {A.}~\bibnamefont {De}}, \bibinfo {author} {\bibfnamefont {B.}~\bibnamefont
  {Ware}}, \bibinfo {author} {\bibfnamefont {W.}~\bibnamefont {Morong}},
  \bibinfo {author} {\bibfnamefont {K.}~\bibnamefont {Collins}}, \bibinfo
  {author} {\bibfnamefont {C.}~\bibnamefont {Monroe}}, \bibinfo {author}
  {\bibfnamefont {Z.}~\bibnamefont {Davoudi}},\ and\ \bibinfo {author}
  {\bibfnamefont {A.~V.}\ \bibnamefont {Gorshkov}},\ }\href
  {https://arxiv.org/abs/2411.10652} {\bibinfo {title} {String-breaking
  dynamics in quantum adiabatic and diabatic processes}} (\bibinfo {year}
  {2024}),\ \Eprint {https://arxiv.org/abs/2411.10652} {arXiv:2411.10652
  [quant-ph]} \BibitemShut {NoStop}%
\bibitem [{\citenamefont {Jacob}\ \emph {et~al.}(2024)\citenamefont {Jacob},
  \citenamefont {Goold}, \citenamefont {Landi},\ and\ \citenamefont
  {Barra}}]{jacob_universal_2024}%
  \BibitemOpen
  \bibfield  {author} {\bibinfo {author} {\bibfnamefont {S.~L.}\ \bibnamefont
  {Jacob}}, \bibinfo {author} {\bibfnamefont {J.}~\bibnamefont {Goold}},
  \bibinfo {author} {\bibfnamefont {G.~T.}\ \bibnamefont {Landi}},\ and\
  \bibinfo {author} {\bibfnamefont {F.}~\bibnamefont {Barra}},\ }\href
  {https://doi.org/10.1103/PhysRevLett.133.207101} {\bibfield  {journal}
  {\bibinfo  {journal} {Phys. Rev. Lett.}\ }\textbf {\bibinfo {volume} {133}},\
  \bibinfo {pages} {207101} (\bibinfo {year} {2024})}\BibitemShut {NoStop}%
\bibitem [{\citenamefont {Jarzynski}(2000)}]{jarzynski_hamiltonian_2000}%
  \BibitemOpen
  \bibfield  {author} {\bibinfo {author} {\bibfnamefont {C.}~\bibnamefont
  {Jarzynski}},\ }\href {https://doi.org/10.1023/A:1018670721277} {\bibfield
  {journal} {\bibinfo  {journal} {Journal of Statistical Physics}\ }\textbf
  {\bibinfo {volume} {98}},\ \bibinfo {pages} {77} (\bibinfo {year}
  {2000})}\BibitemShut {NoStop}%
\bibitem [{\citenamefont {Crooks}(1999)}]{crooks_entropy_1999}%
  \BibitemOpen
  \bibfield  {author} {\bibinfo {author} {\bibfnamefont {G.~E.}\ \bibnamefont
  {Crooks}},\ }\href {https://doi.org/10.1103/PhysRevE.60.2721} {\bibfield
  {journal} {\bibinfo  {journal} {Phys. Rev. E}\ }\textbf {\bibinfo {volume}
  {60}},\ \bibinfo {pages} {2721} (\bibinfo {year} {1999})}\BibitemShut
  {NoStop}%
\bibitem [{\citenamefont {Esposito}\ \emph {et~al.}(2010)\citenamefont
  {Esposito}, \citenamefont {Lindenberg},\ and\ \citenamefont {den
  Broeck}}]{esposito_entropy_2010}%
  \BibitemOpen
  \bibfield  {author} {\bibinfo {author} {\bibfnamefont {M.}~\bibnamefont
  {Esposito}}, \bibinfo {author} {\bibfnamefont {K.}~\bibnamefont
  {Lindenberg}},\ and\ \bibinfo {author} {\bibfnamefont {C.~V.}\ \bibnamefont
  {den Broeck}},\ }\href {https://doi.org/10.1088/1367-2630/12/1/013013}
  {\bibfield  {journal} {\bibinfo  {journal} {New Journal of Physics}\ }\textbf
  {\bibinfo {volume} {12}},\ \bibinfo {pages} {013013} (\bibinfo {year}
  {2010})}\BibitemShut {NoStop}%
\bibitem [{\citenamefont {Korbel}\ and\ \citenamefont
  {Wolpert}(2024)}]{korbel_nonequilibrium_2024}%
  \BibitemOpen
  \bibfield  {author} {\bibinfo {author} {\bibfnamefont {J.}~\bibnamefont
  {Korbel}}\ and\ \bibinfo {author} {\bibfnamefont {D.~H.}\ \bibnamefont
  {Wolpert}},\ }\href {https://doi.org/10.1103/PhysRevResearch.6.013021}
  {\bibfield  {journal} {\bibinfo  {journal} {Phys. Rev. Res.}\ }\textbf
  {\bibinfo {volume} {6}},\ \bibinfo {pages} {013021} (\bibinfo {year}
  {2024})}\BibitemShut {NoStop}%
\bibitem [{\citenamefont {Esposito}\ and\ \citenamefont {Van~den
  Broeck}(2010)}]{esposito_three_2010}%
  \BibitemOpen
  \bibfield  {author} {\bibinfo {author} {\bibfnamefont {M.}~\bibnamefont
  {Esposito}}\ and\ \bibinfo {author} {\bibfnamefont {C.}~\bibnamefont {Van~den
  Broeck}},\ }\href {https://doi.org/10.1103/PhysRevLett.104.090601} {\bibfield
   {journal} {\bibinfo  {journal} {Phys. Rev. Lett.}\ }\textbf {\bibinfo
  {volume} {104}},\ \bibinfo {pages} {090601} (\bibinfo {year}
  {2010})}\BibitemShut {NoStop}%
\bibitem [{\citenamefont {Horowitz}\ and\ \citenamefont
  {Jarzynski}(2007)}]{horowitz_comparison_2007}%
  \BibitemOpen
  \bibfield  {author} {\bibinfo {author} {\bibfnamefont {J.}~\bibnamefont
  {Horowitz}}\ and\ \bibinfo {author} {\bibfnamefont {C.}~\bibnamefont
  {Jarzynski}},\ }\href {https://doi.org/10.1088/1742-5468/2007/11/P11002}
  {\bibfield  {journal} {\bibinfo  {journal} {Journal of Statistical Mechanics:
  Theory and Experiment}\ }\textbf {\bibinfo {volume} {2007}},\ \bibinfo
  {pages} {P11002} (\bibinfo {year} {2007})}\BibitemShut {NoStop}%
\bibitem [{\citenamefont {Bertini}\ \emph {et~al.}(2012)\citenamefont
  {Bertini}, \citenamefont {Gabrielli}, \citenamefont {Jona-Lasinio},\ and\
  \citenamefont {Landim}}]{bertini_thermodynamic_2012}%
  \BibitemOpen
  \bibfield  {author} {\bibinfo {author} {\bibfnamefont {L.}~\bibnamefont
  {Bertini}}, \bibinfo {author} {\bibfnamefont {D.}~\bibnamefont {Gabrielli}},
  \bibinfo {author} {\bibfnamefont {G.}~\bibnamefont {Jona-Lasinio}},\ and\
  \bibinfo {author} {\bibfnamefont {C.}~\bibnamefont {Landim}},\ }\href
  {https://doi.org/10.1007/s10955-012-0624-5} {\bibfield  {journal} {\bibinfo
  {journal} {Journal of Statistical Physics}\ }\textbf {\bibinfo {volume}
  {149}},\ \bibinfo {pages} {773} (\bibinfo {year} {2012})}\BibitemShut
  {NoStop}%
\bibitem [{\citenamefont {Bertini}\ \emph {et~al.}(2002)\citenamefont
  {Bertini}, \citenamefont {De~Sole}, \citenamefont {Gabrielli}, \citenamefont
  {Jona-Lasinio},\ and\ \citenamefont {Landim}}]{bertini_macroscopic_2002}%
  \BibitemOpen
  \bibfield  {author} {\bibinfo {author} {\bibfnamefont {L.}~\bibnamefont
  {Bertini}}, \bibinfo {author} {\bibfnamefont {A.}~\bibnamefont {De~Sole}},
  \bibinfo {author} {\bibfnamefont {D.}~\bibnamefont {Gabrielli}}, \bibinfo
  {author} {\bibfnamefont {G.}~\bibnamefont {Jona-Lasinio}},\ and\ \bibinfo
  {author} {\bibfnamefont {C.}~\bibnamefont {Landim}},\ }\href
  {https://doi.org/10.1023/A:1014525911391} {\bibfield  {journal} {\bibinfo
  {journal} {Journal of Statistical Physics}\ }\textbf {\bibinfo {volume}
  {107}},\ \bibinfo {pages} {635} (\bibinfo {year} {2002})}\BibitemShut
  {NoStop}%
\bibitem [{\citenamefont {Talkner}\ \emph {et~al.}(2007)\citenamefont
  {Talkner}, \citenamefont {Lutz},\ and\ \citenamefont
  {H\"anggi}}]{talkner_fluctuation_2007}%
  \BibitemOpen
  \bibfield  {author} {\bibinfo {author} {\bibfnamefont {P.}~\bibnamefont
  {Talkner}}, \bibinfo {author} {\bibfnamefont {E.}~\bibnamefont {Lutz}},\ and\
  \bibinfo {author} {\bibfnamefont {P.}~\bibnamefont {H\"anggi}},\ }\href
  {https://doi.org/10.1103/PhysRevE.75.050102} {\bibfield  {journal} {\bibinfo
  {journal} {Phys. Rev. E}\ }\textbf {\bibinfo {volume} {75}},\ \bibinfo
  {pages} {050102} (\bibinfo {year} {2007})}\BibitemShut {NoStop}%
\bibitem [{\citenamefont {Funo}\ \emph {et~al.}(2018)\citenamefont {Funo},
  \citenamefont {Ueda},\ and\ \citenamefont {Sagawa}}]{funo_quantum_2018}%
  \BibitemOpen
  \bibfield  {author} {\bibinfo {author} {\bibfnamefont {K.}~\bibnamefont
  {Funo}}, \bibinfo {author} {\bibfnamefont {M.}~\bibnamefont {Ueda}},\ and\
  \bibinfo {author} {\bibfnamefont {T.}~\bibnamefont {Sagawa}},\ }in\ \href
  {https://doi.org/10.1007/978-3-319-99046-0_10} {\emph {\bibinfo {booktitle}
  {Thermodynamics in the {Quantum} {Regime}: {Fundamental} {Aspects} and {New}
  {Directions}}}},\ \bibinfo {series and number} {Fundamental {Theories} of
  {Physics}},\ \bibinfo {editor} {edited by\ \bibinfo {editor} {\bibfnamefont
  {F.}~\bibnamefont {Binder}}, \bibinfo {editor} {\bibfnamefont {L.~A.}\
  \bibnamefont {Correa}}, \bibinfo {editor} {\bibfnamefont {C.}~\bibnamefont
  {Gogolin}}, \bibinfo {editor} {\bibfnamefont {J.}~\bibnamefont {Anders}},\
  and\ \bibinfo {editor} {\bibfnamefont {G.}~\bibnamefont {Adesso}}}\ (\bibinfo
   {publisher} {Springer International Publishing},\ \bibinfo {address}
  {Cham},\ \bibinfo {year} {2018})\ pp.\ \bibinfo {pages}
  {249--273}\BibitemShut {NoStop}%
\bibitem [{\citenamefont {Dalmonte}\ \emph {et~al.}(2022)\citenamefont
  {Dalmonte}, \citenamefont {Eisler}, \citenamefont {Falconi},\ and\
  \citenamefont {Vermersch}}]{dalmonte_entanglement_2022}%
  \BibitemOpen
  \bibfield  {author} {\bibinfo {author} {\bibfnamefont {M.}~\bibnamefont
  {Dalmonte}}, \bibinfo {author} {\bibfnamefont {V.}~\bibnamefont {Eisler}},
  \bibinfo {author} {\bibfnamefont {M.}~\bibnamefont {Falconi}},\ and\ \bibinfo
  {author} {\bibfnamefont {B.}~\bibnamefont {Vermersch}},\ }\href@noop {}
  {\bibfield  {journal} {\bibinfo  {journal} {Annalen der Physik}\ }\textbf
  {\bibinfo {volume} {534}},\ \bibinfo {pages} {2200064} (\bibinfo {year}
  {2022})}\BibitemShut {NoStop}%
\bibitem [{\citenamefont {Elben}\ \emph {et~al.}(2019)\citenamefont {Elben},
  \citenamefont {Vermersch}, \citenamefont {Roos},\ and\ \citenamefont
  {Zoller}}]{elben_statistical_2019}%
  \BibitemOpen
  \bibfield  {author} {\bibinfo {author} {\bibfnamefont {A.}~\bibnamefont
  {Elben}}, \bibinfo {author} {\bibfnamefont {B.}~\bibnamefont {Vermersch}},
  \bibinfo {author} {\bibfnamefont {C.~F.}\ \bibnamefont {Roos}},\ and\
  \bibinfo {author} {\bibfnamefont {P.}~\bibnamefont {Zoller}},\ }\href@noop {}
  {\bibfield  {journal} {\bibinfo  {journal} {Physical Review A}\ }\textbf
  {\bibinfo {volume} {99}},\ \bibinfo {pages} {052323} (\bibinfo {year}
  {2019})}\BibitemShut {NoStop}%
\bibitem [{\citenamefont {Huang}\ \emph {et~al.}(2020)\citenamefont {Huang},
  \citenamefont {Kueng},\ and\ \citenamefont
  {Preskill}}]{huang_predicting_2020}%
  \BibitemOpen
  \bibfield  {author} {\bibinfo {author} {\bibfnamefont {H.-Y.}\ \bibnamefont
  {Huang}}, \bibinfo {author} {\bibfnamefont {R.}~\bibnamefont {Kueng}},\ and\
  \bibinfo {author} {\bibfnamefont {J.}~\bibnamefont {Preskill}},\ }\href@noop
  {} {\bibfield  {journal} {\bibinfo  {journal} {Nature Physics}\ }\textbf
  {\bibinfo {volume} {16}},\ \bibinfo {pages} {1050} (\bibinfo {year}
  {2020})}\BibitemShut {NoStop}%
\bibitem [{\citenamefont {Huang}\ \emph {et~al.}(2022)\citenamefont {Huang},
  \citenamefont {Broughton}, \citenamefont {Cotler}, \citenamefont {Chen},
  \citenamefont {Li}, \citenamefont {Mohseni}, \citenamefont {Neven},
  \citenamefont {Babbush}, \citenamefont {Kueng}, \citenamefont {Preskill}
  \emph {et~al.}}]{huang_quantum_2022}%
  \BibitemOpen
  \bibfield  {author} {\bibinfo {author} {\bibfnamefont {H.-Y.}\ \bibnamefont
  {Huang}}, \bibinfo {author} {\bibfnamefont {M.}~\bibnamefont {Broughton}},
  \bibinfo {author} {\bibfnamefont {J.}~\bibnamefont {Cotler}}, \bibinfo
  {author} {\bibfnamefont {S.}~\bibnamefont {Chen}}, \bibinfo {author}
  {\bibfnamefont {J.}~\bibnamefont {Li}}, \bibinfo {author} {\bibfnamefont
  {M.}~\bibnamefont {Mohseni}}, \bibinfo {author} {\bibfnamefont
  {H.}~\bibnamefont {Neven}}, \bibinfo {author} {\bibfnamefont
  {R.}~\bibnamefont {Babbush}}, \bibinfo {author} {\bibfnamefont
  {R.}~\bibnamefont {Kueng}}, \bibinfo {author} {\bibfnamefont
  {J.}~\bibnamefont {Preskill}}, \emph {et~al.},\ }\href@noop {} {\bibfield
  {journal} {\bibinfo  {journal} {Science}\ }\textbf {\bibinfo {volume}
  {376}},\ \bibinfo {pages} {1182} (\bibinfo {year} {2022})}\BibitemShut
  {NoStop}%
\bibitem [{\citenamefont {Elben}\ \emph {et~al.}(2023)\citenamefont {Elben},
  \citenamefont {Flammia}, \citenamefont {Huang}, \citenamefont {Kueng},
  \citenamefont {Preskill}, \citenamefont {Vermersch},\ and\ \citenamefont
  {Zoller}}]{elben_randomized_2023}%
  \BibitemOpen
  \bibfield  {author} {\bibinfo {author} {\bibfnamefont {A.}~\bibnamefont
  {Elben}}, \bibinfo {author} {\bibfnamefont {S.~T.}\ \bibnamefont {Flammia}},
  \bibinfo {author} {\bibfnamefont {H.-Y.}\ \bibnamefont {Huang}}, \bibinfo
  {author} {\bibfnamefont {R.}~\bibnamefont {Kueng}}, \bibinfo {author}
  {\bibfnamefont {J.}~\bibnamefont {Preskill}}, \bibinfo {author}
  {\bibfnamefont {B.}~\bibnamefont {Vermersch}},\ and\ \bibinfo {author}
  {\bibfnamefont {P.}~\bibnamefont {Zoller}},\ }\href@noop {} {\bibfield
  {journal} {\bibinfo  {journal} {Nature Reviews Physics}\ }\textbf {\bibinfo
  {volume} {5}},\ \bibinfo {pages} {9} (\bibinfo {year} {2023})}\BibitemShut
  {NoStop}%
\bibitem [{\citenamefont {Pichler}\ \emph {et~al.}(2016)\citenamefont
  {Pichler}, \citenamefont {Zhu}, \citenamefont {Seif}, \citenamefont
  {Zoller},\ and\ \citenamefont {Hafezi}}]{pichler_measurement_2016}%
  \BibitemOpen
  \bibfield  {author} {\bibinfo {author} {\bibfnamefont {H.}~\bibnamefont
  {Pichler}}, \bibinfo {author} {\bibfnamefont {G.}~\bibnamefont {Zhu}},
  \bibinfo {author} {\bibfnamefont {A.}~\bibnamefont {Seif}}, \bibinfo {author}
  {\bibfnamefont {P.}~\bibnamefont {Zoller}},\ and\ \bibinfo {author}
  {\bibfnamefont {M.}~\bibnamefont {Hafezi}},\ }\href@noop {} {\bibfield
  {journal} {\bibinfo  {journal} {Physical Review X}\ }\textbf {\bibinfo
  {volume} {6}},\ \bibinfo {pages} {041033} (\bibinfo {year}
  {2016})}\BibitemShut {NoStop}%
\bibitem [{\citenamefont {Dalmonte}\ \emph {et~al.}(2018)\citenamefont
  {Dalmonte}, \citenamefont {Vermersch},\ and\ \citenamefont
  {Zoller}}]{dalmonte_quantum_2018}%
  \BibitemOpen
  \bibfield  {author} {\bibinfo {author} {\bibfnamefont {M.}~\bibnamefont
  {Dalmonte}}, \bibinfo {author} {\bibfnamefont {B.}~\bibnamefont
  {Vermersch}},\ and\ \bibinfo {author} {\bibfnamefont {P.}~\bibnamefont
  {Zoller}},\ }\href@noop {} {\bibfield  {journal} {\bibinfo  {journal} {Nature
  Physics}\ }\textbf {\bibinfo {volume} {14}},\ \bibinfo {pages} {827}
  (\bibinfo {year} {2018})}\BibitemShut {NoStop}%
\bibitem [{\citenamefont {Kokail}\ \emph
  {et~al.}(2021{\natexlab{a}})\citenamefont {Kokail}, \citenamefont {van
  Bijnen}, \citenamefont {Elben}, \citenamefont {Vermersch},\ and\
  \citenamefont {Zoller}}]{kokail_entanglement_2021}%
  \BibitemOpen
  \bibfield  {author} {\bibinfo {author} {\bibfnamefont {C.}~\bibnamefont
  {Kokail}}, \bibinfo {author} {\bibfnamefont {R.}~\bibnamefont {van Bijnen}},
  \bibinfo {author} {\bibfnamefont {A.}~\bibnamefont {Elben}}, \bibinfo
  {author} {\bibfnamefont {B.}~\bibnamefont {Vermersch}},\ and\ \bibinfo
  {author} {\bibfnamefont {P.}~\bibnamefont {Zoller}},\ }\href@noop {}
  {\bibfield  {journal} {\bibinfo  {journal} {Nature Physics}\ }\textbf
  {\bibinfo {volume} {17}},\ \bibinfo {pages} {936} (\bibinfo {year}
  {2021}{\natexlab{a}})}\BibitemShut {NoStop}%
\bibitem [{\citenamefont {Kokail}\ \emph
  {et~al.}(2021{\natexlab{b}})\citenamefont {Kokail}, \citenamefont {Sundar},
  \citenamefont {Zache}, \citenamefont {Elben}, \citenamefont {Vermersch},
  \citenamefont {Dalmonte}, \citenamefont {van Bijnen},\ and\ \citenamefont
  {Zoller}}]{kokail_quantum_2021}%
  \BibitemOpen
  \bibfield  {author} {\bibinfo {author} {\bibfnamefont {C.}~\bibnamefont
  {Kokail}}, \bibinfo {author} {\bibfnamefont {B.}~\bibnamefont {Sundar}},
  \bibinfo {author} {\bibfnamefont {T.~V.}\ \bibnamefont {Zache}}, \bibinfo
  {author} {\bibfnamefont {A.}~\bibnamefont {Elben}}, \bibinfo {author}
  {\bibfnamefont {B.}~\bibnamefont {Vermersch}}, \bibinfo {author}
  {\bibfnamefont {M.}~\bibnamefont {Dalmonte}}, \bibinfo {author}
  {\bibfnamefont {R.}~\bibnamefont {van Bijnen}},\ and\ \bibinfo {author}
  {\bibfnamefont {P.}~\bibnamefont {Zoller}},\ }\href
  {https://doi.org/10.1103/PhysRevLett.127.170501} {\bibfield  {journal}
  {\bibinfo  {journal} {Phys. Rev. Lett.}\ }\textbf {\bibinfo {volume} {127}},\
  \bibinfo {pages} {170501} (\bibinfo {year} {2021}{\natexlab{b}})}\BibitemShut
  {NoStop}%
\bibitem [{\citenamefont {Mueller}\ \emph {et~al.}(2023)\citenamefont
  {Mueller}, \citenamefont {Carolan}, \citenamefont {Connelly}, \citenamefont
  {Davoudi}, \citenamefont {Dumitrescu},\ and\ \citenamefont
  {Yeter-Aydeniz}}]{mueller_quantum_2023}%
  \BibitemOpen
  \bibfield  {author} {\bibinfo {author} {\bibfnamefont {N.}~\bibnamefont
  {Mueller}}, \bibinfo {author} {\bibfnamefont {J.~A.}\ \bibnamefont
  {Carolan}}, \bibinfo {author} {\bibfnamefont {A.}~\bibnamefont {Connelly}},
  \bibinfo {author} {\bibfnamefont {Z.}~\bibnamefont {Davoudi}}, \bibinfo
  {author} {\bibfnamefont {E.~F.}\ \bibnamefont {Dumitrescu}},\ and\ \bibinfo
  {author} {\bibfnamefont {K.}~\bibnamefont {Yeter-Aydeniz}},\ }\href@noop {}
  {\bibfield  {journal} {\bibinfo  {journal} {PRX Quantum}\ }\textbf {\bibinfo
  {volume} {4}},\ \bibinfo {pages} {030323} (\bibinfo {year}
  {2023})}\BibitemShut {NoStop}%
\bibitem [{\citenamefont {Bringewatt}\ \emph {et~al.}(2024)\citenamefont
  {Bringewatt}, \citenamefont {Kunjummen},\ and\ \citenamefont
  {Mueller}}]{bringewatt_randomized_2024}%
  \BibitemOpen
  \bibfield  {author} {\bibinfo {author} {\bibfnamefont {J.}~\bibnamefont
  {Bringewatt}}, \bibinfo {author} {\bibfnamefont {J.}~\bibnamefont
  {Kunjummen}},\ and\ \bibinfo {author} {\bibfnamefont {N.}~\bibnamefont
  {Mueller}},\ }\href@noop {} {\bibfield  {journal} {\bibinfo  {journal}
  {Quantum}\ }\textbf {\bibinfo {volume} {8}},\ \bibinfo {pages} {1300}
  (\bibinfo {year} {2024})}\BibitemShut {NoStop}%
\bibitem [{\citenamefont {Mueller}\ \emph {et~al.}(2024)\citenamefont
  {Mueller}, \citenamefont {Wang}, \citenamefont {Katz}, \citenamefont
  {Davoudi},\ and\ \citenamefont {Cetina}}]{mueller_quantum_2024}%
  \BibitemOpen
  \bibfield  {author} {\bibinfo {author} {\bibfnamefont {N.}~\bibnamefont
  {Mueller}}, \bibinfo {author} {\bibfnamefont {T.}~\bibnamefont {Wang}},
  \bibinfo {author} {\bibfnamefont {O.}~\bibnamefont {Katz}}, \bibinfo {author}
  {\bibfnamefont {Z.}~\bibnamefont {Davoudi}},\ and\ \bibinfo {author}
  {\bibfnamefont {M.}~\bibnamefont {Cetina}},\ }\href@noop {} {\bibfield
  {journal} {\bibinfo  {journal} {arXiv preprint arXiv:2408.00069}\ } (\bibinfo
  {year} {2024})}\BibitemShut {NoStop}%
\bibitem [{\citenamefont {Sakurai}(1993)}]{sakurai_modern_1993}%
  \BibitemOpen
  \bibfield  {author} {\bibinfo {author} {\bibfnamefont {J.~J.}\ \bibnamefont
  {Sakurai}},\ }\href@noop {} {\emph {\bibinfo {title} {Modern Quantum
  Mechanics (Revised Edition)}}},\ \bibinfo {edition} {1st}\ ed.\ (\bibinfo
  {publisher} {Addison Wesley},\ \bibinfo {year} {1993})\BibitemShut {NoStop}%
\bibitem [{\citenamefont {Uhlmann}(1977)}]{uhlmann_relative_1977}%
  \BibitemOpen
  \bibfield  {author} {\bibinfo {author} {\bibfnamefont {A.}~\bibnamefont
  {Uhlmann}},\ }\href {https://doi.org/10.1007/BF01609834} {\bibfield
  {journal} {\bibinfo  {journal} {Communications in Mathematical Physics}\
  }\textbf {\bibinfo {volume} {54}},\ \bibinfo {pages} {21} (\bibinfo {year}
  {1977})}\BibitemShut {NoStop}%
\bibitem [{\citenamefont {Lindblad}(1974)}]{lindblad_expectations_1974}%
  \BibitemOpen
  \bibfield  {author} {\bibinfo {author} {\bibfnamefont {G.}~\bibnamefont
  {Lindblad}},\ }\href {https://doi.org/10.1007/BF01608390} {\bibfield
  {journal} {\bibinfo  {journal} {Communications in Mathematical Physics}\
  }\textbf {\bibinfo {volume} {39}},\ \bibinfo {pages} {111} (\bibinfo {year}
  {1974})}\BibitemShut {NoStop}%
\bibitem [{\citenamefont {Donald}(1986)}]{donald_relative_1986}%
  \BibitemOpen
  \bibfield  {author} {\bibinfo {author} {\bibfnamefont {M.~J.}\ \bibnamefont
  {Donald}},\ }\href {https://doi.org/10.1007/BF01212339} {\bibfield  {journal}
  {\bibinfo  {journal} {Communications in Mathematical Physics}\ }\textbf
  {\bibinfo {volume} {105}},\ \bibinfo {pages} {13} (\bibinfo {year}
  {1986})}\BibitemShut {NoStop}%
\bibitem [{Note9()}]{Note9}%
  \BibitemOpen
  \bibinfo {note} {The commutation relations do not suffice for deriving
  $W_{\protect \mathrm {diff}} = W_{H^*} = W_{E^*}$ for an interaction quench.
  The reason appears to be an interplay between partial traces, the
  tensor-product form of the interaction quench's initial state, and the
  interaction term's not factoring into system and reservoir
  factors.}\BibitemShut {Stop}%
\end{thebibliography}%

\onecolumngrid
\appendix
\section{Weak-coupling limit} \label{app:weakcoupling}
In the weak-coupling limit, the interactions between $S$ and $R$, while nonvanishing, are small enough to be neglected in calculations of partition functions and free energies. Therefore, we model the weak-coupling limit by setting $V_{\sur} = 0$ in \cref{eq:hsurgen}:
\begin{equation}\label{eq:hsur_weak_coupling}
    \h_{\sur} = \h_S + \h_R \, .
\end{equation}
In this limit, we demonstrate, the main text's three internal-energy definitions are equivalent. 

Under the assumption in \cref{eq:hsur_weak_coupling}, one obtains
\begin{equation}\label{eq:hstareqhs} 
    \hso 
     \coloneqq -\frac{1}{\beta} \ln{ \left( \frac{\Tr_R \left( e^{-\beta \left( \h_S + \h_R \right)} \right) }{ \Tr_R \left( e^{-\beta \h_R} \right) } \right) } = -\frac{1}{\beta} \ln{ \left( e^{-\beta \h_S} \right)}
    = \h_S \,
\end{equation}
and
\begin{equation}
    \eso \coloneqq
    \partial_{\beta} \left( \beta \hso \right)
    =\partial_{\beta} \left( \beta \h_S \right) 
    = \h_S \, 
    \label{eq:estareqhs} .
\end{equation}
Thus, both $\hso$ and $\eso$ reduce to the system Hamiltonian $\h_S$ in the weak-coupling limit. Consequently, by \cref{eq:uhstar,eq:uestar},
\begin{equation}\label{eq:uhstarequestar}
    U_{H^*} = U_{E^*} = \Tr_S \left( \h_S \; \rhoo_S \right) \, ,
\end{equation}
for any state $\rhoo_S$. 

We now additionally assume that $\sur$ is in a global Gibbs state. As a result, $\rhoo_S = \pio_S^0$, $U_{H^*} = U_{E^*} = U_S^0$, and 
\begin{subequations}\label[pluralequation]{eq:utotequs0}
\begin{align}
    U_{\tot} & = \Tr_{SR} \left( \left[ \h_S \otimes \hat{\mathbbm{1}}_R + \hat{\mathbbm{1}}_S \otimes \h_R \right] \left[ \pio_S^0 \otimes \pio_R^0 \right] \right) - \Tr_R \left( \h_R \; \pio_R^0 \right)
    \\
    & = \Tr_S \left(\h_S \; \pio_S^0 \right) + \Tr_R \left( \h_R \; \pio_R^0 \right) - \Tr_R \left( \h_R \; \pio_R^0 \right) 
    \\ 
    & = \Tr_S \left(\h_S \; \pio_S^0 \right)
    = U_S^0 \, . 
\end{align}
\end{subequations}
Thus,
\begin{equation}\label{eq:equalu}
    U_{\tot} = U_{H^*} = U_{E^*} = U_S^0 \, .
\end{equation}

Three internal-energy definitions would be equivalent in the weak-coupling limit if one defined $U_{\tot}$ differently from \cref{eq:utot}. Define $U_R \coloneqq \Tr_R(\h_R\rho_R)$ and $\widetilde{U}_{\tot} \coloneqq U_{\sur} - U_R$. If $\rho_{\sur}$ is any product state $\rho_S \otimes \rho_R$, then $\widetilde{U}_{\tot} = U_{H^*} = U_{E^*}$. Since $\widetilde{U}_{\tot}$ is nonstandard, we use \cref{eq:utot} in this manuscript.

\section{Equivalence of \texorpdfstring{$U_{\mathrm{diff}}$}{utot} and \texorpdfstring{$U_{E^{*}}$}{uestar} when \texorpdfstring{$S \cup R$}{sur} is in a global Gibbs state} \label{app:utotequestar}
In this appendix, we show that $ U_{E^{*}} = U_{\tot}$ when $\sur$ is in a global Gibbs state. The classical analog was proved in Ref.~\cite{seifert_first_2016}, then used in Ref.~\cite{miller_hamiltonian_2018} to derive quantum fluctuation theorems. We do not assume, in this appendix, that the coupling is weak.

We begin by rewriting~\cref{eq:estar}:
\begin{subequations}\label[pluralequation]{eq:partialsimestar}
\begin{align} 
    \eso & = \partial_{\beta} \left( \beta \hso \right)  
    \\ 
    & = \partial_{\beta} \left[ -\ln{\left( \frac{\Tr_R \left( e^{-\beta \h_{\sur}} \right) }{\Tr_R \left( e^{-\beta \h_R }\right) } \right)} \; \right] 
    \\ 
    & = -\frac{\partial_{\beta}\left[ \Tr_R \left( e^{-\beta \h_{\sur}}\right) \right] }{ \Tr_R \left( e^{-\beta \h_{\sur}}\right) } + \frac{ \partial_{\beta} \left[ \Tr_R \left( e^{-\beta \h_R}\right) \right] }{\Tr_R \left( e^{-\beta \h_R} \right)} \, .
\end{align}
\end{subequations}
Consider applying the derivative's definition to the first term:
\begin{subequations}\label[pluralequation]{eq:pardervhsur}
\begin{align}
    \partial_{\beta} \left[ \Tr_R \left( e^{-\beta \h_{\sur}}\right) \right] & = \lim_{\delta \beta \to 0} \frac{\Tr_R \left( e^{-\left( \beta +\delta \beta \right)\h_{\sur} }\right) -\Tr_R \left( e^{-\beta \h_{\sur}}\right) }{\delta \beta} 
    \\
    & = \lim_{\delta \beta \to 0} \Tr_{R} \left( e^{-\beta \h_{\sur}}  \frac{\left( e^{-\delta \beta \h_{\sur}} -1 \right) } {\delta \beta} \right) 
    \\
    & = -\Tr_R \left( \h_{\sur} \; e^{-\beta \h_{\sur}} \right) \, .
\end{align}
\end{subequations}
The final term in \cref{eq:partialsimestar} simplifies similarly. Therefore,
\begin{equation}\label{eq:simpestar}
    \eso  = \frac{\Tr_R \left( \h_{\sur} \; e^{-\beta \h_{\sur}} \right) }{\Tr_R \left( e^{-\beta \h_{\sur}}\right)}  - \frac{\Tr_R \left( \h_R \; e^{-\beta \h_R} \right)}{ \Tr_R \left( e^{-\beta \h_R}\right)} \, .
\end{equation}
To calculate $U_{E^*}$, one takes the expectation value of $\eso$ in the state $\pio_S$:
\begin{equation}
    U_{E^*} = \Tr_{S} \left( \eso \; \pio_S \right) \, .
\end{equation}
Substituting $\pio_S = \Tr_R \left(\pio_{\sur} \right)$, and $\hat{E}^*_S$ from \cref{eq:simpestar}, gives
\begin{equation} \label{eq:uestarsimp}
    U_{E^*} = \frac{1}{Z_{\sur}}\Tr_S \bm{\big(} \Tr_R \big( \h_{\sur} \; e^{-\beta \h_{\sur}} \big) \bm{\big)} - \frac{1}{Z_R }\Tr_S \bm{\big(} \pio_S \; \Tr_R \big( \h_R e^{-\beta \h_R} \big) \bm{\big)}  \, .    
\end{equation}
By distributing and evaluating the traces, one obtains
\begin{subequations}
\begin{align}
        U_{E^*} & = \Tr_{SR} \left( \h_{\sur} \; \pio_{\sur} \right) - \Tr_S \left( \pio_S \right) \Tr_R \left( \h_R \; \pio_R \right) \\
        & = \Tr_{SR} \left( \h_{\sur} \; \pio_{\sur} \right) - U_R^0 \, .
\end{align}
\end{subequations}
Therefore, when $\rhoo_\sur=\pio_\sur$,
\begin{equation}\label{eq:utotequestar}
    U_{E^*} = U_{\tot} \, .
\end{equation}
%

\section{Operator commutativity and equal work quantities in a system quench}
\label{app:commutation}
In this appendix, we show that the commutation relations in Eqs.~\eqref{eq:commutationrels} suffice for establishing the equalities $W_{\tot} = W_{H^*} = W_{E^*}$ for a system quench. We motivate these equalities by observing that the corresponding classical work definitions are equivalent (for a system quench)~\cite{jarzynski_stochastic_2017}, and that the commutation of observables under multiplication is often a signal of classicality.

Recall that a system-quench process proceeds as follows. $\sur$ begins in the Gibbs state $\pio_{\sur}^{\mi}$ at $t=0^-$. At $t=0$, one quenches the system Hamiltonian from $\h_S^{\mi}$ to $\h_S^{\f}$. From $t=0^+$ to $t_\mathrm{f}$, $\sur$ evolves under $\h_\sur^{\f}$. 

First, we show that $W_{\tot} = W_{H^*}$ if the commutation relations \eqref{eq:commutationrels} hold. Substituting the $\hso$ definition [\cref{eq:hstar}] into the $W_{H^*}$ definition~[\cref{eq:whstaroperator}], one obtains
\begin{equation}\label{eq:whstarpartialsimp1}
    W_{H^*} = -\frac{1}{\beta} \Tr_S \bm{\Big(} \ln{ \left( \Tr_R \left( e^{-\beta \h_{\sur}^{\f}} \right) \right)} \pio_S^{\mi} \bm{\Big)} + \frac{1}{\beta} \Tr_S \bm{\Big(} \ln{ \left( \Tr_R \left( e^{-\beta \h_{\sur}^{\mi}} \right) \right)} \pio_S^{\mi} \bm{\Big)} \, .
\end{equation}
The expression simplifies because $\h_S^{\mi}$ and $\h_S^{\f}$ commute with $\hat{V}_{\sur}$:
\begin{subequations}\label[pluralequation]{eq:whstarpartialsimp2}
\begin{align}
     W_{H^*} = & 
     -\frac{1}{\beta} \Tr_S \bm{\Big(} \ln{ \left( e^{-\beta \h_S^{B}} \Tr_R \left( e^{-\beta (\h_R+ \hat{V}_{\sur})} \right) \right)} \pio_S^{\mi} \bm{\Big)} +\frac{1}{\beta} \Tr_S \bm{\Big(} \ln{ \left( e^{-\beta \h_S^{A}} \Tr_R \left( e^{-\beta (\h_R+ \hat{V}_{\sur})} \right) \right)} \pio_S^{\mi} \bm{\Big)}
    \\
    = & -\frac{1}{\beta} \Tr_S \bm{\Big(} \ln{ \left( e^{-\beta \h_S^{B}} \right)} \pio_S^{\mi} \bm{\Big)}
    +\frac{1}{\beta} \Tr_S \bm{\Big(} \ln{ \left( e^{-\beta \h_S^{A}} \right)} \pio_S^{\mi} \bm{\Big)} \, . 
\end{align}
\end{subequations}

We substitute $\pio_S^{\mi} = \Tr_R (\pio_{\sur}^{\mi})$, rearrange the traces,  and recall that 
$\h_S^{\f} - \h_S^{\mi} = \h_{\sur}^{\f} - \h_{\sur}^{\mi}$:
\begin{equation}\label{eq:whstareqwtot}
    W_{H^*} = \Tr_{SR} \left( \left[ \h_{\sur}^{\f} - \h_{\sur}^{\mi} \right] \pio_{\sur}^{\mi} \right) = W_{\tot} \, . 
\end{equation}

To obtain a similar expression for $W_{E^*}$, we apply the definitions of $\hso$ [\cref{eq:hstar}] and $\eso$ [\cref{eq:estar}], as well as the commutation relations~[Eqs.~\eqref{eq:commutationrels}]. The difference $\esof - \esoi$ becomes
\begin{subequations}\label[pluralequation]{eq:estardiff}
\begin{align}
    \esof - \esoi  & = \partial_{\beta} \left( \beta \hsof - \beta \hsoi \right) \label{eq:estardiff1}\\ 
    & = - \partial_{\beta} \left[ \ln\bigg( \Tr_R \left( e^{-\beta \h_{\sur}^{\f}} \right) \bigg) - \ln\bigg( \Tr_R \left( e^{-\beta \h_{\sur}^{\mi}} \right) \bigg)  \right] \label{eq:estardiff2}\\ 
    & = - \partial_{\beta} \Biggl [ \ln\bigg( e^{-\beta \h_S^{B}} \Tr_R \left( e^{-\beta (\h_R+ \hat{V}_{\sur})}  \right) \bigg) -\ln\bigg( e^{-\beta \h_S^{A}} \Tr_R \left( e^{-\beta (\h_R+ \hat{V}_{\sur})}  \right) \bigg) \Biggl ] \label{eq:estardiff3}\\
    & = \h_S^{\f} - \h_S^{\mi} \, \label{eq:estardiff4}.
\end{align}
\end{subequations}
Inserting \cref{eq:estardiff4} into \cref{eq:westaroperator}, with $\pio_{S}^{A} = \Tr_R \left( \pio_{\sur}^{\mi} \right)$, gives
%
\begin{equation}\label{eq:westareqwtot}
    W_{E^*} = \Tr_{SR} \left( \left[ \h_{\sur}^{\f} - \h_{\sur}^{\mi} \right] \pio_{\sur}^{\mi}  \right) = W_{\tot} \, .
\end{equation}

Thus, $W_{\tot} = W_{H^*} = W_{E^*}$ when a system quench obeys the commutation relations \eqref{eq:commutationrels}. In contrast, even when the commutation relations hold, the three work quantities differ if the interaction term changes abruptly~\footnote{The commutation relations do not suffice for deriving $W_{\tot} = W_{H^*} = W_{E^*}$ for an interaction quench. The reason appears to be an interplay between partial traces, the tensor-product form of the interaction quench's initial state, and the interaction term's not factoring into system and reservoir factors.}. We speculate that this observation extends to the analogous classical setup.

If $W_\tot = W_{H^*} = W_{E^*}$, $W_{E^*}$ satisfies the second law due to Eqs.~\eqref{eq:whstargeqf} [and, separately, due to Eqs.~\eqref{eq:wtotgeqf}]: $W_{E^*} \geq \Delta F_S$.

\section{Explicit expressions for work and heat exchanged in the two-spin model} \label{app:wandhforspin}
We analytically derive expressions for the work, heat, entropy, and free energy in the two-spin model of \cref{sec:spinmodel}. 
We use these expressions to generate~\cref{fig:toy_quench}.

The spin Hamiltonian $\h_{\sur}$ [\cref{eq:toyham}] can be diagonalized easily. In  terms of the eigenbasis of $\h_{\sur}$, the Gibbs state is
\begin{equation}\label{eq:explicitpisur}
    \pio_{\sur} = \frac{1}{Z_{\sur}} \text{diag} \left( e^{-\beta (\gamma - \eta_+)},e^{- \beta (\gamma + \eta_+ ) },e^{\beta (\gamma + \eta_-)},e^{ \beta (\gamma - \eta_- ) } \right) \, .
\end{equation}
We have defined $\eta_{\pm} \coloneqq \sqrt{{ \left(\frac{\epsilon}{2} \pm \frac{\alpha}{2}\right)}^2 + \chi^2 }$, where $\epsilon$, $\alpha$, and $\chi$ are the Hamiltonian parameters. The partition function is
\begin{equation}\label{eq:explicitztot}
    Z_{\sur} = 2 e^{-\beta \gamma} \cosh{ \left( \beta \eta_+ \right)} + 2 e^{\beta \gamma} \cosh{ \left( \beta \eta_- \right)} \, .
\end{equation}

To calculate the system's thermal state, we trace out the reservoir from $\pio_{\sur}$. In terms of the $\sigma_S^{z}$ eigenbasis, 
\begin{equation}\label{eq:explicitpis}
    \pio_S = \frac{1}{Z_{\sur}}\begin{pmatrix}
            X_- & 0 \\
            0 & X_+
        \end{pmatrix} \ , 
\end{equation}
wherein
\begin{equation}\label{eq:explicitxy}
    X_\pm = e^{-\beta \gamma} \left[ \cosh{ \left( \beta \eta_+ \right)} \pm \frac{a_+}{\eta_+} \sinh{ \left( \beta \eta_+ \right)} \right] + e^{\beta \gamma} \left[ \cosh{ \left(\beta \eta_- \right)} \pm \frac{a_-}{\eta_-} \sinh{\left( \beta \eta_- \right)} \right] \, 
\end{equation}
and $a_{\pm} = (\epsilon \pm \alpha)/2$. Using the same basis, we also calculate $\hso$ and $\eso$:
\begin{equation}\label{eq:explicithstar}
    \hso = -\frac{1}{\beta} \begin{pmatrix}
        \ln{\left( \frac{X_-}{2\cosh{\left( \frac{\beta \alpha}{2} \right)}} \right)} & 0 \\
        0 & \ln{\left( \frac{X_+}{2\cosh{\left( \frac{\beta \alpha}{2} \right)}} \right)}
    \end{pmatrix} \,  ,
\end{equation}
and 
\begin{equation}\label{eq:explicitestar}
    \eso = \begin{pmatrix}
        - \frac{\partial_{\beta}X_-}{X_-}  + \frac{\alpha}{2} \tanh{\left( \frac{\beta \alpha}{2} \right)} & 0 \\ 
        0 & - \frac{\partial_{\beta}X_+}{X_+}  + \frac{\alpha}{2} \tanh{\left( \frac{\beta \alpha}{2} \right)} 
    \end{pmatrix} \, .
\end{equation}

\subsection{System quench}\label{sec:systemquench}
During the system quench, the system-Hamiltonian parameter $\epsilon$ in \cref{eq:toyham} changes from $\epsilon_{\mi}$ to $\epsilon_\f$ [see \Cref{tab:quench} and \cref{fig:quench}(a)]. Let us define $a_+^{\mi} \coloneqq a_+ \lvert_{ \epsilon=\epsilon_{\mi} }$. We analogously define other quantities that carry $\mi$ or $\f$ superscripts.

\Cref{eq:wtotoperator,eq:whstaroperator,eq:westaroperator} show the three work definitions. The first evaluates to
\begin{equation}\label{eq:qp1wtot}
    W_{\tot} =  \frac{\left( \epsilon_{\mi} - \epsilon_{\f}\right)}{Z_{\sur}^{\mi}} \left[ \frac{a_+^{\mi}}{\eta_+^{\mi}} e^{-\beta \gamma} \sinh{\left( \beta \eta_+^{\mi} \right)}  + \frac{a_-^{\mi}}{\eta_-^{\mi}} e^{\beta \gamma} \sinh{ \left( \beta \eta_-^{\mi} \right)}\right] \, .
\end{equation}
We substitute \cref{eq:explicitpis,eq:explicithstar} into \cref{eq:whstaroperator}, with superscripts $A$ and $B$ denoting the parameter values $\epsilon_{\mi}$ and $\epsilon_{\f}$:
\begin{equation}\label{eq:qp1whstar}
    W_{H^*} = \frac{-1}{\beta Z_{\sur}^{\mi}}\left[ X_-^{\mi} \ln{\left( \frac{X_-^{\f}}{X_-^{\mi}} \right)} + X_+^{\mi} \ln{\left( \frac{X_+^{\f}}{X_+^{\mi}} \right)} \right] \ .
\end{equation}
Analogously, we calculate $W_{E^*}$ using \cref{eq:explicitpis,eq:explicitestar,eq:westaroperator}: 
\begin{equation}\label{eq:qp1westar}
    W_{E^*} = \frac{1}{Z_{\sur}^{\mi}} \Big( \partial_{\beta}X_-^{\mi} - \frac{X_-^{\mi}}{X_-^{\f}}\partial_{\beta}X_-^{\f}  + \partial_{\beta}X_+^{\mi} - \frac{X_+^{\mi}}{X_+^{\f}}\partial_{\beta}X_+^{\f} \Big) \, . 
\end{equation}

One can calculate the heat similarly:
\begin{align}\label{eq:qp1qtot}
    Q_{\tot}  & = \frac{1}{Z_{\sur}^{\f}} \left\{ 2\gamma \left[ e^{-\beta \gamma} \cosh{\left( \beta \eta_+^{\f} \right)} - e^{\beta \gamma} \cosh{ \left( \beta \eta_-^{\f} \right)} \right] -2\eta_+^{\f} e^{-\beta \gamma} \sinh{\left( \beta \eta_+^{\f} \right)} - 2\eta_-^{\f} e^{\beta \gamma} \sinh{\left( \beta \eta_-^{\f} \right)} \right\} \nonumber \\ 
    & + \frac{1}{Z_{\sur}^{\mi}} \Big\{ 2 \gamma \left[ - e^{-\beta \gamma} \cosh{\left( \beta \eta_+^{\mi} \right)} + e^{\beta \gamma } \cosh{\left( \beta \eta_-^{\mi}\right)}\right] + 2 e^{-\beta \gamma} \sinh{\left( \beta \eta_+^{\mi}\right)} \left( \frac{a_+^{\f} a_+^{\mi} + \chi^2}{\eta_+^{\mi}} \right) +  2 e^{ \beta \gamma} \sinh{\left( \beta \eta_-^{\mi}\right)} \left( \frac{a_-^{\f} a_-^{\mi} + \chi^2}{\eta_-^{\mi}} \right)\Big\} \,  , 
\end{align}
\begin{equation}
    Q_{H^*} =  \frac{1}{\beta} \Bigg[ \left( \frac{X_-^{\mi}}{Z_{\sur}^{\mi}} - \frac{X_-^{\f}}{Z_{\sur}^{\f}} \right) \ln{X_-^{\f}}  + \left( \frac{X_+^{\mi}}{Z_{\sur}^{\mi}} - \frac{X_+^{\f}}{Z_{\sur}^{\f}} \right) \ln{X_+^{\f}} \Bigg] \, ,
\end{equation}
and 
\begin{equation}\label{eq:qp1qestar}
    Q_{E^*} = \left( \frac{X_-^{\mi}}{Z_{\sur}^{\mi}} - \frac{X_-^{\f}}{Z_{\sur}^{\f}} \right) \frac{\partial_{\beta} X_-^{\f}}{X_-^{\f}} + \left( \frac{X_+^{\mi}}{Z_{\sur}^{\mi}} - \frac{X_+^{\f}}{Z_{\sur}^{\f}} \right) \frac{\partial_{\beta} X_+^{\f}}{X_+^{\f}} \, .
\end{equation}
During a system quench, the system's free energy [\cref{eq:F_S}] changes by an amount
\begin{equation}
    \Delta F_S = -\frac{1}{\beta} \ln{\left( \frac{ e^{-\beta \gamma} \cosh{ \left( \beta \eta_+^{\f} \right) } + e^{\beta \gamma} \cosh{ \left( \beta \eta_-^{\f} \right)} }{e^{-\beta \gamma} \cosh{ \left( \beta \eta_+^{\mi} \right) } + e^{\beta \gamma} \cosh{\left( \beta \eta_-^{\mi} \right)}} \right)} \, .
\end{equation}

The three internal-energy definitions lead to three entropy definitions via \cref{eq:genentropy}:
\begin{equation}
   \Delta \mathcal{S}_{\tot}
   = \Delta \mathcal{S}_{E^*} = -\beta \Biggl( \frac{\partial_{\beta} X_-^{\f} + \partial_{\beta}X_+^{\f}}{Z_{\sur}^{\f}} 
   - \frac{\partial_{\beta}X_-^{\mi} + \partial_{\beta}X_+^{\mi}}{Z_{\sur}^{\mi}} \Biggr) + \ln{\left( \frac{Z_{\sur}^{\f}}{Z_{\sur}^{\mi}}\right)} \,  ,
\end{equation}
and
\begin{equation}
   \Delta \mathcal{S}_{H^*} = \frac{X_-^{\f}}{Z_{\sur}^{\f}} \ln{\left( \frac{Z_{\sur}^{\f}}{X_-^{\f}} \right)} - \frac{X_-^{\mi}}{Z_{\sur}^{\mi}} \ln{\left( \frac{Z_{\sur}^{\mi}}{X_-^{\mi}} \right)} + \frac{X_+^{\f}}{Z_{\sur}^{\f}} \ln{\left( \frac{Z_{\sur}^{\f}}{X_+^{\f}} \right)} - \frac{X_+^{\mi}}{Z_{\sur}^{\mi}} \ln{\left( \frac{Z_{\sur}^{\mi}}{X_+^{\mi}} \right)} \, .
\end{equation}
Since $\sur$ begins and ends in a Gibbs state,
$\Delta U_{\tot} = \Delta U_{E^*} $ implies that $ \Delta S_{\tot} = \Delta S_{E^*}$ [\cref{eq:genentropy,eq:utotequestar}].

\subsection{Interaction quench}
During an interaction quench, the system-reservoir interaction changes abruptly [see \Cref{tab:quench} and \cref{fig:quench}(b)]. The coupling parameters, $\gamma$ and $\chi$, change from zero to $\gamma_{\f}$ and $\chi_{\f}$ at $t=0$. Subsequently, $\sur$ evolves to a Gibbs state of the Hamiltonian $\h_\sur^B$ [\cref{eq:toyham}]. The initial state $\pio_S^0 \otimes \pio_R^0$ contains the factors
\begin{equation}\label{eq:qp2piso}
     \pio_S^0 = \frac{1}{2 \cosh{ \left( {\beta \epsilon}/{2} \right) }} \begin{pmatrix}
        \epsilon /2 & 0 \\
        0 & -\epsilon /2
        \end{pmatrix} \,
\end{equation}
and 
\begin{equation}\label{eq:qp2piro}
    \pio_R^{0} = \frac{1}{2 \cosh{ \left( {\beta \alpha}/{2} \right) }} \begin{pmatrix}
        \alpha /2 & 0  \\
        0 & -\alpha /2
        \end{pmatrix} \, ,
\end{equation}
in terms of the $\sigma_S^{z}$ and $\sigma_R^{z}$ eigenbases, respectively. At $t=0^-$, $S$ and $R$ do not interact; hence $\gamma_{\mi} = \chi_{\mi} = 0$, and $\hsoi = \esoi = \h_S^{\mi}$ (see \cref{app:weakcoupling}). When $t > 0$, $\hsof$ and $\esof$ have the forms in \cref{eq:explicithstar,eq:explicitestar}, with $\gamma_{\f}, \chi_{\f} \neq 0$.~\Cref{eq:explicitpisur} specifies the final $\sur$ state, $\pio_{\sur}^\f$; and \cref{eq:explicitpis}, the final $S$ state, $\pio_S^\f$. The final total partition function is
\begin{equation}
    Z_{\sur}^{\f} = 2 e^{-\beta \gamma} \cosh{ \left( \beta \eta_1 \right)} + 2 e^{\beta \gamma} \cosh{ \left(\beta \eta_2 \right)} \, . 
\end{equation} 
Here and below, to prevent clutter, we drop subscripts and denote the coupling parameters by
$\gamma_{\f}=\gamma$ and $\chi_{\f}=\chi$. Evaluating \cref{eq:wtotoperator,eq:whstaroperator,eq:westaroperator}, we arrive at the following expressions for work:
\begin{equation}\label{eq:qp2wtot}
    W_{\tot} = \frac{\epsilon}{2} \tanh{\left( \frac{\beta \epsilon}{2} \right)} + \frac{\alpha}{2} \tanh{\left( \frac{\beta \alpha}{2} \right)} - \frac{2 a_+ \sinh{\left( \beta a_+ \right)} - 2 \gamma \cosh{\left(\beta a_+\right)} + 2 a_- \sinh{\left(\beta a_-\right)} + 2 \gamma \cosh{\left(\beta a_-\right)}}{ 4 \cosh{\left( \beta \epsilon /2 \right)} \cosh{\left( \beta \alpha /2 \right)}} \, , 
\end{equation}
\begin{equation}\label{eq:qp2whstar}
    W_{H^*} = -\frac{1}{\beta} \Bigg[ \frac{e^{-\beta \epsilon /2} \ln{({X_-})} + e^{\beta \epsilon /2} \ln{({X_+})}}{2 \cosh{\left( \beta \epsilon /2\right)}}  - \ln{\left( 2 \cosh{\left( \frac{\beta \alpha}{2}  \right)} \right)} \Bigg] 
    + \frac{\epsilon}{2} \tanh{\left( \frac{\beta \epsilon}{2} \right)} \, ,
\end{equation}
and
\begin{equation}\label{eq:qp2westar}
    W_{E^*} = \frac{\epsilon}{2} \tanh{\left( \frac{\beta \epsilon}{2} \right)} + \frac{\alpha}{2} \tanh{\left( \frac{\beta \alpha}{2} \right)} -\frac{1}{2 \cosh{\left( \beta \epsilon /2 \right)}} \left[ e^{-\beta \epsilon /2} \left( \frac{\partial_{\beta} {X_-}}{{X_-}}\right) + e^{\beta \epsilon /2} \left( \frac{\partial_{\beta} {X_+}}{{X_+}}\right)\right] \, .
\end{equation}

Similarly, heat expressions follow from \cref{eq:qtotoperator,eq:qhstaroperator,eq:qestaroperator}:
\begin{align}
    Q_{\tot} & = \; \frac{1}{Z_{\sur}^{\f}} \left\{ 2 \gamma \left[ e^{-\beta \gamma} \cosh{\left( \beta \eta_+\right)} - e^{\beta \gamma} \cosh{\left(\beta \eta_-\right)} \right] - 2 \eta_+ e^{-\beta \gamma} \sinh{\left(\beta \eta_+\right)} - 2 \eta_- e^{\beta \gamma} \sinh{\left(\beta \eta_-\right)} \right\} \nonumber \\ 
    &\qquad \qquad \: + \frac{2 a_+ \sinh{\left(\beta a_+\right)} - 2 \gamma \cosh{\left(\beta a_+\right)} + 2 a_- \sinh{\left(\beta a_-\right)} + 2 \gamma \cosh{\left(\beta a_-\right)}}{ 4 \cosh{\left( \beta \epsilon /2 \right)} \cosh{\left( \beta \alpha /2 \right)}} \, ,
\end{align}
\begin{equation}
    Q_{H^*} = \frac{1}{\beta} \Biggl\{ \;\left[ \frac{e^{-\beta \epsilon /2}}{2 \cosh{\left( \beta \epsilon /2 \right)}} - \frac{{X_-}}{Z_{\sur}^{\f}} \right] \ln{({X_-})} + \left[ \frac{e^{\beta \epsilon /2}}{2 \cosh{\left( \beta \epsilon /2 \right)}} - \frac{{X_+}}{Z_{\sur}^{\f}} \right] \ln{({X_+})} \Biggl\} \, ,
\end{equation}
and
\begin{equation}
    Q_{E^*} = \left( \frac{e^{-\beta \epsilon / 2}}{2 {X_-} \cosh{\left(\beta \epsilon / 2\right)}} - \frac{1}{Z_{\sur}^{\f}}\right)\partial_{\beta}{X_-} +\left( \frac{e^{\beta \epsilon / 2}}{2 {X_+} \cosh{\left(\beta \epsilon / 2\right)}} - \frac{1}{Z_{\sur}^{\f}}\right)\partial_{\beta}{X_+} \, .
\end{equation}

The system's free energy changes by an amount
\begin{equation}
    \Delta F_S = -\frac{1}{\beta} \ln{\left( \frac{e^{-\beta \gamma} \cosh{ \left(\beta \eta_+ \right)} + e^{\beta \gamma} \cosh{\left(\beta \eta_-\right)} }{2 \cosh{\left( \beta \epsilon /2 \right)} \cosh{\left( \beta \alpha /2 \right)}}  \right)} \, .
\end{equation}

We compute the entropy change as we did for a system quench process (\Cref{sec:systemquench}), obtaining
\begin{equation}
    \Delta \mathcal{S}_{\tot} = \Delta \mathcal{S}_{E^*} = \frac{\beta \epsilon}{2} \tanh{\left( \frac{\beta \epsilon}{2} \right)} + \frac{\beta \alpha}{2} \tanh{\left( \frac{\beta \alpha}{2} \right)} - \frac{\beta}{Z_{\sur}^{\f}} \left( \partial_{\beta}{X_-} + \partial_{\beta}{X_+} \right) + \ln{\left[ \frac{Z_{\sur}^{\f}}{4 \cosh{\left(\beta \epsilon /2\right)} \cosh{\left(\beta \alpha /2\right)}} \right]} \, 
\end{equation}
and
\begin{equation}
    \Delta \mathcal{S}_{H^*} = \frac{\beta \epsilon}{2} \tanh{\left( \frac{\beta \epsilon}{2} \right)} -\frac{1}{Z_{\sur}^{\f}} \left( {X_-} \ln{{X_-}} + {X_+} \ln{{X_+}} \right) + \ln{\left[ 2\cosh{\left( \frac{\beta \alpha}  {2}\right)} \right]} + \ln{\left[ \frac{Z_{\sur}^{\f}}{4 \cosh{\left( \beta \epsilon /2 \right)} \cosh{\left( \beta \alpha /2 \right)}}\right]} \, .
\end{equation}

\subsection{LGT-type two-spin model}

In \cref{sec:spinmodel}, we introduced an LGT-type two-spin model [\cref{eq:toylgtham}]. One can generalize to this model~\Cref{sec:systemquench}'s expressions for work, heat, free energy, and entropy in the context of a system quench. One modifies the interaction term from $\gamma \sigma_{S}^{z} \sigma_{R}^{z}  + \chi \sigma_{S}^{x} \sigma_{R}^{x}$ to $ k ( \mathbbm{1}_{\sur} - \sigma_{S}^{z} \sigma_{R}^{z} ) + \chi \sigma_{S}^{x} \sigma_{R}^{x}$ and takes the limit as $k \rightarrow \infty$. \Cref{eq:toylgtham} shows $\hat{H}_{\sur}$, in which the system term has a strength $\epsilon/2$. Consider switching $\epsilon$ instantaneously from $\epsilon_{\mi}=0$ to $\epsilon_{\f}$ [\cref{fig:quench}(a)]. $\alpha$ and $\chi$ remain fixed. 

\begin{figure}[btp]
   \centering
   \includegraphics[scale=0.75]{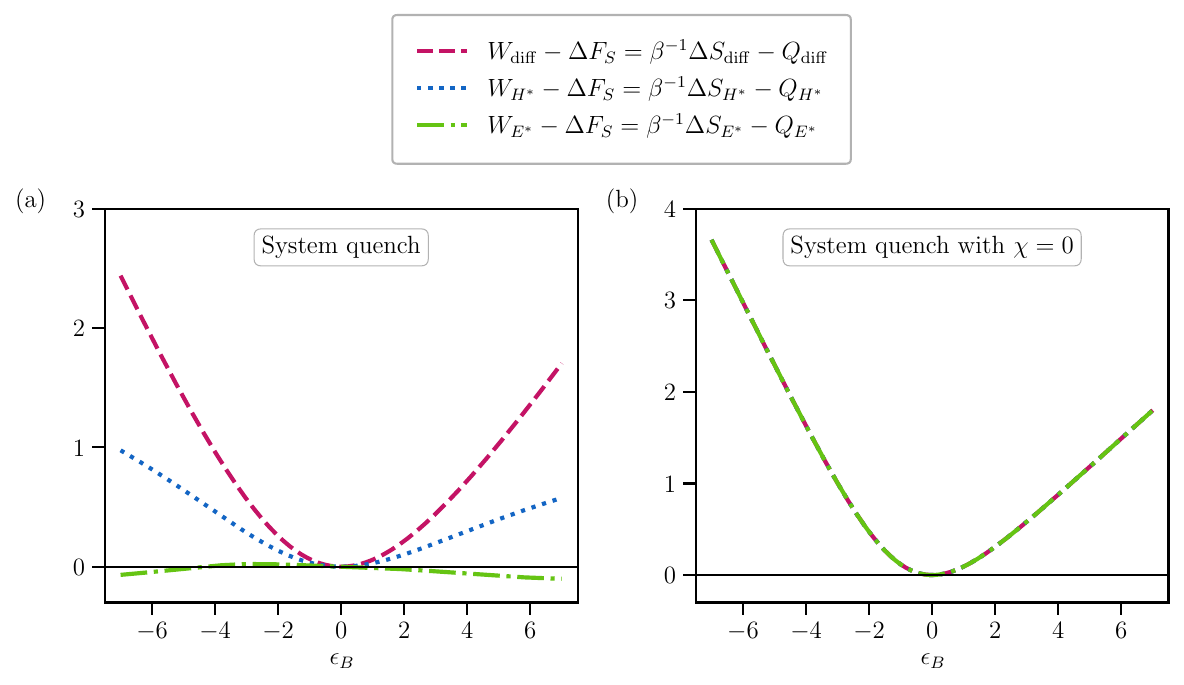}
    \caption{\textit{Thermodynamic quantities of the LGT-type two-spin model undergoing a system-quench process.} The $\epsilon$ in \cref{eq:toylgtham} changes instantaneously from $\epsilon_{\mi} =0$ to $\epsilon_{\f}$. We fix $\alpha=0.8$ and $\chi = 1.8$. The dissipated work is plotted as a function of $\epsilon_{\f}$. (a) During a system quench, $W_{\tot}$ and $W_{H^*}$ satisfy the second law [\cref{eq:genwsecondlaw}]; consequently, so do $Q_{\tot}$ and $Q_{H^*}$ [\cref{eq:genqsecondlaw}]. $W_{E^*}$ and $Q_{E^*}$ do not. (b) We set $\chi = 0$ in~\cref{eq:toylgtham}, such that $\h_S$ commutes with $\hat{V}_{\sur}$ before and after the quench [Eqs.~\eqref{eq:commutationrels}]. The work dissipated during the system quench is plotted as a function of $\epsilon_{\f}$. The three curves coincide, as expected [\cref{eq:WWW}].}
    \label{fig:lgt_toy_quench}
\end{figure}

\Cref{fig:lgt_toy_quench} (a) displays the dissipated work as a function of $\epsilon_{\f}$. The plot shows that $W_\tot$ and $W_{H^*}$ obey the second law, while $W_{E^*}$ does not. However, when $\chi=0$, two behaviors stand out. First, all three dissipated-work quantities equal each other, in agreement with \cref{eq:WWW}. Second, the quantities obey the second law.

\end{document}